\RequirePackage{fix-cm}

\documentclass[twocolumn]{svjour3}          
\smartqed  
\usepackage{graphicx}
\usepackage[utf8]{inputenc}
\usepackage[english]{babel}
\usepackage{makeidx}
\usepackage{graphicx}
\usepackage[hidelinks]{hyperref}

\usepackage{amsmath}
\usepackage[inline]{enumitem}
\usepackage{multirow}
\usepackage{subcaption}
\usepackage{caption}
\usepackage{color}
\usepackage{booktabs}
\usepackage{siunitx}
\usepackage{algpseudocode}
\usepackage{algorithmicx}
\usepackage{array}
\usepackage{float}
\usepackage{todonotes}
\usepackage[linesnumbered,ruled,vlined]{algorithm2e}
\usepackage{amsfonts}
\usepackage{threeparttable}
\usepackage{ragged2e}
\usepackage{etoolbox}
\usepackage{natbib}
\usepackage[normalem]{ulem}
\DeclareUnicodeCharacter{2265}{ }
\DeclareUnicodeCharacter{03BC}{ }
\usepackage{xcolor}

\usepackage{multirow}
\usepackage{colortbl}
\usepackage{array}

\usepackage{hyperref}
\usepackage{balance}
\usepackage{arydshln}

\usepackage{color}
\usepackage{tabularray}
\usepackage[utf8]{inputenc}

\usepackage{flushend}

\newcommand{\RVV}[1]{{\color{blue}{#1}}}

\makeatletter 
\pretocmd\@bibitem{\color{black}\csname keycolor#1\endcsname}{}{\fail}
\newcommand\citecolor[1]{\@namedef{keycolor#1}{\color{black}}}
\makeatother
\citecolor{_4}
\citecolor{opixray2}
\citecolor{wang1}
\citecolor{wang2}
\citecolor{najla1}
\citecolor{najla2}

\begin{document}

\title{A Comprehensive Review of Artificial Intelligence Applications in Major Retinal Conditions}


\author{Hina Raja \and
        Taimur Hassan \textsuperscript{$\star$} \and
        Bilal Hassan \and
        Muhammad Usman Akram \and
        Hira Raja \and
        Alaa A Abd-alrazaq\and
        Siamak Yousefi \and
        Naoufel Werghi
}

\institute{ Hina Raja \at Department of Ophthalmology, University of Tennessee Health Science 
              Center, Memphis, USA \\
               \email{h.raja@uthsc.edu} \\ \and
             Taimur Hassan (\textsuperscript{$\star$}Corresponding Author) \at
              Department of Electrical, Computer, and Biomedical Engineering, Abu Dhabi University, Abu Dhabi, UAE \\
              \email{taimur.hassan@adu.ac.ae}
              \and
           Bilal Hassan \at
              Department of Electrical Engineering and Computer Science, Khalifa University, Abu Dhabi, United Arab Emirates. \\
              \email{bilal.hassan@ku.ac.ae}           
           \and          
           Muhammad Usman Akram \at Department of Computer and Software Engineering, National University of Sciences and Technology, Pakistan\\
           \email {usman.akram@ceme.nust.edu.pk}
              \and
           Hira Raja \at Margalla Institute of Health Sciences, Rawalpindi, Pakistan\\
           \email{dhanyalhira@gmail.com}
              \and
            Alaa A. Abd-alrazaq \at AI Center for Precision Health, Weill Cornell Medicine-Qatar, Doha, Qatar\\ 
           \email{aaa4027@qatar-med.cornell.edu}
              \and
             Siamak Yousefi \at Department of Ophthalmology, University of Tennessee Health Science 
              Center, Memphis, USA \\
               \email{siamak.yousefi@uthsc.edu}
              \and 
              Naoufel Werghi \at Center for Cyber-Physical Systems (C2PS), Department of Electrical Engineering and Computer Science, Khalifa University, Abu Dhabi, United Arab Emirates\\
              \email{naoufel.werghi@ku.ac.ae} 
}

\date{Received: date / Accepted: date}

\maketitle

\begin{abstract}
{This paper provides a systematic survey of retinal diseases that cause visual impairments or blindness, emphasizing the importance of early detection for effective treatment. It covers both clinical and automated approaches for detecting retinal disease, focusing on studies from the past decade. The survey evaluates various algorithms for identifying structural abnormalities and diagnosing retinal diseases, and it identifies future research directions based on a critical analysis of existing literature. This comprehensive study, which reviews both clinical and automated detection methods using different modalities, appears to be unique in its scope. Additionally, the survey serves as a helpful guide for researchers interested in digital retinopathy.}
\end{abstract}


\section{Introduction}
\begin{figure}[t]
	\centering
	\includegraphics[width=1\linewidth]{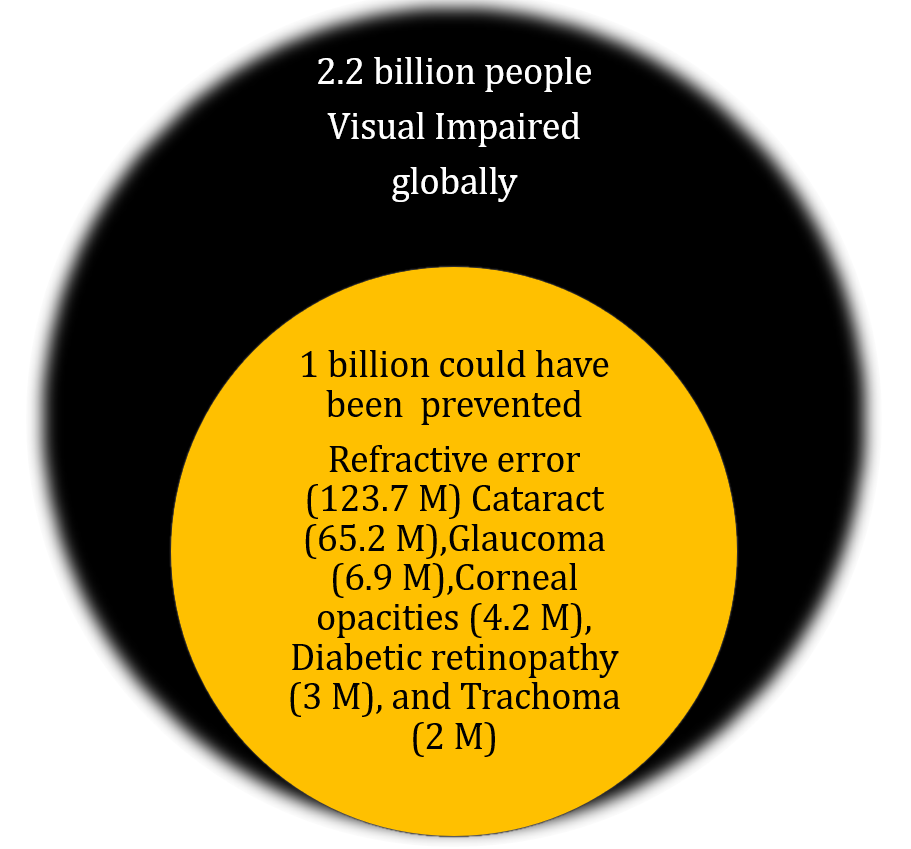}
	\caption{\small Global prevalence of people with vision impairment and preventable diseases, as depicted by WHO statistics.}
	\label{f}
\end{figure}

\noindent Vision is the most important of all the senses and plays a key part in everyone's life. According to a report by the World Health Organization (WHO) in 2012, the number of individuals with visual impairment (VI) was estimated to be 285 million \RVV{\citep{ref3a}}. Among the total population of 285 million individuals, a significant proportion of 246 million were observed to have low vision (LW), while 39 million were identified as being blind. The population of visually impaired individuals experienced an increase to 2.2 billion in the year 2019, with approximately one billion cases potentially preventable through timely and accurate diagnosis \RVV{\citep{ref4}} (Figure \ref{f}). The leading causes of vision impairment and blindness are retinal diseases, which include trachoma (2 M), diabetic retinopathy (3 M), corneal opacities (4.2 M), glaucoma (6.9 M), cataract (65.2 M), and refractive error (123.7 M). The retinal disease slowly progresses; initially, there are often no symptoms, so it is difficult for most of the subjects to detect the early symptoms. According to the report, fifty percent of the patients in the United States were found to be uninformed about their ocular condition \RVV{\citep{tham2014global}}. As the early symptoms are left undetected so these VI subjects lead towards permanent blindness. The goal of the global community is to uplift eye care facilities by providing early detection of retinal diseases to halt the progression through appropriate treatment in order to preserve vision. Due to the high correlation between retinal diseases and blindness, a lot of clinical research is going on for screening retinal diseases in the early stages. Due to the subjective nature and time expenditure of manually extracting the retinal lesion, automated methods are used to aid ophthalmologists. Improvements to these automated procedures are the subject of ongoing research. 
\noindent In this paper, we are discussing clinical pathophysiologies of retinal diseases and various modalities which have been used to detect them. Optical coherence tomography (OCT), Optical coherence tomography angiography (OCTA), color fundus photography (CFP), fundus fluorescein angiography (FFA), and ultrawide-field photography (UWFP) modalities have been used for diagnosis of retinal abnormalities. CFP has been widely employed for the initial screening of various retinal abnormalities. However, OCT is extensively used by ophthalmologists for the detection of ocular disorders and to monitor their progression. In addition to this, a comprehensive review of automated studies related to the detection of retinal diseases is presented.

\subsection{Existing Surveys}  
\noindent Retinal image analysis is a wide research area where clinicians and researchers work together to devise techniques for the early detection of retinopathies. We critically reviewed survey studies related to retinal analysis in the context of clinical and automated literature. However, most of the review articles are either disease-specific or modality-specific. The next section summarizes each category of review papers found in the literature.

\subsubsection{Disease-specific survey papers}
\noindent Nicholson et al. \RVV{\citep{nicholson2013central}} performed a systematic review to analyze the pathogenesis of central serous chorioretinopathy (CSR). The study \RVV{\citep{Khalil2014}} presented a survey to detect glaucoma changes through fundus images. Preprocessing, feature extraction, feature selection, and machine learning (ML) techniques were discussed. Gupta et al. \RVV{\citep{gupta2015survey}} reported a survey study that analyzes the automated techniques for DR diagnosis. A comparison was made between the algorithm that detects the various structural changes through fundus images. A total of 13 studies were included in the survey. The review study \RVV{\citep{das2016evidence}} was presented in the literature that analyzed diabetic macular edema (DME) management in Indian subjects. Muramatsu et al. \RVV{\citep{muramatsu2018survey}} presented the survey for the treatment of DME in Japanese subjects. The clinical and technical review was presented for glaucoma diagnosis through fundus and OCT images \RVV{\citep{Naveed}}. Pead et al.\RVV{\citep{pead2019review}} presented a review study that evaluated the ML and deep learning (DL) techniques for automated drusen detection in the context of AMD. The paper included only those studies which detected the drusen in color fundus photography. A total of 14 articles were reviewed and only compared the ML and DL methods presented in those studies. Araki et al. \RVV{\citep{araki2019central}} presented a survey to analyze the effect of steroids on Japanese CSR subjects. Another clinical review \RVV{\citep{van2019central}} investigated the different treatments related to CSC, which included photodynamic therapy, laser treatment, and pharmacology. The survey \RVV{\citep{lakshmi2021survey}} was conducted over the period of five years, from 2016 to 2021, to investigate the automated techniques, which includes ML and DL approaches, for the detection of DR in fundus and OCT images. A total of 114 papers were comprehensively reviewed from the open literature. Another review study \RVV{\citep{abdullah2021review}} was presented that compared the automated ML techniques to detect the structural changes in fundus images. In addition to this, the author discussed the various fundus-related datasets (public and private). Rubina et al. \RVV{\citep{sarki2020automatic}} comprehensively reviewed state-of-the-art approaches for the detection of diabetic and glaucomatous changes through fundus images. Image processing, ML, and DL techniques were explored. The author also reported available datasets. 
The paper \RVV{\citep{bala2021review}} presented the clinical and technical survey for glaucoma diagnosis. It reported  DL techniques for detecting pathological changes in fundus and OCT images. The study \RVV{\citep{shahriari2022artificial}} discussed how artificial intelligence (AI) is being used to screen, diagnose, and categorize DME.

\subsubsection{Modality-specific survey papers}
\begin{figure}[t]
	\centering
	\includegraphics[width=1\linewidth]{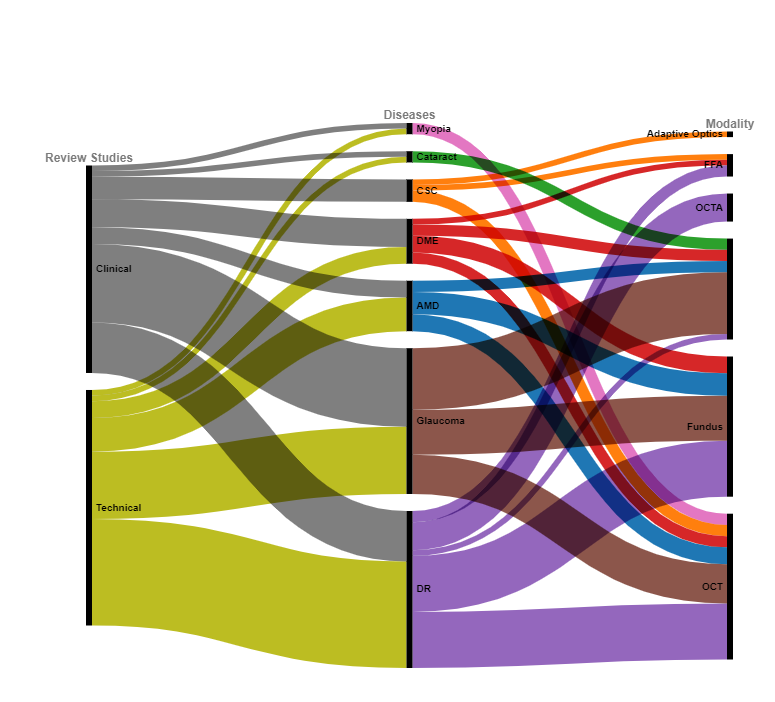}
	\caption{Chord diagram showing the interactions between categories in a systematic review of clinical and technical studies across multiple diseases and modalities. The diagram highlights the relationships between subcategories of diseases (e.g., glaucoma, DR, AMD, DME, Cataract, and CSR) and modalities (e.g., Fundus, OCT, OCTA, Adaptive optics, and FFA). The thickness of the chords represents the strength of the connections between the categories, with thicker chords indicating stronger connections." }
	\label{v1}
\end{figure}

\noindent Michael et al. \RVV{\citep{Abramoffreview}} presented a clinical review of the retinal imaging trends.  Besides this, the paper summarized the most prevalent causes of blindness, which include AMD, DR, and glaucoma. The review was about 2-D fundus imaging and 3-D OCT imaging techniques. Another study \RVV{\citep{das2018survey}} in the literature presented the clinical review of fundus images for detecting retinal diseases. The study \RVV{\citep{kafieh2013review}} is modality specific, where image segmentation methods were reviewed for processing the retinal OCT images.  The OCT segmentation approaches were classified into five categories, such A.scan, B.scan, active contour, AI methods, 3D graphs, and 3D OCT volumetric. Baghaie et al. \RVV{\citep{baghaie2015state}} reported the major issues related to OCT image analysis. More specifically, different techniques for noise reduction, image segmentation, and registration were discussed. Usman et al. \RVV{\citep{usman2017computer}} provided an exhaustive review of various class image processing and computer vision techniques for detecting glaucoma, DR, and pathological myopia. The authors also reported the causes, symptoms, and pathological alterations of these diseases in OCT images, which can aid in the development of an automated system for the detection of retinal disorders. The precision of algorithms determined performance after an exhaustive examination and evaluation of various methods. Khan \RVV{\citep{khan2019review}} presented a survey that is also modalities-specific. The survey comprehends the automated techniques for extracting retinal vessels in fundus images. The techniques are categorized into supervised and unsupervised groups. Supervised approaches are further classified into ensemble classification and neural network-based approaches. However, unsupervised techniques are grouped into four classes: matched filtering, mathematical morphological, multi-scale-based techniques, and region-growing methods. A valuable comparison was made among the techniques which were reported on the publicly available datasets. In the article \RVV{\citep{nuzzi2021impact}}, a clinical review  was reported on the state-of-the-art applications for AI in ophthalmology, which helps clinicians to have an overview of growing trends. The paper \RVV{\citep{badar2020application}} was modality specific, focusing on DL techniques for retinal analysis through fundus images. The review includes automated disease classification methods based on retinal pathological landmarks. The methods were evaluated using accuracy, F score, sensitivity, specificity, and area under ROC curve on publicly available datasets Angiography has gained popularity in the field of ophthalmology for the diagnosis of ocular diseases. Boned \RVV{\citep{boned2022optical}} presented the survey study related to OCTA in diabetes subjects. Deep learning techniques were reported for the detection of retinal vascularization. Stolte et al. \RVV{\citep{stolte2020survey}} performed a comprehensive survey for DR diagnosis covering the clinical and technical aspects. The paper also described the publicly available datasets of the fundus and OCT modalities. In addition to this, ML and DL frameworks were reviewed for the detection and classification of DR. However, fundus-related literature was more critically reviewed as compared to OCT and OCTA modalities. A systematic review of clinical and technical studies across many disorders and modalities, represented as a chord diagram (Figure \ref{v1}) illustrating the relationship between different categories.

\noindent There are other review studies reported in the literature that are either modality specific \RVV{\citep{amini2016classification} \citep{salehi2022retinal} \citep{raja2022glaucoma} \citep{im2022prevalence} \citep{ye2023applications} \citep{vujosevic2023optical} \citep{khalili2023optical} \citep{pavithra2023computer}}, or disease-specific \RVV{\citep{keenan2022deep}\citep{jimenez2022validation} \citep{aboobakar2022genetics}\citep{ong2022perspectives} \citep{mauschitz2022age} \citep{chou2022screening} \citep{patel2022recent} \citep{fea2022glaucoma}\citep{yuksel2022ophthalmic} \citep{wang2022genetic} \citep{huang2022hyperreflective}\citep{chauhan2022current} \citep{srivastava2023artificial} \citep{iannucci2023childhood} \citep{saeed2023preserflo}}, but a single study is not to be considered as a comprehensive survey for retinal analysis of various diseases. 

\begin{table*}
\caption{Summary of review studies related to retinal diseases. The abbreviation are: CL:Clinical, TEC:Technical,  CVAS: Cardiovascular, SJG: Surgery, MAG: Management NeuroD: Neurodegeneration, TRT: Treatment, PMED: Precision Medicine, CAT: Cataract, and GLA: Glaucoma }
\label{tab1} 
\centering
\begin{tabular}{lllll}
\hline\noalign{\smallskip}
Study & Modality &Timeline  & Disease & Category \\
\noalign{\smallskip}\hline\noalign{\smallskip}

\RVV{\cite{Abramoffreview}}  & Fundus, OCT & till 2010  & Glaucoma, AMD, DR &  TEC\\

\RVV{\cite{kafieh2013review}}  & OCT & 1997 to 2012  & - & CL, TEC \\

\RVV{\cite{nicholson2013central}}& FFA, OCT, Adaptive Optics & till 2012  & CSR & CL \\

\RVV{\cite{gupta2015survey}} & Fundus & 2009-2013  & DR & TEC \\

\RVV{\cite{baghaie2015state}}& OCT & 1980 to 2015  & - &  TEC \\

\RVV{\cite{das2016evidence}} & FFA, OCT & till 2015 & DME & CL   \\

\RVV{\cite{usman2017computer}}  & OCT & till 2015 & GLA,DR, Myopia & CL, TEC \\

\RVV{\cite{muramatsu2018survey}} & - & till 2017   & DME & CL   \\

\RVV{\cite{khan2019review}} & Fundus & till 2017 & GLA,DR,AMD & TEC \\

\RVV{\cite{pead2019review}} & Fundus & 2018   & AMD & TEC    \\

\RVV{\cite{araki2019central}} & - & till 2018   & CSR & CL   \\

\RVV{\cite{van2019central}} & - & till 2019   & CSR & CL   \\

\RVV{\cite{stolte2020survey}} &  Fundus, OCT, OCTA & till 2020 &  DR & CL, TEC \\

\RVV{\cite{lakshmi2021survey}}&  Fundus, OCT &2016-2021 &  DR & TEC \\

\RVV{\cite{abdullah2021review}} & Fundus & till 2021  & GLA & TEC \\

\RVV{\cite{sarki2020automatic}}& Fundus & 2016-2020  & GLA, DR & TEC \\

\RVV{\cite{bala2021review}} & Fundus, OCT & till 2021  & GLA  & CL, TEC \\

\RVV{\cite{badar2020application}} & Fundus & till 2018  & GLA, AMD, DR  &  TEC \\

\RVV{\cite{nuzzi2021impact}}&  - & till 2021  & GLA, AMD, CAT, DR  &  CL \\

\RVV{\cite{boned2022optical}} & OCTA & till 2021  &  DR  &  TEC \\

\RVV{\cite{elsharkawy2022role}} & Fundus,FFA, OCT, OCTA & till 2021  &  DR &  CL, TEC \\

\RVV{\cite{salehi2022retinal}}& OCT & till 2021  &  AMD &   CL \\
\RVV{\cite{chou2022screening}} & - &  2021  &  GLA &   CL \\
\RVV{\cite{aboobakar2022genetics}} & - &  2021  &  GLA &   CL (Gene) \\
\RVV{\cite{shahriari2022artificial}} & Fundus, OCT & till July 2021  &  DME &  TEC \\
\RVV{\cite{mauschitz2022age}} & - & Oct 2021  &  AMD &   Clinical \\
\RVV{\cite{patel2022recent}} & - &  2021  &  GLA &   Clinical (TRT) \\
\RVV{\cite{raja2022glaucoma}} & OCT & till 2021  &  GLA  &  CL, TEC \\
\RVV{\cite{fea2022glaucoma}} & - &  2021  &  GLA &   CL (TRT) \\
\RVV{\cite{yuksel2022ophthalmic}}  & - &  2020  &  GLA &   CL\\
\RVV{\cite{wang2022genetic}} & - &  2020  &  GLA &   CL (Gene)\\
\RVV{\cite{huang2022hyperreflective}}& - &  2021  &  DME &   CL (TRT) \\
\RVV{\cite{im2022prevalence}}& - &  2021  &  DME &   CL \\
\RVV{\cite{chauhan2022current}}& - &  2021  &  DME &   CL (TRT) \\
\RVV{\cite{pavithra2023computer}} & Fundus &2022 & DME & TEC\\
\RVV{\cite{srivastava2023artificial}}&-& 2022& GLA,DR,AMD, CAT& TEC\\
\RVV{\cite{sharma2023developments}} &-& 2021& GLA &   CL (MAG)\\
\RVV{\cite{iannucci2023childhood}}&-& -& Uveitic GLA &   CL (MAG)\\
\RVV{\cite{fea2023precision}} &-& Aug 2022 & GLA &   CL (PMED)\\

\RVV{\cite{saeed2023preserflo}} &-& - & GLA &   CL (SJG)\\
\RVV{\cite{ye2023applications}}&OCT, Fundus& - & DR GLA & TEC  \\
\RVV{\cite{vujosevic2023optical}} &OCT& - &-  & CL (NeuroD)  \\
\RVV{\cite{khalili2023optical}} &OCTA & Nov 2020&- & CL \\
\noalign{\smallskip}\hline
\end{tabular}
\end{table*}

\vspace{-0.3cm}
\subsection{Our contribution}
\noindent Following are the major contributions of this survey.
\begin{enumerate}

\item To the best of our knowledge, this is the first study that presents a comprehensive review of clinical and automated literature related to various retinal diseases, which are major causes of blindness.

\item This paper presents a detailed survey of fundus and OCT modalities for retinal analysis, reporting almost all the significant aspects.

\item To the best of our knowledge, this review is the novel that targets most retinal diseases such as AMD, glaucoma, DR, and DME.

\end{enumerate}

\noindent The rest of the paper is organized as section II presents the methods used for conducting this review study, section III describes the anatomy of the human eye, and section IV presents the significance of retinal diseases toward blindness. Section V discusses the different modalities used to detect retinal disorders. Sections VI and VII highlight the pathological changes and clinical severity of major retinal diseases, respectively. The clinical literature related to retinal disorders is discussed in section VIII. Whereas section IX presents a review of automated algorithms for detecting different biomarkers and classification of retinal disorders based on image processing techniques, ML, and DL techniques. Section X discusses the advanced DL techniques and related work found in the literature. The publicly available dataset for fundus and OCT have summarised in section XI. Finally, future directions are discussed in section XI.

\section{Methods}
\subsection{Timeline}
\noindent We included studies published only in the last decade (i.e., January 2013- January 2023).
\vspace{-0.3cm}
\subsection{Eligibility Criteria}
\noindent The following descriptions serve as eligibility criteria for the studies that were included in the review: (1) studies include clinical and experimental findings related to ocular diseases (diabetic retinopathy (DR), glaucoma, age-related macular degeneration (AMD), macular edema (ME)), (2) studies presented an automated algorithm for detection of different retinal lesions, (3) segmentation and classification techniques, (4) image processing techniques, classic machine learning techniques, deep learning models, (5) modalities fundus, FFA, OCT, and OCTA). The main exclusion criteria were studies that were not related to the above-mentioned disease and data from conference reports, communications, or letters.
\vspace{-0.3cm}
\subsection{Search Strategy}
\noindent The studies were retrieved from secondary sources, which included internet sources, reports,  conferences, and journal articles. Sampling is purposive, aimed at having a review study on diagnosing and detecting the retinal lesion. The review was performed by searching articles on retinal diseases from PubMed, Science Direct, IEEE Xplore Digital Library, Springer Link, and Google Scholar. The search was carried out with a combination of different keywords, which included retinal diseases, diabetic retinopathy (DR), glaucoma, age-related macular degeneration (AMD), macular edema (ME), automated detection, machine learning model, deep learning model, and advanced deep learning techniques. The search criteria were intentionally broad in order to encompass all studies that could potentially meet the eligibility criteria. Original peer-reviewed articles were considered regardless of publication status. 
\vspace{-0.3cm}
\subsection{Study selection process}
\begin{figure}[h]
	\centering
	\includegraphics[width=1\linewidth]{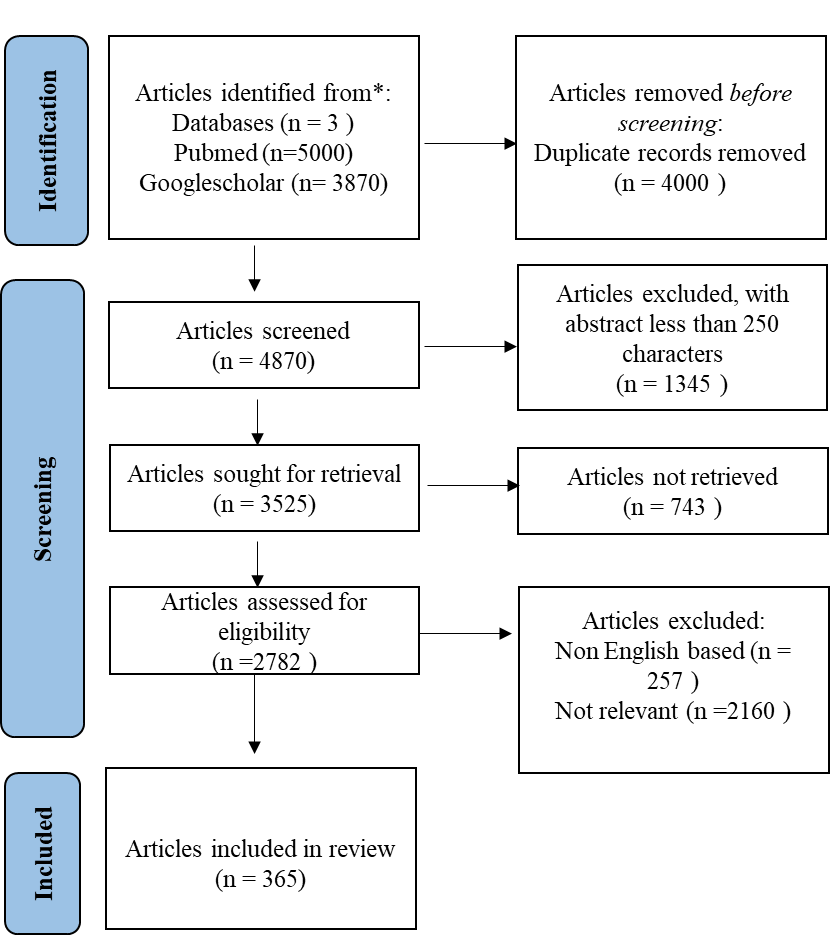}
	\caption{\small PRISMA Flowchart for the Inclusion of studies in a  review study of retinal diseases. The flowchart outlines the process of study selection and inclusion in the review, including the number of articles retrieved, screened, excluded, and included at each step.}
	\label{PRISMA}
\end{figure}

\noindent Articles found in the primary search were evaluated for eligibility to be included in the review based on their relevance to the research question or topic. We follow the review protocol of Systematic Reviews and Meta-analysis (PRISMA), shown in table \ref{PRISMA}. 
\vspace{-0.3cm}
\subsection{Data extraction and Synthesis}
\noindent Following a thorough reading and summary of the chosen publications, the key points and arguments from each were extracted and synthesized in a separate file. The following findings were extracted from selected studies: pathological association from clinical or experimental articles, techniques, datasets, and results from technical literature. Results include evaluation metrics such as accuracy, sensitivity, specificity, precision, recall, F1 score, and intersection over union. However, some of the discussed studies have limited data, as we could not access the full paper.

\section{Eye anatomy}
\begin{figure}[t]
	\centering
	\includegraphics[width=8cm,height=10cm,keepaspectratio]{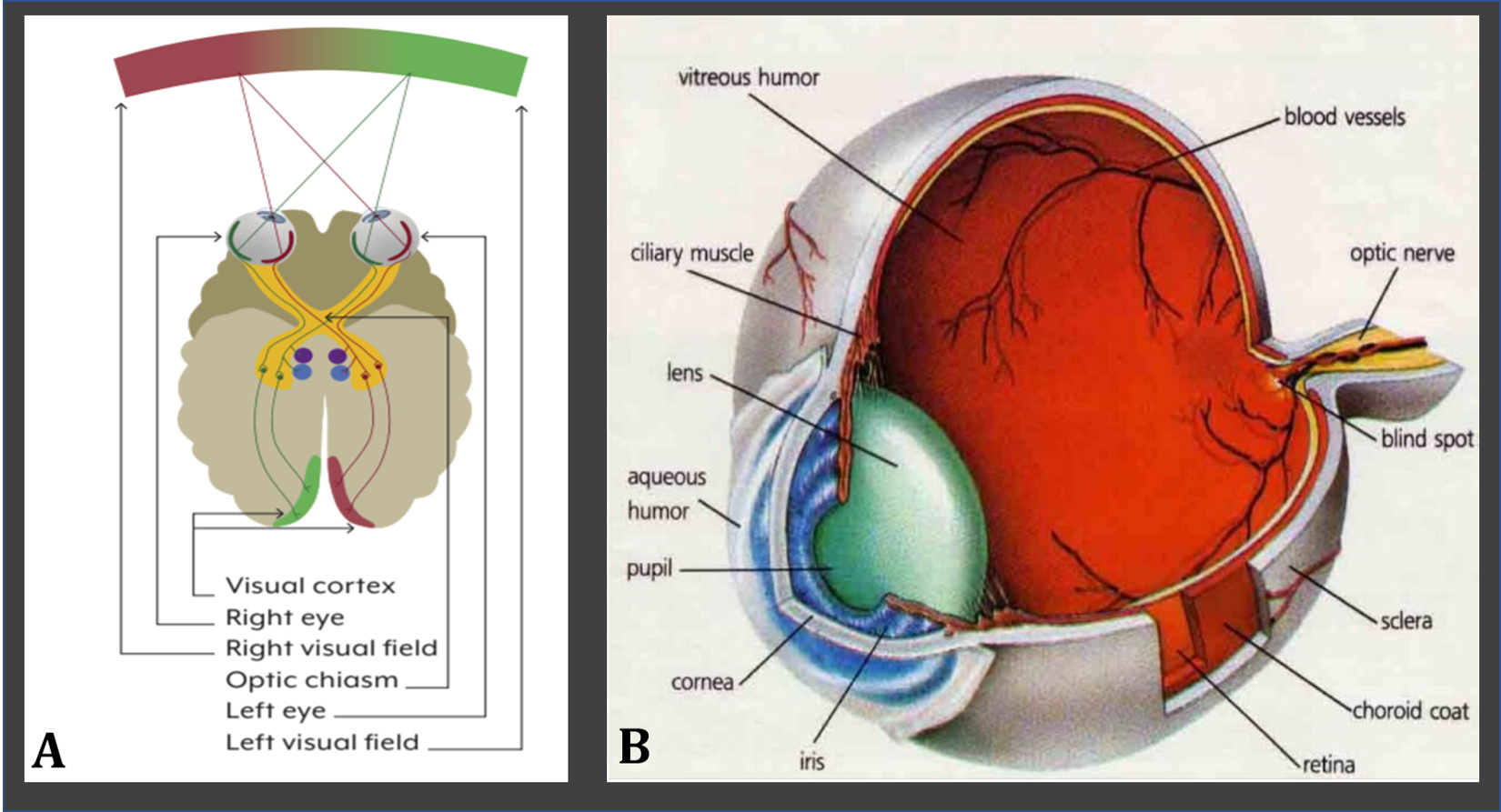}
	\caption{\small A. Visual system of the human eye. It consists of the eye, optic nerve,e, and visual cortex \RVV{\citep{ref2}}, B. The structure of the human eye consists of three layers \RVV{\cite{ref3}}.}
	\label{f1}
\end{figure}

\noindent Vision is the most preponderant sense and plays a pivotal role. The human visual system consists of the sensor organ eye and parts of the central nervous system, which are the optic nerve, optic track, and visual cortex. The eye is a sensory organ that receives visual data and sends it to the brain. Eye is the most complicated structure in the human body; it is spherical in shape and consists of three layers \RVV{\citep{ref1}}. The outermost layer of eye is made up of the sclera and cornea (see Figure \ref{f1}). The sclera is composed of connective tissues, which helps in maintaining the eye shape and it also provides protection to the whole eyeball. The anterior most part of an eye is cornea, which is a protective transparent membrane that covers the iris and pupil. The human cornea has an average horizontal and vertical diameter of 11.5mm and 10.5mm, respectively. Beneath the sclera, a choroid lies that contains blood vessels that provide oxygen and nourishment to the whole eyeball \RVV{\citep{hassan2016AO}}. Moreover, the third and innermost layer of an eyeball is the retina, which contains light-sensitive tissues responsible for producing vision. The retina comprises two main regions, i.e., the macular region (also known as the macula of the retina) and the peripheral region \RVV{\citep{hassan2018JOMS}}. Light enters the eye through the pupil, is focused by the biconvex lens, and lands on the retina, where the macular region uses rod and cone cells in the macular center (called the fovea) to produce central vision. The peripheral region is responsible for producing side vision \RVV{\citep{Raja2020TBME}}. The other parts of the eye's middle layer are the choroid, the ciliary body, and the iris. The ciliary body provides support to the lens and also produces the aqueous humor. Also, the moment of the pupil is regulated by the iris, which controls the amount of light that gets into the eye through contraction and relaxation process \RVV{\citep{hassan2015Review}}. Furthermore, the retina has ten layers that help translate visual data into the electrical signals that are sent to the brain. The electrical signals are transmitted from the retina to the brain through the optic nerve situated near the optic nerve head (ONH) region of the peripheral retina.

\section{Significance of retina \& retinal diseases towards blindness}
\begin{figure}[t]
	\centering
	\includegraphics[width=8cm,height=10cm,keepaspectratio]{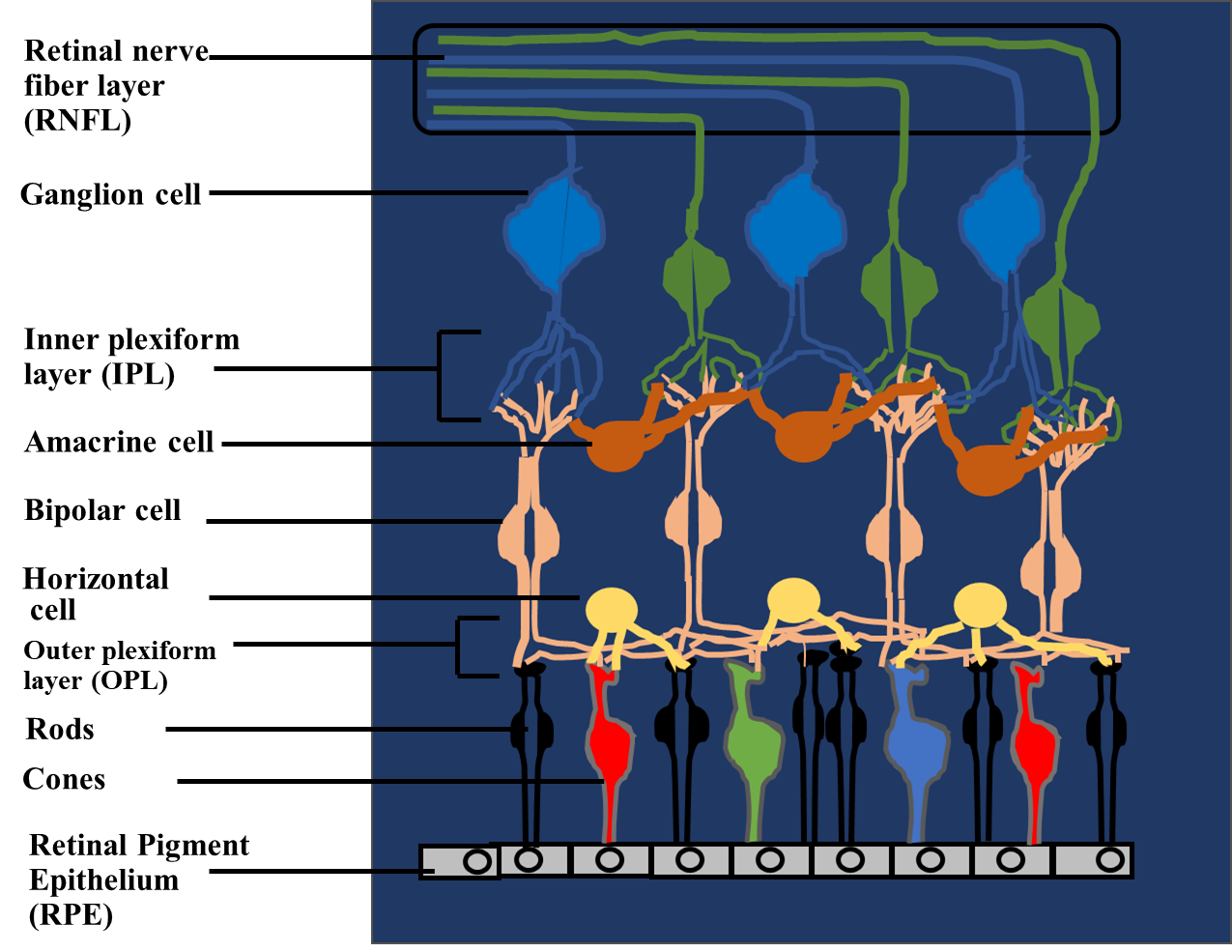}
	\caption{Systemic view of retinal layers of the human eye.}
	\label{f2.2}
\end{figure}

\noindent The retina is the light-sensitive innermost layer of the eye, which transmits visual information to the brain for interpretation. These neural impulses reach the brain, where they are combined with stored knowledge to generate a complete picture of the object's context. Because the retina and optic nerve protruded from the developing brain, it is regarded to be a member of the brain's central nervous system (CNS). Because of this, it is the only region of the central nervous system that may be visualized without resorting to invasive procedures. There are ten layers to the retina (see Figure \ref{f2.2}), and each layer is responsible for a certain function, such as transforming light into electrical signals. The inner limiting membrane (ILM), which is made up of astrocytes and müller cells, is the first retinal layer beneath the vitreous body. Retinal ganglion cells (RGCs) with axons make up the retina's second layer, the retinal nerve fiber (RNFL). The 1.5 million retinal ganglion cell axons in the human eye converge at the optic nerve head (ONH), travel through the inner and outer neural canals, and finally exit the eye and enter the brain \RVV{\citep{ref16}}. Lamina cribrosa (LC) and bruch opening membrane (BMO) refer to the inner and outer neural canals, respectively. The LC is an inmost layer of the ONH and, thus a posterior sclera. It is a network of capillaries that provide nourishment to RGCs and a 3D network of elastic porous connective tissues \RVV{\citep{Park_2013(57)}}. The fenestrated trabeculae create a pathway for the egress of RGC axons and vascular tissue. The ganglion cell layer (GCL) follows the RNFL and contains the bodies of ganglion cells. The inner plexiform layer (IPL) is situated posterior to the GCL and encompasses the synaptic connections between the dendrites of the ganglion, amacrine, and bipolar cells. The fifth position within the ocular anatomy is occupied by the inner nuclear layer (INL). It consists of the cell bodies of amacrine, bipolar, and horizontal cells. The next layer of the retina is the outer plexiform layer (OPL), which consists of a dense network of neuronal synapses between the dendrites of horizontal, bipolar cells (from INL) and photoreceptor cells. The outer nuclear layer (ONL) consists of the rod and cone nuclei responsible for visual phototransduction. The human retina contains approximately 7 million cones and 75-150 million rods. Cones correspond to photopic vision, whereas rods are responsible for scotopic vision. Cones are concentrated in the fovea, while rods are distributed throughout the retina, with the exception of the fovea. Cones and rods undergo a chemical transformation that transmits electrical impulses to the nerves. Initially, signals travel through bipolar and horizontal cells, then amacrine and ganglion cells, and finally, optic nerve fibers to the brain. These neural layers are responsible for processing the incoming picture data. The rod is the source of the signals, while the cones are the unprocessed data from individual points that are used to identify more complex features, including shapes, colors, contrasts, and motion.  In contrast, photoreceptor cells' nucleus and inner segments are separated by an outer limiting membrane (OLM). Each photoreceptor cell's inner and outer segments can be found in the IO/OS layer. Retinal pigment epithelium (RPE) is located between IO/OS and choroid and is the outermost layer of the retina. 
\section{Retinal Imaging Modalities}
\noindent Several examinations have been performed to identify and track the advancement of ocular disorders. The perimeter has been utilized to assess visual field (VF) impairments. Fundoscopy is a primary diagnostic tool employed by ophthalmologists to examine the eye. The key areas of interest that can be evaluated through this technique include the ONH, macula, peripheral retina, and central retina. The fundus image, however, does not reveal any information about the retinal layers' pathological changes. Imaging techniques like OCT/OCTA, scanning laser polarimetry, and confocal scanning laser ophthalmoscopy are utilized to evaluate the retinal layers, macula, and ONH quantitatively. OCT provides assessments of both the retinal layers and the optic disc, in contrast to CSLO, SLP, and other methods that test only the optic disc. However, OCT has seen widespread usage as an imaging tool for identifying structural retinal abnormalities and tracking the number of ocular illnesses. A variety of imaging modalities are available for the screening and diagnosis of ocular illnesses, and the following sections provide an introduction to them.

\subsection{Fundus Imagery}
\noindent Fundus photography is used to analyze the rear of eye known as fundus \RVV{\citep{tran2012construction}}. In fundus photography, specialized fundus cameras are utilized that are comprised of an intricate microscope and it is connected to a flash-enabled camera. The principle of fundus cameras is based on the concept of monocular indirect ophthalmoscopy \RVV{\citep{refFUNDUS}}. A fundus camera shows a magnified view of the fundus region. A typical camera shows 30 to 50° of the retinal area with a 2.5x magnification. This relationship can be changed with zoom or auxiliary lenses from 15°, which gives a 5x magnification, to 140°, which gives a 0.5x magnification with a wide-angle lens. The objective lens of the camera focuses the light from the observer's retina using a set of lenses formed like a doughnut with a central aperture and an annulus in the center. The light that is reflected from the retina goes through the hole in the doughnut that isn't lit up by the lighting system. As each system's light path is separate, the resulting image contains very few reflected rays. The light that forms a picture travels onward into the modest magnification telescopic eyepiece. When the shutter release is activated, a mirror moves into the light path, reflecting the flash bulb's light into the subject's eye. At the same time, a mirror drops in front of the observation telescope, focusing the incoming light onto the film or digital CCD. 

\noindent Fundus photography can be conducted utilizing chromatic filters or specialized contrast agents such as fluorescein and indocyanine green \RVV{\cite{refFUNDUS1}}. Color fundus photograph (CFP) is acquired when the retina is illuminated by white light. A filter is utilized during the process of red-free fundus photography so that superficial lesions and certain vascular anomalies inside the retina and the surrounding tissue can be observed more clearly. The wavelengths of light between 540 and 570 nanometers are blocked by using a green filter. This provides a stronger contrast for viewing retinal blood vessels and accompanying hemorrhages, yellowish lesions such as exudates and drusen, and abnormalities in the nerve fiber layer and epiretinal membranes. This technique can be used to evaluate the course of DR by better monitoring intraretinal microvascular anomalies and disc and other areas of neovascularization. In addition, red-free photography is frequently utilized as a baseline photo before angiography is performed \RVV{\cite{lim2020different}}. By injecting a fluorescent dye into a patient's bloodstream, an angiographer can take pictures and make recordings of the blood flow within the retina and surrounding tissue. When exposed to light of a certain wavelength (excitation color), this dye fluoresces in a distinct color. The autofluorescent light is then isolated by a barrier filter, preventing the passage of any other light. Sodium fluorescein angiograms (known as FAF) are used to visualize retinal vascular disease; they require a blue excitation and fluoresce a yellow light of around 490nm 530nm, respectively. CME and DR are two of the many eye diseases for which FAF is commonly utilized. Indocyanine Green Angiography, or ICG, is utilized primarily to analyze deeper choroidal diseases. It uses a near-infrared diode laser of 805 nm and barrier filters that allow light of 500nm and 810nm. Choroidal vascular outpouching can be observed with ICG in a variety of diseases, including CSC, ocular malignancies, and hyperpermeable vessels.\\
\noindent Structures such as the macula, OD, and central retina are easily discernible on a fundus photograph. Fundus images are widely used by ophthalmologists for initial screening of retinal diseases, such as DR, glaucoma, AMD, and DME \RVV{\cite{son2022light}}. As the fundus photograph is two-dimensional, it does not provide a detailed analysis of retinal layers. For detecting and diagnosing common eye illnesses, fundus photography can be used to track the effects of anti-malarial treatment on patients by documenting any changes in the fundus during routine screening. Fundus photography had a remarkable period of development and evolution over the last decade. The fundus camera market is very competitive, with brands including Welch Allyn, Kowa, Zeiss, Digisight, Topcon, Canon, Nidek, CSO, CenterVue, Ezer, and Optos.  Amazing advancements in telecommunications and smartphone technology have made ophthalmology screening in remote areas a practical option. Therefore, recently portable fundus cameras are gaining popularity for initial screening of retinal diseases \RVV{\citep{yao2022developing}}. These portable cameras are beneficial in rural areas and especially in developing countries where ophthalmologist to patient ratio is low and where access to healthcare facilities are limited \RVV{\citep{panwar2016fundus}}.

\subsection{Fundus Autofluorescence (FAF) }
\noindent Fundus autofluorescence (FAF) is a non-invasive imaging modality that has gained popularity in research and clinical applications due to its capacity to map naturally and pathologically occurring fluorophores in the posterior pole of the eye. FAF was first used for in vivo fundus imaging in 1995, when Delori et al. \RVV{\citep{delori1995vivo}} characterized the intrinsic autofluorescent properties of the human retina using a novel fundus spectrophotometer. The fluorescence utilized the characteristics of lipofuscin within the RPE to produce an image. Fluorophores stored as lipofuscin in postmitotic RPE cells' lysosomal storage bodies provide the majority of the excitation light, while photopigments in the outer photoreceptor segments filter it. Lipofuscin is made up of many different bisretinoids, such as A2E, A2PE, isoA2E, and A2-DHP-PE, and is a result of the lysosomal degradation of photoreceptor outer segments \RVV{\citep{cameronfundus}}. Bisretinoids, depending on the lipofuscin's chemistry, absorb blue light with a peak excitation wavelength of 470 nm and produce yellow-green light with a peak wavelength of 600 nm when exposed to a light source. After a detector has recorded the emission signals, the image can be interpreted as a density map of lipofuscin, with darker areas denoting lower density \RVV{\citep{schmitz2021fundus}}. In addition, the detected signal can be disrupted by any fluorophore or filter present in the light path, even under healthy conditions. Abnormal patterns of autofluorescence (AF) on FAF imaging can serve as indicators for retinal disease since numerous retinal disorders frequently lead to RPE dysfunction and buildup of lipofuscin. Blood is able to substantially absorb the blue (488nm) or green (514nm) light that is generally employed in FAF imaging, therefore blood vessels appear black in a normal fundus without retinal disease. Since there is no retinal pigment epithelium (RPE) or lipofuscin in the optic nerve, it may appear dark depending on the imaging device. The high concentration of light-absorbing xanthophyll pigment in the fovea makes it visible as a spot of hypo-AF with blue or green short-wavelength FAF. Abnormal areas of hyper-autofluoresence (AF) develop when there is either an abnormally high concentration of lipofuscin or other substances with a comparable autofluorescent spectrum, or abnormally high fluorescence transmission. Notable causes of hyper-AF include: optic disc drusen, loss of macular photopigment, presence of subretinal autofluorescent material, and increased RPE lipofuscin.  

\noindent Particularly with progressive nuclear cataracts and when employing short-wavelength light (i.e., blue light) for excitation, the lens is a significant confounder of the signal. Actually, the brownish-yellowing of the lens, which usually develops with age, is a strong absorber of blue excitation light. It blocks the excitation light on its way to the posterior pole and makes its own fluorescence light, which is then scattered in the light path and interferes with the FAF detection of signals from the ocular fundus. Blocking can also be seen with retinal hemorrhages, but the signal at the site of the hemorrhages may become much stronger over time as a result of the metabolic breakdown of blood chemicals \RVV{\citep{sawa2006autofluorescence}}. At the fovea and parafovea, a rise in the optical density of melanin and the buildup of macular pigment in the inner retinal layers when blue light is used to excite the eye reduce the FAF signal \RVV{\citep{delori1995vivo}}. When the RPE degenerates and loses its fluorophores, more fluorophores at the top of the choroid may add to the FAF signal. This could be because of melanolipofuscin at this anatomical level or because of other minor fluorophores like elastin and collagen in the walls of choroidal blood vessels and the sclera. Factors contributing to increased FAF include the bleaching phenomenon and loss of photopigment, which causes decreased absorbance anterior to the RPE level \RVV{\citep{bindewald2023blue}}. 

\noindent Many different types of FAF devices have become commercially available as a result of technological advancements; these devices vary in their excitation wavelength (e.g., green short-wavelength, blue short-wavelength), the type of imaging they employ (e.g., confocal scanning laser ophthalmoscope, ultra-widefield confocal scanning laser ophthalmoscope, broad line fundus imaging, fundus camera), and other factors. 
The basic mechanism is the same as that used in fluorescein or indocyanine green angiography (i.e., molecules are excited by light in a specific wavelength range, and the emitted light has a longer wavelength than the excitation light). Dye-based angiography detects foreign molecules, whereas FAF imaging relies on the visualization of endogenous fluorophores. Because the necessary molecules for FAF imaging are endogenous, there is no need to inject a contrast agent into a vein.

\subsection{Fluorescein Angiography}
\noindent Fluorescein angiography (FA) has been the gold standard for the identification of retinal vascular abnormalities since 1961 \RVV{\citep{weinhaus1995comparison}}. Capillaries and deep retinal vasculature are not easily visible, however, even using traditional cameras or even the most cutting-edge scanning laser ophthalmoscope \RVV{\citep{keane2014retinal}}. Although FA is mainly used for qualitative evaluation of retinal perfusion, the principle of dye dilution tracking has been described for quantifying blood flow. A dye dilution curve is generated by plotting the fluorescein dye concentration in the blood at a given time point against time. In 1965, Hickam and Frayser \RVV{\citep{hickam1965photographic}} published the first work utilizing this concept to estimate retinal blood flow. From a series of fluorescein photographs taken 1.5 seconds apart, they determined the mean retinal circulation time, providing a rough estimate of retinal perfusion. A special fundus camera with excitation and barrier filters is needed for FA. High contrast images of the early stages of the angiography are obtained by rapidly injecting fluorescein dye intravenously, typically through an antecubital vein \RVV{\citep{hansen2013retinal}}. A blue excitation filter is used to modify the white light from a flash. When exposed to blue light (465-490 nm), free fluorescein molecules absorb the energy and fluoresce, emitting yellow-green light (520-530 nm). Excited fluorescein emits light between 520 and 530 nm, which a barrier filter can block. The imaging process begins immediately following the injection and typically lasts for ten minutes. Photographs can be shot digitally or on 35mm film.

\subsection{OCT Imagery}
\begin{figure}[t]
	\centering
	\includegraphics[width=8cm,height=10cm,keepaspectratio]{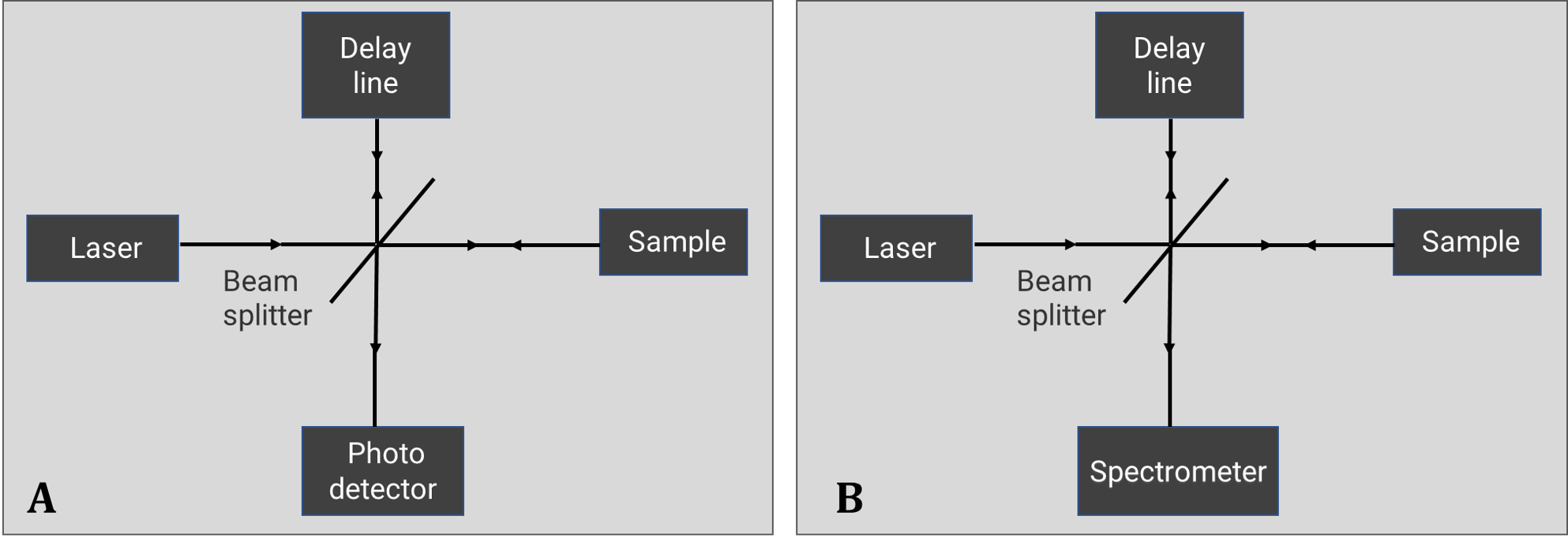}
	\caption{Illustration of time domain and spectral domain OCT principle. (A) In TD-OCT, the reference mirror moves across a distance. In (B) SD-OCT, the reference mirror remains fixed, and a spectrometer detects the spectral variation.}
	\label{f6}
\end{figure}

\noindent OCT is a noninvasive imaging modality that employs low-coherence light to produce a higher-resolution cross-sectional image. \RVV{\citep{ref20} \citep{ref21}}. The OCT method is extensively utilized in ophthalmology, but it also has numerous clinical applications in cardiology, dermatology, oncology, and gastrointestinal \cite{ref22}. In dermatology, it holds great promise for the early diagnosis of skin cancer and other illnesses. Intravascular OCT also helps detect atherosclerotic lesions in their earliest stages. Future investigations in several clinical domains are likely to include OCT. \\
OCT imaging is rapidly becoming an essential tool for getting a 3-D image of the retina and is, therefore, the most common application for OCT. The Michaelson Interferometer is the apparatus used to acquire an OCT image; it measures the sample's spatial location not through the passage of time but by means of light waves (near the infrared spectrum). The utilization of a super luminescent diode as a source is favored due to its broadband spectrum, which is conducive to the acquisition of deeper structures. The coherence length of the emitted light is a determining factor for the resolution of the OCT machine.  Tomography is a method that involves the acquisition of 2D or 3D images of an object through the use of a penetrating wave, which allows for imaging by sections or slices. The fundamental concept underlying OCT technology is rooted in the interference phenomena of light. Specifically, when the depth of a sample corresponds to the time delay of flight of the referencing mirror, they undergo coherent interference, resulting in the production of a detectable signal. This phenomenon is commonly referred to as coherence. An axial scan (A-scan) was generated by calculating the time delay of the reflected light wave from the sample. The reflected signals' intensities are represented using a color scale that mimics the colors of a rainbow. The process of determining color is predicated on the sequence of reflectivity, wherein the color spectrum progresses from black to white, with a corresponding escalation in reflectivity \RVV{\citep{ref26}}. The acquisition of numerous axial scans at consecutive lateral positions can produce a two-dimensional representation, commonly referred to as a B-scan, that depicts a cross-sectional view of the specimen.  The 2D map estimates the microstructure's transverse position and depth \RVV{\citep{ref20}}. Time-domain optical coherence tomography (TD-OCT) is the standard OCT technique. The depth range is sampled one point at a time by shifting the location of the reference mirror in order to create a longitudinal scan (A-scan). However, in order to achieve an axial scan, the reference mirror must be displaced mechanically by one cycle at a time. The utilization of Fourier transformation facilitated the transition from the conventional TD-OCT technique to the spectral domain OCT (SD-OCT) implementation, owing to its technical proficiency. In order to quantify the spectral modulations caused by interference between the reference reflection and the sample reflection, a spectrometer has been utilized in place of a single detector, and the reference mirror has been kept fixed \RVV{\citep{ref27}}. Additionally, a Fourier transform was used to convert the spectrum modulations to depth information in order to capture an axial scan. Schematics of both TD-OCT and SD-OCT are shown in Figure \ref{f6} (A and B),  respectively. 
Topcon 3D-OCT 2000 (FD-OCT), Cirrus HD-OCT (FD-OCT), Carl Zeiss Meditec, and Heidelberg Engineering's Spectralis OCT are the commercially available OCT devices. 

\subsection{Optical Coherence Tomography Angiography}
\noindent Optical coherence tomography angiography (OCTA) is a noninvasive imaging modality; it is dye-free OCT-based that provides volumetric visualization of retinal and choroidal vasculature (CNV) \RVV{\citep{park2016characterization}}. OCTA relies on repeated scans of the same area to identify movement. Therefore, time-domain systems with lower speeds prevented the creation of OCTA. Fourier-domain OCT systems initially introduced a factor 50 improvement in scanning speed for OCT, allowing for advanced OCT applications. In 2006, Makita et al. \RVV{\cite{makita2006optical}} initially described OCTA utilizing an SD-OCT device with a spectral resolution of 18.7 kHz. With further improvement of OCT hardware and advancement in data processing techniques, higher quality OCT angiograms could be generated with fewer image artifacts.

\noindent OCTA is a novel variant of OCT; it has the capability to generate 3D angiograms of the retina and choroid with high resolution. Additionally, it can detect subretinal neovascular blood vessels. To analyze structural changes in various retinal illnesses, ophthalmologists are increasingly turning to OCTA, which is rapidly becoming one of the most important diagnostic tools in the field. Successive OCT-B scans are rapidly acquired and subsequently compared to generate 3D angiograms of the retinal and choroidal vasculature. The decorrelation signal is created by the movement of blood cells from one scan to the next. 
We currently know that four vascular plexuses provide blood to the retina. The blood flow from the central retinal artery enters the superficial capillary plexus (SCP), which subsequently divides into the intermediate capillary plexus (ICP) and the deep capillary plexus (DCP). SCPs are found in the ganglion, inner plexiform, and RNFL. In contrast to the DCPs, which are below the inner nuclear layer, the ICPs are situated above it \RVV{\citep{campbell2017detailed}}. Photoreceptors and the outer plexiform layers do not have any blood vessels. Radial peripapillary capillary plexus (RPC) is the fourth retinal plexus, and it is aligned with the axons of the nerve fiber layer. When compared to the DCP, the RPC lacks a lobular structure. OCTA's 3D and higher resolution for visualizing the eye's microvasculature are two of its main benefits over FA. This permits the capillary plexus to be dissected into its superficial and deep components and the latter to be shown in exquisite detail. OCTA is helpful in detecting retinal diseases like CNV \RVV{\citep{roisman2017oct}}, changes in flow around the optic disc in glaucoma \cite{igarashi2017optical}, capillary dropouts in DR \RVV{\cite{schaal2019vascular}}, macular malformations in telangiectasia and perfusion loss in vessels occlusions \RVV{\cite{suzuki2016microvascular}}. To effectively evaluate the retinal and choroidal vasculature and distinguish between healthy and diseased retina, accurate structural image segmentation is required. Scanning results typically include depth information on vascular anomalies such retinal or choroidal neovascularization (CNV) by coordinating the 3D flow data, provided as 2D images, with the structural data. 
\noindent Table \ref{imag_m} shows the summarised the above-discussed imaging modalities.

\begin{table*}
\centering
\caption{Summary of imaging modalities used in clinical practice for the screening, diagnosis, and progression monitoring of various ocular diseases. The abbreviation are:INV: Invasive, and NINV: Non-invasive.  }
\arrayrulecolor{black}
\label{imag_m}
\begin{tabular}{l|l|l|l|l|l} 
\hline
\textbf{Modality} & \multicolumn{1}{l!{\color{black}\vrule}}{\textbf{Technique}}                                                                               & \multicolumn{1}{l!{\color{black}\vrule}}{\textbf{Information}}                                                                                                                            & \begin{tabular}[c]{@{}l@{}}\textbf{INV/}\\\textbf{NINV}\end{tabular} & \textbf{Application}                                                                                                                                                & \textbf{Devices}                                                                                         \\ 
\cline{1-2}\arrayrulecolor{black}\cline{3-3}\arrayrulecolor{black}\cline{4-6}
\textbf{Fundus}   & \multicolumn{1}{l!{\color{black}\vrule}}{\begin{tabular}[c]{@{}l@{}}Wide-angle \\photograph\\of the retina using \\specialized cameras\end{tabular}} & \multicolumn{1}{l!{\color{black}\vrule}}{\begin{tabular}[c]{@{}l@{}}\\2D view \\of the retina, \\visualization\\of retinal structures, \\blood vessels, \\OD, and macula\end{tabular}} & NINV                                                                      & \begin{tabular}[c]{@{}l@{}}Screening, documenting \\retinal diseases, monitoring\\progression, general eye \\examination\end{tabular}                               & \begin{tabular}[c]{@{}l@{}}Optos Daytona, \\Topcon Triton, \\Zeiss Visucam,\\Nidek AFC-330\end{tabular}         \\ 
\hline
\textbf{OCT}              & \begin{tabular}[c]{@{}l@{}}Low-coherence\\interferometry\end{tabular}                                                                              & \begin{tabular}[c]{@{}l@{}}High-resolution, \\three-dimensional \\imaging \\of retinal layers\end{tabular}                                                                                           & NINV                                                                      & \begin{tabular}[c]{@{}l@{}}Diagnosing and monitoring \\retinal diseases\end{tabular}                                                                                & \begin{tabular}[c]{@{}l@{}}Cirrus HD-OCT, \\Spectralis OCT, \\Topcon 3D OCT, \\Zeiss Cirrus\end{tabular}        \\ 
\hline
\textbf{OCTA}             & \begin{tabular}[c]{@{}l@{}}Combination of OCT\\and motion contrast \\imaging\end{tabular}                                                          & \begin{tabular}[c]{@{}l@{}}Visualization of \\retinal\\and choroidal blood \\flow,\\microvascular \\networks\end{tabular}                                                                                & NINV                                                                      & \begin{tabular}[c]{@{}l@{}}Diagnosing and monitoring\\retinal vascular diseases, \\including DR, retinal vein \\occlusion,\\and macular degeneration\end{tabular}   & \begin{tabular}[c]{@{}l@{}}AngioVue, Triton \\OCT Angio, \\XR Avanti\\ AngioVue, \\Cirrus HD-OCT\end{tabular}     \\ 
\hline
\textbf{FA}               & \begin{tabular}[c]{@{}l@{}}Intravenous injection\\~of fluorescent dye\end{tabular}                                                                 & \begin{tabular}[c]{@{}l@{}}Dynamic imaging \\of retinal\\~circulation,\\ blood flow, \\leakage\end{tabular}                                                                                            & INV                                                                          & \begin{tabular}[c]{@{}l@{}}Diagnosing and monitoring \\retinal vascular diseases, \\including DR, retinal vein \\occlusion, and macular \\degeneration\end{tabular} & \begin{tabular}[c]{@{}l@{}}Spectralis\\ HRA+OCT, \\Optos Daytona, \\Topcon TRC-50DX\end{tabular}                  \\ 
\hline
\textbf{FAF}              & \begin{tabular}[c]{@{}l@{}}Detection of natural\\autofluorescence \\emitted by molecules \\in the RPE\end{tabular}                                 & \begin{tabular}[c]{@{}l@{}}Visualization of \\metabolic \\health and \\integrity of the \\RPE\end{tabular}                                                                                             & NINV                                                                      & \begin{tabular}[c]{@{}l@{}}Diagnosing and monitoring \\retinal disorders involving \\RPE, including AMD, CSR, \\and inherited retinal \\dystrophies\end{tabular}      & \begin{tabular}[c]{@{}l@{}}Spectralis\\ HRA+OCT,\\Optos Daytona, \\Heidelberg Retina \\Angiograph 2\end{tabular}  \\
\hline
\end{tabular}
\end{table*}
\vspace{-0.3cm}
\section{Major retinal diseases}

\subsection{Diabetic Retinopathy}

\begin{figure}[h]
	\centering
	\includegraphics[width=9cm,height=12cm,keepaspectratio]{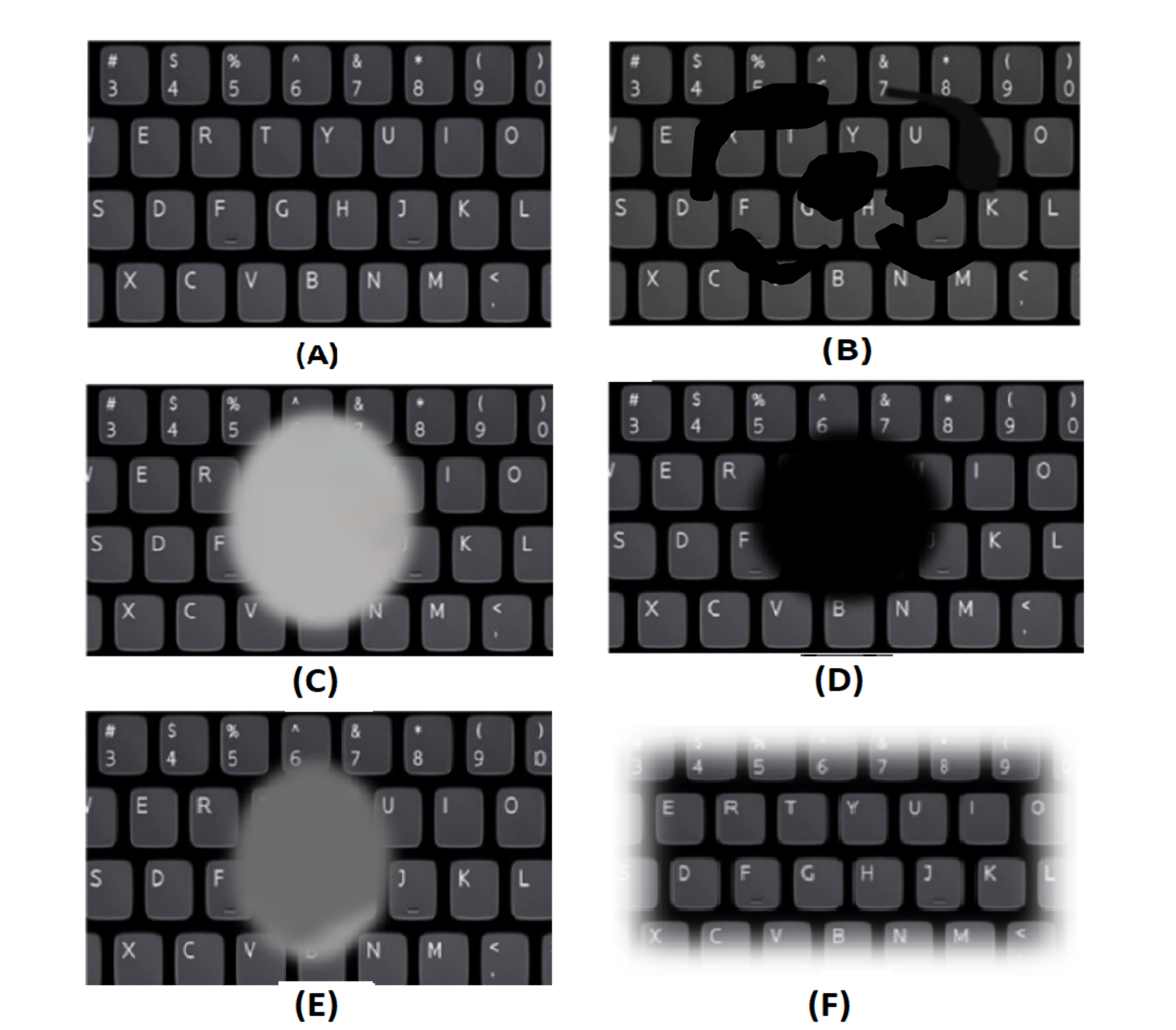}
	\caption{Visual field of retinal diseases, (A) normal vision, (B) DR VF, (C)DME affected vision, (D)  AMD subject's field of view  (E) CSCR affected vision (F) Glaucoma affected vision.}
	\label{Vision}
\end{figure}

\begin{figure}[h]
	\centering
	\includegraphics[width=5cm,height=5cm,keepaspectratio]{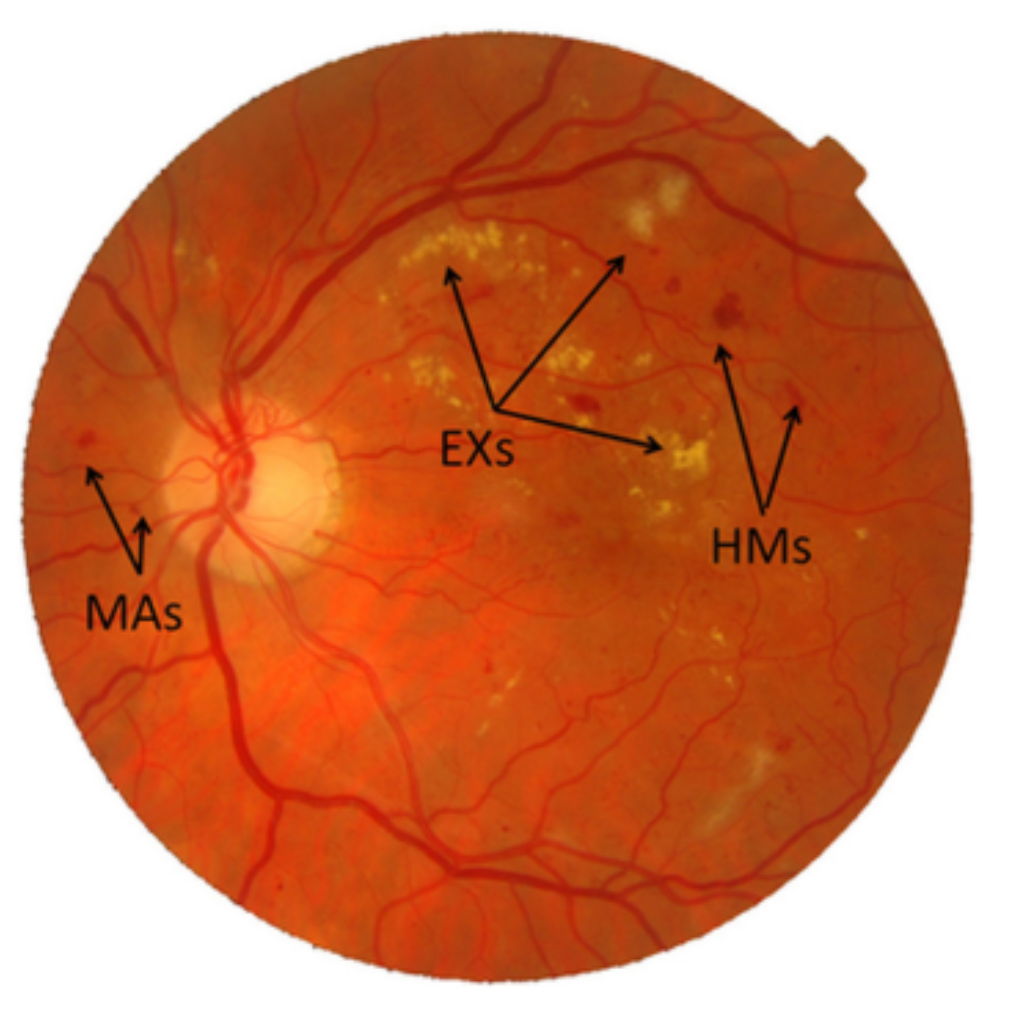}
	\caption{\small A fundus scan of DR subject, showing MAs, EXs, and HMs.}
	\label{dr_1}
\end{figure}

\noindent Diabetic Retinopathy (DR) is a pathological condition of the eye that results from elevated levels of insulin in the bloodstream, leading to abnormalities in the retina \cite{klein1984visual}. DR is the leading cause of visual impairment and blindness. DR is a chronic and degenerative ailment that poses a significant challenge due to its asymptomatic nature during the early stages of the disease. Figure \ref{Vision} (B) shows the visual field of the DR subject. The determination of the severity of DR is contingent upon the quantity and classifications of lesions that are observable on the retinal surface. The human retina comprises diverse constituents, including blood vessels, the fovea, the macula, and the optic disc (OD). DR is commonly categorized into two stages: non-proliferative DR (NPDR) and proliferative DR (PDR). NPDR is characterized by the impairment of blood vessels within the retina, leading to the leakage of fluid onto the retinal surface \RVV{\citep{crick2003textbook}}. This results in the swelling and moistening of the retina. NPDR may present with various manifestations of retinopathy, including microaneurysms (MAs), hemorrhages (HMs), exudates (both hard and soft), and inter-retinal microvascular abnormalities (IRMA) \RVV{\citep{robert}}. PDR is a severe form of DR in which new aberrant blood vessels sprout in various parts of the retina, potentially causing complete blindness. As it is shown in Figure \ref{dr_1}, the NPDR lesions can be either MAs, HMs, or EXs. MAs are the earliest detectable indication of DR, and they form when fluid leaks out of the retina's tiny blood capillaries. Their size is smaller, they have a round form, and they are red in color. The breakdown of MA walls results in HMs. Blot HMs are bigger red lesions, while hemorrhages seem like bright red dots \RVV{\citep{sjolie1997retinopathy}}. EXs are yellow spots on the retina caused by blood leakage containing lipids and proteins. If the lipid accumulation is on or close to the macula, it can result in permanent blindness. Both MAs and HMs are classified as dark lesions, while EXs are considered brilliant lesions \RVV{\citep{robert}}.  

\subsection{Diabetic Macular Edema}
\begin{figure}[h]
	\centering
	\includegraphics[width=8cm,height=12cm,keepaspectratio]{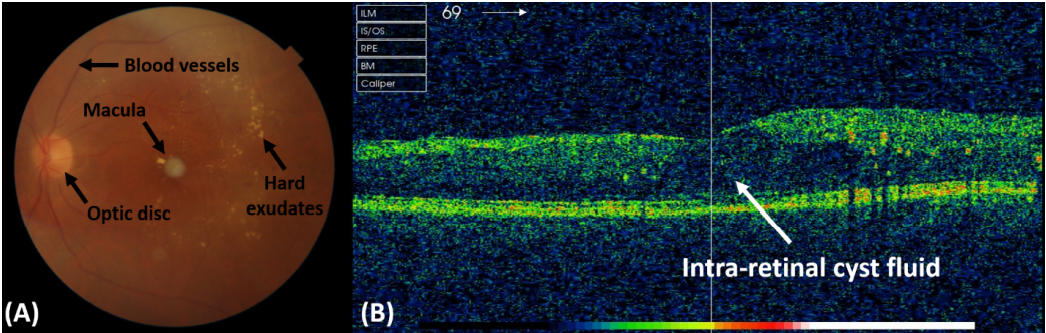}
	\caption{\small (A) Fundus scan with hard exudates, (B) corresponding foveal B-scan with intra-retinal cyst fluid}
	\label{dme2}
\end{figure}

\noindent Diabetic Macular Edema is the most severe form of maculopathy, mainly caused by diabetes. As can be seen in Figure \ref{Vision}, a person's central vision is affected by DME, and it becomes increasingly difficult to regain lost eyesight as the disease progresses. Hyperglycemia causes thinning of the blood vessels in the retina, resulting in blood leakage within the retina and subsequent formation of cyst segments, as depicted in Figure \ref{dme2} (A). During advanced stages, the presence of yellowish lipids, commonly referred to as hard exudates, becomes apparent on the superficial layers of the retina. This manifestation can be visualized on fundus images, as depicted in Figure \ref{dme2} (A).

\subsection{Age-related Macular Degeneration}
\begin{figure}[h]
	\centering
	\includegraphics[width=8cm,height=10cm,keepaspectratio]{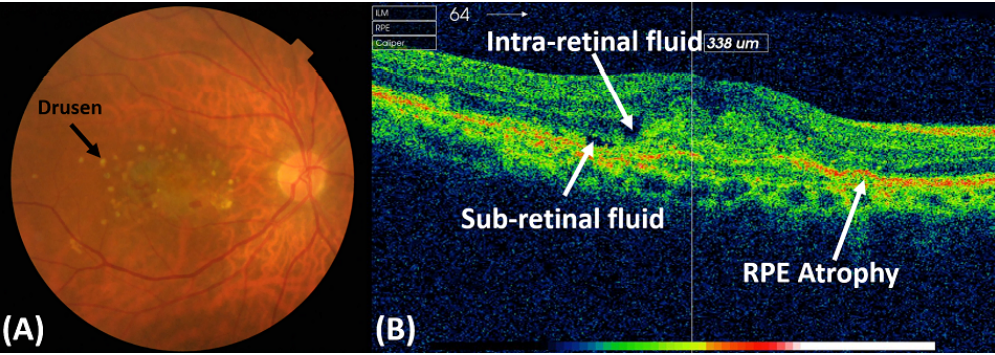}
	\caption{\small  (A) fundus scan with drusen, (B) corresponding foveal B-scan with intra-retinal fluid (IRF), sub-retinal fluid (SRF), and RPE atrophy}
	\label{amd1}
\end{figure}

\noindent Age-related Macular Degeneration (AMD) is a chronic medical condition that typically impacts both eyes and arises from a metabolic disorder \RVV{\citep{de2020age}}. The condition manifests within the macula, a region of the ocular apparatus that holds particular significance in the process of visual acuity. The etiology of this particular type of maculopathy, which ranks second in terms of prevalence, remains incompletely elucidated. Experts believe that macular degeneration develops when there is an issue with the extremely high-energy metabolic processes that occur in the retina's sensory cells. The body has evolved to handle these reactions and eliminate the metabolic byproducts. If the body is unable to process these compounds, however, they accumulate in the form of drusen. The retina doesn't get enough oxygen and nutrients because of these deposits. Drusen growth behind the retina causes age-related macular degeneration, which typically affects the elderly. Due to the RPE layer thinning or atrophying because of these drusen, central vision blurs, straight lines appear vivid and blind spots develop in the central visual field as the condition progresses, as seen in Figure \ref{Vision} (D). Pathological AMD symptoms on fundus and OCT images are depicted in Figure \ref{amd1}. 
AMD can cause significant visual deficits or even irreversible loss of central vision, but it does not cause blindness on its own \RVV{\citep{Whitney}}. The clinical classification of AMD divides the condition into two subtypes: dry AMD and wet AMD. Under the retina, drusen can form when a patient has dry AMD, also known as non-exudative AMD. In the early stages of the disease, small drusen deposits do not impair vision; nevertheless, they do promote RPE atrophy and the creation of scars, both of which contribute to the gradual dimming and distortion of central vision as the disease advances. If dry AMD isn't treated, it might progress to wet AMD, also known as exudative AMD. Wet AMD, also known as choroidal neovascularization, occurs when aberrant blood vessels in the choroid leak fluid and blood into the retina near the macula. Fluid leakage causes peripheral blind spots and a wavy appearance of straight lines.

\vspace{-0.8cm}
\subsection{Glaucoma}
\begin{figure}[h]
	\centering
	\includegraphics[width=8cm,height=10cm,keepaspectratio]{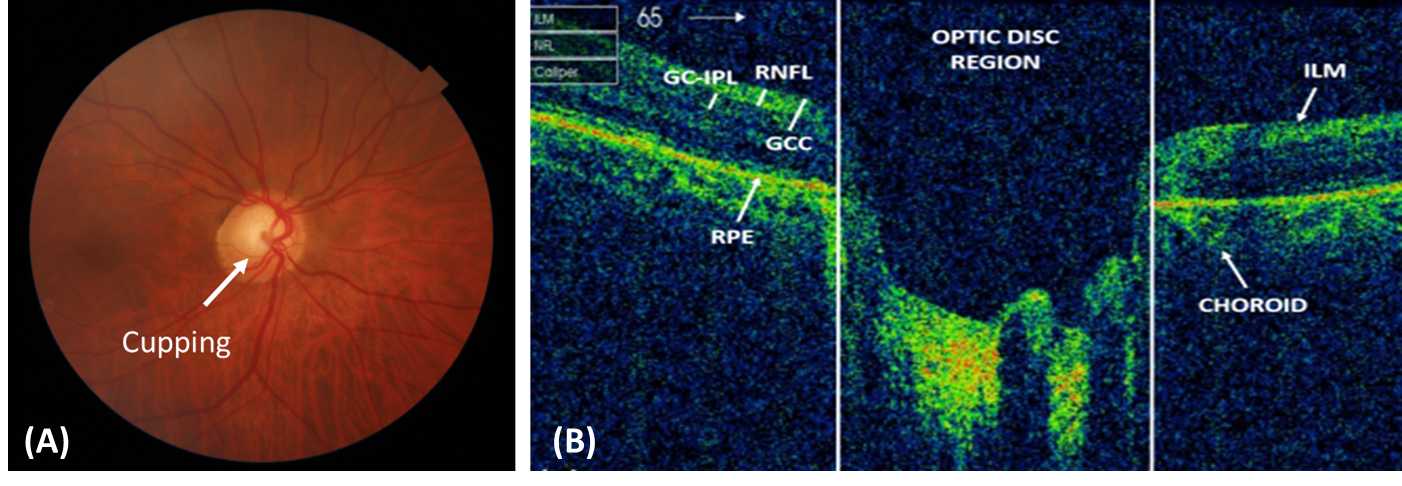}
	\caption{\small The structural changes due to glaucoma are shown in the fundus and OCT scans.}
	\label{g2}
\end{figure}

\noindent Glaucoma is a multifaceted and intricate ocular condition that can result in irreversible vision loss if not addressed. The condition is typically attributed to elevated intraocular pressure (IOP) exceeding 24mm, although it can manifest in eyes with IOP levels within the normal range of less than 20mm. Figure \ref{Vision}(F) depicts the gradual onset and eventual symptomatology of glaucoma. Almost 40\% of retinal neurons were lost before there was any detectable VF loss. The ciliary body secretes aqueous humor into the anterior chamber of the eye for its sustenance. The aqueous humor constantly creates and releases fluid through the trabecular meshwork in the anterior chamber angle to maintain a constant IOP \RVV{\citep{ref18}}. The uveoscleral pathway is responsible for a minor amount of drainage. In the adult population, there is a nearly equivalent ratio of drainage between both tracks. The primary route of drainage in the process of aging is via the trabecular meshwork. The drainage pathways exhibit not only a passive mechanical function but also involve dynamic physiological processes. The elevation of intraocular pressure within the anterior chamber is attributed to the obstruction of fluid outflow or a narrowing of the angle at the point of drainage. When there is an obstruction in the trabecular meshwork, fluid accumulates in the anterior chamber, resulting in increased pressure on the posterior chamber. The nerve fibers are pressurized by the vitreous body, leading to the eventual loss of ganglion cells. This results in the thinning of the ganglion cell complex (GCC) and the enlargement of the optic cup, as depicted in the linked Figure \ref{g2}. The detection of glaucoma is facilitated by utilizing the thickness profiles of the RNFL, GCL, and IPL layers, which are encompassed by GCC as illustrated in Figure \ref{g2} (B).  \\

\vspace{-0.8cm}
\section{Clinical severity grading of retinal diseases}

\subsection{Diabetic Retinopathy}
\noindent DR is a chronic eye condition that worsens over time. NPDR describes the first stages of the disease, while PDR describes its more advanced phases \RVV{\citep{Vicente}}. Tiny regions of enlargement in the retinal blood vessels indicate mild NPDR, the initial stage of DR. Microaneurysms are the medical term for these enlarged regions. The macula can expand if even a small amount of fluid leaks into the retina at this stage. When the enlargement of blood vessels is moderate, it begins to obstruct blood flow to the retina, so impairing its ability to receive adequate nutrition. As a result, fluids and blood collect in the macula. When a larger number of blood vessels in the retina become clogged, blood flow to the retina is drastically reduced. Severe NPDR is characterized by the initiation of angiogenesis signaling in the retina, prompting the growth of new blood vessels. PDR occurs when the illness has progressed to an advanced stage and new blood vessels have begun to grow in the retina. The danger of fluid leakage is increased due to the fragility of these blood arteries. This can lead to a variety of vision issues, including blurriness, a diminished field of vision, and even complete blindness.
\vspace{-0.3cm}
\subsection{Macular Edema}
\begin{figure}[h]
	\centering
	\includegraphics[width=8cm,height=10cm,keepaspectratio]{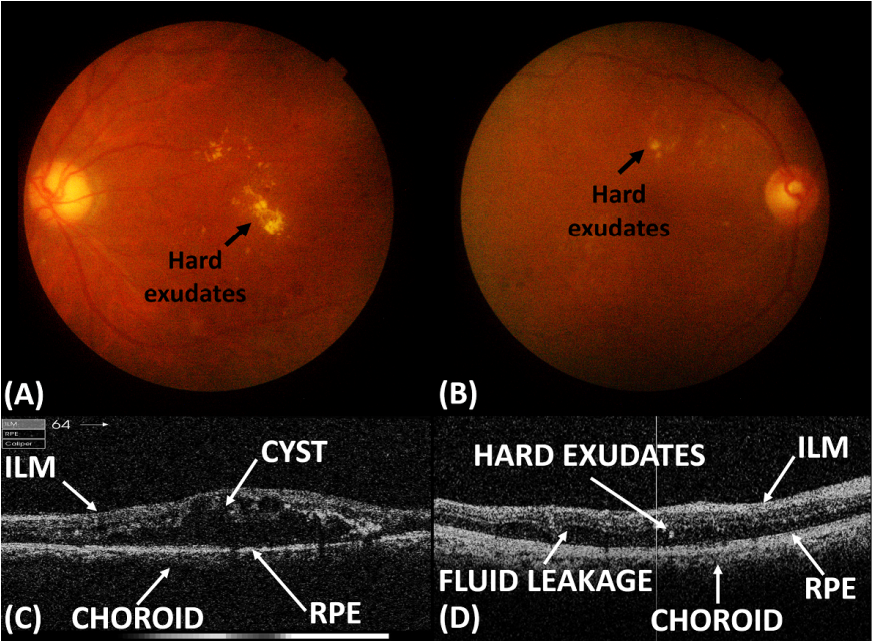}
	\caption{ (A) Fundus scan of CSME subject, (B) fundus scan of non-CSME subject, (C) foveal B-scan of CSME affected subject, (D) foveal B-scan of non-CSME subject}
	\label{dme3}
\end{figure}

\noindent Cystoid macular edema (CME) can occur in non-diabetic patients as a result of macular fluid accumulation, which causes swelling and thickening of the macula. CME is typically caused by conditions such as retinal vein occlusion (RVO), uveitis, cataract, and laser surgeries \RVV{\citep{Randall}}. The clinical grading of DME is based on the extent of retinal thickness, as established by the early treatment diabetic retinopathy study (ETDRS), and is categorized into two stages. Edema that occurs within a 500-micrometer diameter of the fovea is deemed severe due to its significant impact on visual impairment or blindness. The medical term used to describe this stage is clinically significant macular edema (CSME). Non-clinically significant macular edema (non-CSME) refers to the thickening of the retina beyond the specified limit. The visual representation depicted in Figure \ref{dme3} displays the graded fundus and OCT scans of both CSME and non-CSME cases.

\vspace{-0.3cm}
\subsection{Age-related Macular Degeneration}
\begin{figure}[h]
	\centering
	\includegraphics[width=8cm,height=10cm,keepaspectratio]{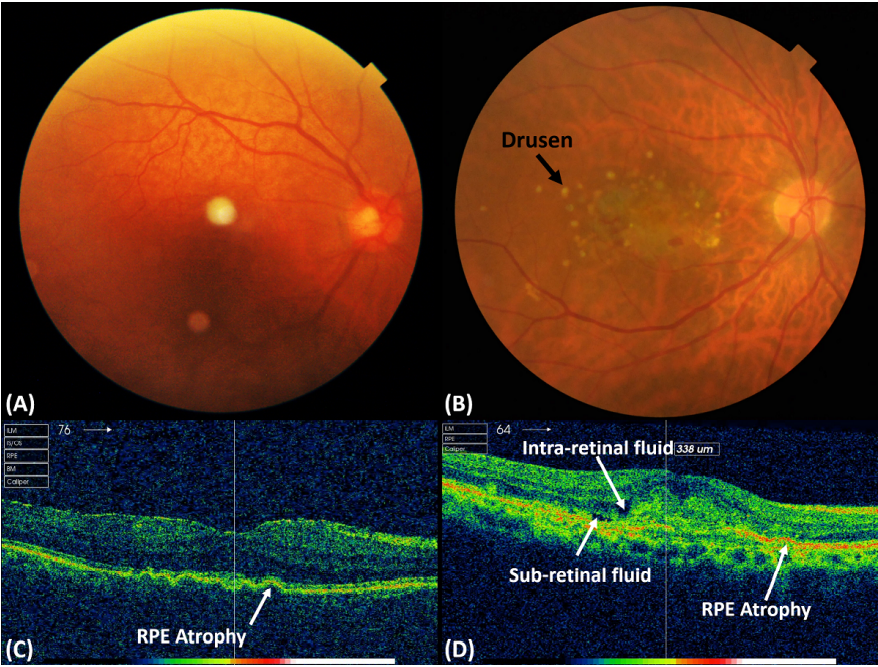}
	\caption{(A) fundus scan of dry AMD subject, (B) fundus scan of wet AMD subject, (C) foveal B-scan of dry AMD affected subject, (D) foveal B-scan of wet AMD subject}
	\label{amd2}
\end{figure}

\noindent The prevalence of AMD rises with age, affecting roughly 1 in 100 people aged 65–75 and 10–20 percent of those aged 85 and more \RVV{\citep{Whitney}}. Macular degeneration is the leading cause of significant vision loss among older persons in industrialized countries.
\noindent In the first of AMD's three stages, when the drusen deposits are still very small, and there has been no loss of pigment, normal vision is maintained. Some persons may experience minor vision loss at the intermediate stage of AMD when big drusen and/or pigment alterations are present. Wet or dry macular degeneration in its advanced stages is the leading cause of blindness in the elderly. Dry AMD is less severe than wet AMD and has a considerably slower progression rate; therefore, it is less likely to result in blindness or other vision impairments. People with advanced AMD typically suffer from wet AMD. The duration of the disease progression towards advanced-stage AMD leading to visual impairment is contingent upon several factors, encompassing the magnitude of the deposits that have accumulated in the retina. Approximately 1-3\% of individuals with small drusen encounter visual impairments within a span of five years, while roughly 50\% of those with larger drusen develop advanced AMD and suffer from vision deterioration within the same time frame. Wet AMD develops from dry AMD and rapidly worsens if untreated, but its progression can be halted or reduced by a number of treatments. Patients with advanced AMD can struggle to read or recognize familiar faces. Central vision loss is a possible long-term complication of this condition \RVV{\citep{Whitney}}. Figure \ref{amd2} shows the fundus and OCT images of AMD subjects.

\vspace{-0.3cm}
\subsection{Glaucoma}
\begin{figure}[h]
	\centering
	\includegraphics[width=9cm,height=10cm,keepaspectratio]{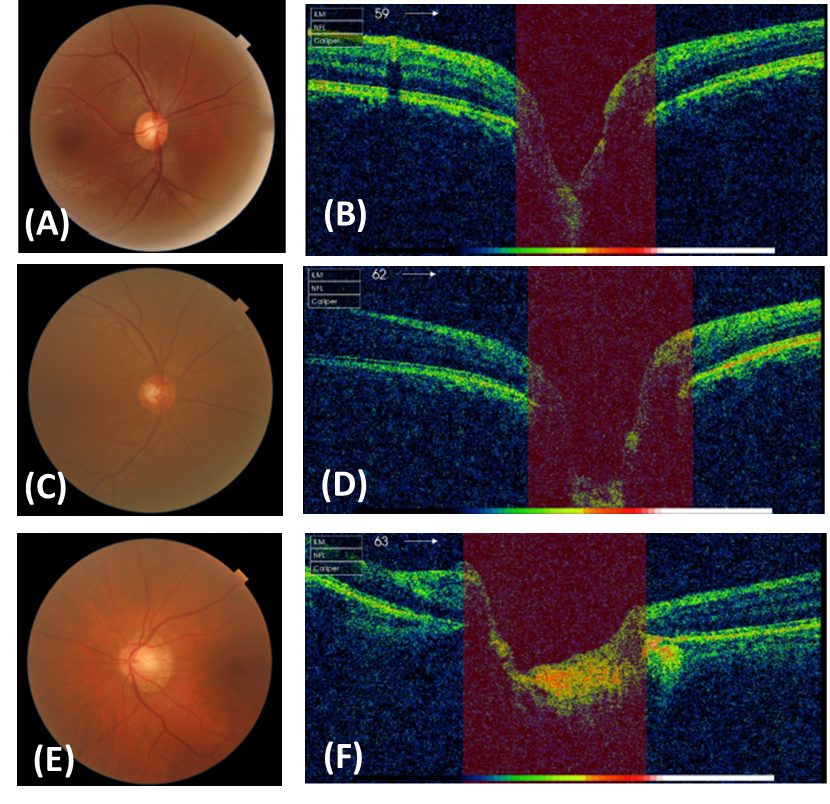}
	\caption{\small Fundus and OCT scans of different stages of glaucoma (A), (B) fundus  and OCT scans of health subject, (C),(D) Scan of early stage glaucoma subject, and (E)(F) Scan of advance stage glaucomatous eye}
	\label{g_stage}
\end{figure}

\noindent Individuals who are aged 40 years or older, have a prolonged history of steroid medication usage, and possess a familial predisposition are at an elevated risk of developing glaucoma \RVV{\citep{ref19}}. Subjects with smaller optic nerves and thinner corneas are more prone to developing glaucoma. Individuals of African and Asian descent are considered to be at an elevated risk for the development of glaucoma. Furthermore, diabetes, hypertension, migraine headaches, and inadequate blood flow may serve as underlying factors. While there are several subtypes of glaucoma, angle closure glaucoma (ACG) and open-angle glaucoma (OAG) are the most prevalent. While Asians are more likely to develop ACG, Africans are more likely to develop OAG. Each subtype of glaucoma has its own unique set of symptoms and indications. Until VF abnormalities manifest, however, OAG is asymptomatic and causes little discomfort. Pain in the eyes, severe headache, nausea, red eyes, and rapid visual disruption are some of the earliest signs of ACG. Perimeter vision loss happens in the first stage of glaucoma and lies dormant until VF abnormalities become obvious. Furthermore, early symptoms may include double vision and rainbow-colored rings surrounding bright lights. The structural alterations detected by fundus and OCT scans at various glaucoma stages are depicted in Figure \ref{g_stage}. Figure \ref{g_stage} (A, B) depicts normal people, while (C, D) displays scans from patients with early-stage glaucoma. As glaucoma progresses, thinning of RNFL occurs and results in enlargement of the cup. This cupping process can easily be observed in Figure \ref{g_stage} (C, D, E, F). If left untreated, the disease progresses over time and results in blindness. Table \ref{stages} shows the different stages of ocular diseases.
\vspace{-0.3cm}

\begin{table*}
\centering
\caption{Summarizing the different stages of retinal diseases  and their symptoms }
\arrayrulecolor{black}
\begin{tabular}{l|l!{\color{black}\vrule}l!{\color{black}\vrule}l} 
\hline
\textbf{Disease}                   & \textbf{Stages}          & \textbf{Symptoms}                                                                                                                                 & \textbf{Risk factors}                                                                                                                                                             \\ 
\cline{1-2}\arrayrulecolor{black}\cline{3-3}\arrayrulecolor{black}\cline{4-4}
\multirow{4}{*}{\textbf{Diabetes}} & Early NPDR               & \multicolumn{1}{l|}{\begin{tabular}[c]{@{}l@{}}\\Microaneurysm\end{tabular}}                                                                      & \multirow{4}{*}{\begin{tabular}[c]{@{}l@{}}Family history, \\African-American, Hispanic, \\Native American or \\Asian-American, \\Having overweight/obesity, \\Age\end{tabular}}    \\ 
\cline{2-3}
                                   & Moderate NPDR            & \begin{tabular}[c]{@{}l@{}}\\Multiple microaneurysms, \\dot-and-blot hemorrhages, \\cotton wool spots\end{tabular}                                &                                                                                                                                                                                   \\ 
\cline{2-3}
                                   & Severe NPDR              & \begin{tabular}[c]{@{}l@{}}\\Microaneurysms, haemorrhages, exudates \\(hard and soft), and inter-retinal\\ microvascular abnormalities\end{tabular} &                                                                                                                                                                                   \\ 
\cline{2-3}
                                   & PDR                      & \begin{tabular}[c]{@{}l@{}}\\Neovascularization, \\blurriness, reduced field of vision, and \\even blindness\end{tabular}                           &                                                                                                                                                                                   \\ 
\hline
\multirow{4}{*}{\textbf{AMD}}      & Early AMD                & \begin{tabular}[c]{@{}l@{}}\\Medium-sized drusen (≥64µm), \\No pigment changes, \\No vision loss\end{tabular}                                     & \multirow{4}{*}{\begin{tabular}[c]{@{}l@{}}Age, \\Smooking, \\Family history\end{tabular}}                                                                                        \\ 
\cline{2-3}
                                   & Intermediate
  stage AMD & \begin{tabular}[c]{@{}l@{}}\\Large drusen (≥173µm), \\No pigment changes, \\Mild vision loss\end{tabular}                                         &                                                                                                                                                                                   \\ 
\cline{2-3}
                                   & Dry /non-exudative AMD   & \begin{tabular}[c]{@{}l@{}}\\RPE atrophy, \\Blurred central vision\end{tabular}                                                                   &                                                                                                                                                                                   \\ 
\cline{2-3}
                                   & Wet AMD                  & \begin{tabular}[c]{@{}l@{}}\\Neovascular atrophy, \\Straight lines \& \\blind spots in the central vision\end{tabular}                              &                                                                                                                                                                                   \\ 
\arrayrulecolor{black}\cline{1-1}\arrayrulecolor{black}\cline{2-3}\arrayrulecolor{black}\cline{4-4}
\multirow{3}{*}{\textbf{Glaucoma}} & Early                    & \begin{tabular}[c]{@{}l@{}}\\No Symptoms \\Loss of Ganglion cells \\peripheral vision loss \\CDR\textgreater{}0.5\end{tabular}                    & \multirow{3}{*}{\begin{tabular}[c]{@{}l@{}}Black, Asian or Hispanic \\Family history,  \\Age, \\medical conditions, \\such as diabetes, heart disease,\\high blood\end{tabular}}  \\ 
\arrayrulecolor{black}\cline{2-3}
                                   & Moderate                 & \begin{tabular}[c]{@{}l@{}}\\0.5\textless{}CDR\textgreater{}0.7, \\Blurred vision and rainbow-coloured\\ circles around bright lights\end{tabular}  &                                                                                                                                                                                   \\ 
\cline{2-3}
                                   & Advanced                 & \begin{tabular}[c]{@{}l@{}}\\CDR\textgreater{}0.7 \\Blindness\end{tabular}                                                                        &                                                                                                                                                                                   \\
\hline
\end{tabular}
\label{stages}
\end{table*}

\vspace{-0.3cm}
\section{Clinical studies: Review}
\noindent The international health care organizations aim to provide an action-oriented, results-driven approach for advancing health equity by improving the quality of care provided to minority and other underserved communities \RVV{\citep{hill2021social}}. However, healthcare organizations have increasingly acknowledged the presence of healthcare disparities across race/ethnicity and socioeconomic status, but significantly fewer have made health equity for diverse patients a proper priority \RVV{\citep{chin2016creating}}. The lack of financial incentives is a major barrier to achieving health equity. Now the focus of healthcare organizations is to report clinical findings based on race, ethnicity, and socioeconomic status in order to provide preventive care and primary care facilities all over the world. Social determinants of health (SDOH) have emerged as a primary focus of intervention in the quest for health equity as the healthcare system shifts toward a greater emphasis on population health outcomes and value-based treatment. Clinical ophthalmology studies have recently shifted their focus to SDOH in order to better understand and promote community health improvement prospects. Extensive research in clinical settings has uncovered crucial diagnostic characteristics, risk factors, phenotypic, therapy, and drug management strategies for retinal diseases. However, we are covering the latest research articles related to DR, AMD, macular edema, and glaucoma.
\vspace{-0.3cm}
\subsection{Diabetic Retinopathy}
\noindent There are additional social and economic costs as a result of the diabetes patient's inability to work. Understanding and reducing the effects of SDOH in diabetes is a top priority due to disease incidence, economic expenses, and a disproportionate population burden \RVV{\citep{haire2019next} \citep{hill2013scientific}}. Hill et al. \RVV{\citep{hill2021social}}  presented a systemic review, discussed associations of SDOH and diabetes risk and outcomes, as well as the results of programs designed to improve SDOH and its effect on diabetes outcomes. Initially, the article provided a brief introduction to key terms and  various SDOH frameworks. The review focuses mostly on research conducted in the United States on individuals with diabetes. Recommendation for diabetic research in order to improve the healthcare facilities. 
The incidence of blindness caused by DR among people aged 20 to 60 years old is rising rapidly over the world. The review study \RVV{\citep{mistry2022centennial}} examined the development of diabetes in children and adolescents over the past century and its implications for future advancements in the discipline. The study assessed etiologic factors in a birth cohort and technology use among children and evaluated the drug management of type 2 diabetes in adolescents. Another study \RVV{\citep{pacaud2016description}} presented in the literature that used the international SWEET (Better control in Pediatric and Adolescent Diabetes: working to create CEnTers of Reference) database  to characterize the population of children with various forms of diabetes (non-type 1). It was concluded that Type 2 diabetes is more common but still difficult to diagnose in SWEET centers. However, better management and outcomes for patients with uncommon kinds of diabetes may be achieved through collaboration with SWEET centers sharing clinical information and outcomes. The review study \RVV{\citep{tosur2022precision}} summarized the history of maturity-onset diabetes of the young (MODY), the most common types of MODY, the clinical features of MODY, and some tips for diagnosing and treating MODY. 

\noindent The lipopolysaccharide (LPS) found on the outer membrane of gram-negative bacteria is responsible for triggering the host's immune system and leading to systemic inflammation and cellular apoptosis. Patients with advanced diabetes have been reported to have high serum LPS levels, most likely as a result of intestinal permeability and dysbiosis. Consequently, there is substantial indication that systemic LPS challenge is closely linked to the prognosis of DR. Even though the underlying molecular mechanisms are not yet fully explored, LPS-related events in the retina may render DR's vasculopathy and neurodegeneration severe. Xinran et al. \RVV{\citep{qin2022role}} presented a review while focusing on how LPS affects the development of DR, especially how it affects the blood-retina barrier and how it affects glial activation. In the end, they summarise the recent improvements in therapeutic strategies for blocking the effects of LPS, which could be used to treat DR with good clinical promise. It has been suggested that intestinal dysbiosis plays a contributing role in the development of type 2 diabetes (T2D) \RVV{\citep{sharma2019gut}}. The review study \RVV{\citep{yang2021role}} provides an overview of the gut microbiota in T2D and associated diseases, focusing on its possible features and molecular pathways in relation to intestinal barrier breakdown, metabolic abnormalities, and chronic inflammation. The author concluded by summarising a therapeutic strategy for improving the malignant progression of type 2 diabetes and related disorders through intestinal microecology, with an emphasis on influencing gut bacteria. The goal of study \RVV{\citep{pasini2019effects}} was to find out how long-term exercise affects the gut flora and leaky gut in people with stable T2D. Exercise helps to control blood sugar levels by changing the gut microbiota and its functions. This data indicates an extra way exercise works and suggests that boosting gut flora could be a key part of tailor-made treatments for T2D. The putative roles of pyroptosis-signaling pathways in the pathophysiology and impact of DR development are discussed in detail in the review study \RVV{\citep{al2021role}}. The review reveals briefly the pharmacological drugs might be useful in the future treatment and management of DR. \\
\noindent The vascular endothelial growth factor (VEGF) family consists of the five ligands for the VEGF receptor (VEGFR) (VEGF-A, -B, -C, -D, and the placental growth factor [PlGF]). However, VEGF-A binds VEGFR1 and VEGFR2, while VEGF-B and PlGF only bind VEGFR1. Even though a lot of research has been done on VEGFR2 to Figure out what its main role is in retinal diseases, recent work has shown that VEGFR1 and its family of ligands are also important and play a role in microinflammatory cascades, vascular permeability, and angiogenesis in the retina \RVV{\citep{uemura2021vegfr1}}. VEGFR1 signaling alone leads to the pathological changes seen in DR, retinopathy of prematurity, retinal vascular occlusions, and AMD. Anti-VEGF medicines have shown remarkable clinical efficacy in various diseases, and their effect on modulating VEGFR1 signaling remains a fertile area for future research. Upregulation of VEGF-A in the diabetic eye has been linked to DR progression. The study \RVV{\citep{singh2019advances}} presented a review of anti-VEGF treatments for DR that have been approved for use in the USA. An improvement of 2 steps on the DR severity scale developed for the Early Treatment Diabetic Retinopathy Study is regarded clinically meaningful. After One year of medication with ranibizumab or aflibercept, about one-third of individuals with DR and DME obtain this level of improvement. Another study \RVV{\citep{huang2022circfat1}} presents novel concepts for the prevention and treatment of DR.

\noindent There is research going on to find the association of diabetes and its risk factors with other medical conditions. Kun et al. \RVV{\citep{xiong2022risk}} investigated the IOP changes and acute angle closure (AAC) risk in diabetic patients after pupil dilatation. Diabetic patients were at a reduced risk of acquiring AAC after pupil dilatation. Increased post-dilation IOP was associated with lower pre-IOP. The study  \RVV{\citep{kjaersgaard2022relationship}} presented to determine whether or not DR is linked to and indicative of primary open-angle glaucoma (POAG). No significant links were found between DR and either the prevalence or incidence of POAG. The purpose of the study \RVV{\citep{vergroesen2022association}} was to assess whether or not diabetes medication is linked to the prevalent eye disorders of AMD, OAG, and cataract, as well as to evaluate these diseases' cumulative lifetime risks in a large cohort study. The findings of cohort analysis indicate that diabetes medication was not connected with cataracts, despite the fact that diabetes itself was definitely associated with cataracts. Metformin treatment was associated with a lower risk of OAG, and other diabetes medications were associated with a lower risk of AMD. To demonstrate the efficacy of the treatment, interventional clinical trials are required. The other studies \RVV{\citep{cui2022role},\citep{zhang2022influence}, \citep{jiang2022association}, \citep{yongpeng2022association}, \citep{cao2022high}, \citep{kulshrestha2022axial}, \citep{peled2022myopia}, \citep{eton2022call}} that found in the published research that investigates the connection between diabetes and other ocular disorders.
\noindent Researchers have been looking into the effects of commonly occurring comorbidities like diabetes as a result of the rapidly spreading coronavirus disease 2019 (COVID-19) pandemic. Although diabetes does not appear to raise the incidence of COVID-19 infection, it has been proven that hyperglycemia of any degree predisposes to worse outcomes, including more severe respiratory involvement, ICU admissions, the requirement for ventilators, and mortality. Infection with COVID-19 has also been linked to the development of new-onset diabetes and hyperglycemia, as well as a worsening of glycemic control in pre-existing diabetes \RVV{\citep{xiong2022risk}}. Previously, researchers hypothesized that this was related to the virus damaging the pancreas directly, the body's stress response to infection, and the use of diabetogenic medicines such corticosteroids to treat severe COVID-19 infections. Patients diagnosed with mild COVID-19 may continue to take the majority of diabetes drugs while switching to insulin is the treatment of choice for those diagnosed with severe conditions. 
Diabetes and periodontal disease both exhibit the same pattern of inflammation. Both of these diseases, if not addressed, can cause a cytokine storm, which spreads pro-inflammatory substances all over the body \RVV{\citep{stoica2022diabetes}}. Periodontitis has recently been considered to be the sixth complication of diabetes, and the most current studies point to a relationship between these two disorders that cannot be denied. Recent scientific research suggests that better glucose control in diabetes patients may be possible if their periodontal health is managed by appropriate and timely medication. New evidence of central visual system damage in diabetes patients was revealed in the recently published study \RVV{\citep{chen2022new}}. Diabetes can cause damage to the peripheral sensory organs and the central visual system, which can result in decreased color vision.
\noindent Adhesive capsulitis (AC) occurs more frequently and lasts longer in diabetic patients compared to patients with idiopathic AC. Joshua et al. presented a study \RVV{\citep{gordon2022evaluating}}, the goal was to find out how gene expression is different in AC with and without diabetes mellitus. The study of RNA-sequencing data showed that 66 genes were significantly expressed between nondiabetic patients and diabetic patients with AC. Still, only three genes were differentially expressed between control patients with and without diabetes. In addition, 286 genes were found to have differential expression in patients with idiopathic AC, while 61 genes were found to have differential expression in patients with diabetic AC. The newly expressed genes provide an explanation for the dissimilarities in disease progression and provide potential therapeutic targets that could lead to alternative treatment strategies for the two groups. This study presented the use of ribonucleic acid (RNA) sequencing and analytics to examine gene expression in alveolar bone in health and diabetes subjects. The study \RVV{\citep{zhu2020gene}} presented to investigate the candidate genes involved in the T2D. The Gene Expression Omnibus (GEO) database was used to get the gene expression profile GSE26168. Differentially expressed genes were obtained using the online tool GEO2R. Metascape was used for annotation, visualization, and comprehensive discovery, to perform the Kyoto Encyclopedia of Genes and Genomes (KEGG) pathway and Gene Ontology (GO) term enrichment analysis. Cytoscape was used to identify prospective genes and important pathways for building the protein-protein interaction (PPI) network of DEGs. A total of 981 differentially expressed genes (DEGs) were identified in T2D, including 301 upregulated and 680 downregulated genes. Six potential genes (PIK3R1, RAC1, GNG3, GNAI1, CDC42, and ITGB1) were selected based on the DEGs' PPI network. There are other studies investigation the genes affecting diabetes \RVV{\citep{sufyan2021identifying}\citep{lei2021bioinformatics} \citep{pujar2022identification} \citep{prashanth2021investigation} \citep{dieter2021impact} \citep{chen2022diabetic} \citep{oraby2022microrna}}.
\noindent The study \RVV{\citep{nair2022effectual}} aims to find a strong new set of symptoms so that DR screening can be done automatically. For the automated DR detection, a new symptomatic instrument based on information-driven profound learning was made and tested. The system used a shading technique on fundus images and ranked them based on whether or not they had DR, allowing for the easy identification of medically relevant cases for referral. While the complication of DR has been extensively studied, but less attention has been given to the impact of diabetes on ocular surface health. While diabetic keratopathy can be a serious threat to one's eyesight, it can also be used as a diagnostic and therapeutic tool for other diabetic systemic problems. In this review article \RVV{\citep{bu2022ocular}}, the current knowledge of diabetic ocular surface illness, which includes neuropathy, dry eye, and other corneal morphological alterations, was discussed. They also addressed several topics that have received less attention in the existing literature. This involves problems of the ocular surface in pre-diabetic stages as well as variances in the pathology of the ocular surface between human diabetics and animal models of diabetes. In addition to this, the author highlighted that recent breakthroughs have been made in experimental models of diabetic ocular surface problems. Finally, the most recent approaches to the diagnosis, therapy, and monitoring of ocular surface diseases caused by diabetes were analyzed. The future research prospects were described, the recent development of a technique known as protein microarrays, which has the potential to be utilized in the diagnosis and management of diabetic ocular surface disease.
Traditional dilated ophthalmoscopy has been used for the initial screening of diabetic symptoms. However, as DR is a progressive disease, fundus doesn't provide details of structural changes in the retina. Technological improvements in retinal imaging have allowed for more accurate diagnosis and treatment of DR. The review study \RVV{\citep{saleh2022role}} discussed the several imaging techniques that can be used to diagnose, detect, and grade DR. Vivian et al.\RVV{\citep{schreur2022imaging}} reviewed the current imaging modalities (CFC, OCT, OCTA, FFA, UWFP) for the DR diagnosis. It was suggested that integrating data from multiple imaging techniques could lead to more precise diagnosis, treatment planning, and monitoring of disease.\\ 
\textbf{Summary}\\
\noindent DR is the leading cause of blindness in people of working age, and it is getting worse as the number of people with diabetes rises. Imaging modalities include CFC, OCT, and OCTA, which are used for DR screening and diagnosis. Existing treatments for DR focus on inflammation, angiogenesis, and oxidation, but they don't work well enough to cure the disease completely. Researchers are working on finding new genes and risk factors for DR.  

\subsection{Diabetic Macular Edema}
\noindent DME can develop at any stage of DR and poses a severe threat to the patient's vision. DME has multiple pathways and cytokines involved in its development. The study \RVV{\citep{o2022factors}} determined risk factors for the development of DME and described its types that occur in eyes affected by PDR. Observational and retrospective case series of patients diagnosed with PDR, in which the patient's medical history, demographic information, ophthalmologic history, OCT, and FA image features were analyzed. The condition of DME, as well as the vitreomacular interface (VMI), was assessed using OCT images. DME was broken down into two categories, non-center-involving DME (NCI-DME) and center-involving DME (CI-DME). The result showed no significant association between VMI status and DME in this exploratory investigation of diabetic individuals with PDR; however, VMI status, younger age, and the presence of epiretinal membrane may be linked with CI-DME. Wang et al. \RVV{\citep{wang2022prevalence}} compared the risk factors for DR and DME in people with early-onset diabetes (40 years) and people with late-onset diabetes (less than 40 years). The prevalence of any DR and DME in the EOD patients was 67.0\% (95\% CI: 60.3–73.7\%) and 39.3\% (95\% CI: 32.1–46.5\%), respectively, which were both substantially greater than that in the LOD patients ( DME: 14.4\%, 12.7–16.1\%, p 0.001; DR: 41.9\%, 39.6–44.2\%, p 0.001). It was concluded that in the group of people with T2D, the prevalence of both DR and DME was apparently higher in patients who had diabetes before the age of 40 compared to those who developed diabetes at a later age.

\noindent Retinal anatomy and vasculature can be imaged using OCT and OCTA; however, OCT has been commonly used to track progression during treatment \RVV{\citep{kwan2019imaging}}. The study \RVV{\citep{suciu2022interleaved}} assessed the clinical, laboratory, ophthalmologic exam, and OCT parameters to explore the interconnections between them to lead to innovative pathogenetic ideas and novel treatment approaches. Patients with DME and T2D had significantly higher OCT quantitative biomarker values, whereas patients with DME and T1DM had more severe effects on systemic and laboratory biomarkers. Correlations between OCT imaging biomarkers and intraocular cytokines and VEGF therapy responsiveness suggested a potential connection to underlying pathways and their relevance to DME prognosis \RVV{\citep{abraham2021aqueous}}. Sil et al. \RVV{\citep{kar2021multi}} determined the specific spatial compartments that contribute most relevant features for predicting therapeutic response using SD-OCT images and texture-based dermoscopic features within individual fluid and retinal tissue partitions from these images. Rebounders and non-rebounders to anti-VEGF therapy were most easily distinguished by texture-based radiomics features related to the IRF subcompartment. In DME, the response to VEGF therapy has been linked to texture patterns from OCT images, and computational imaging biomarkers (CIBs) of arterial tortuosity from UWFA. Radiogenomic analysis was used to determine the link between underlying cytokines, OCT, and UWFA based DME CIBs \RVV{\citep{kar2022computational}}. The study grouped eyes with similar imaging phenotypes based on OCT CIBs and UWFA that showed similar therapy response patterns and cytokine expression. A strong link between VEGF and UWFA-derived leakage morphologic and vessel vascularity features was found. OCT and OCTA are able to detect and monitor edema while also providing a comprehensive inspection of the morphological retinal changes. The study \RVV{\citep{li2022hyperreflective}} examined the OCT/OCTA morphological features and patient profiles. The correlations between best-corrected visual acuity (BCVA) and the various morphological features of both modalities were analyzed using linear mixed models. Small, round lesions seen on OCT are called hyperreflective foci (HF), and their cause is currently unknown. It was suggested that hyperreflective material might be a predictor of BCVA and a possible biomarker of dyslipidemia in DME. The results showed that hard exudates are primarily responsible for HF, while microglial cell activation is only partially responsible for hyperreflective dots (HRDs).
\noindent Laura et al. \RVV{\citep{rubsam2021behavior}} investigated the pathogenesis of HF by comparing the number and location of HF episodes in DME patients before and after treatment. The study was presented \RVV{\citep{liu2019hyperreflective}} to examine the dynamic alterations of HF in Chinese patients with DME receiving intravitreal conbercept medication. Reduction of outer retinal HFs, as well as total retinal HFs, was positively correlated with improvement in BCVA (r = 0.40, p = 0.043; r = 0.393, p = 0.04). Anti-VEGF medications were considered to be the primary therapeutic option for DME. Tatsuya et al. \RVV{\citep{yoshitake2020hyperreflective}} evaluated the association between HF in the outer retinal layers and functional performance in DME patients who had undergone intravitreal ranibizumab (IVR) injections. In order to determine whether or not HF were related to treatment response in DME following anti-VEGF medication, the retrospective study \RVV{\citep{he2022bench}} was conducted. The correlations between changes in BCVA and HF, as well as changes in other biomarkers from OCT, were analyzed. Post-treatment HF count was associated with baseline hemoglobin A1C and the presence of hard exudate (p=0.001 and p<0.001, respectively). It was conducted that  baseline OCT indicators, along with the amount of HF in the outer retina, may be utilized to predict the therapeutic response in DME after anti-VEGF medication. According to structural investigations, it was revealed that eyes with HF in the outer layers had a thinner central subfield and a better ellipsoid zone of photoreceptors compared to eyes without such lesions. Outer retinal HF at baseline was declared as a reliable indicator of the therapeutic success of IVR injections for DME.
It was observed patients with advanced DME or wet AMD who received a single intravitreal injection (IVT) of up to 10 μg of UBX1325 experienced no adverse events during the 24-week study period. Treated patients showed statistically significant improvements in disease-related outcomes, including BCVA, CST, IR, and SR fluid. Cellular senescence is thought to play a role in the pathophysiology of the retinal microvascular network, which is the primary cause of disease in DME and wet AMD. The new small molecule Bcl-xL inhibitor UBX1325 was shown to be an effective senolytic agent. The prospective study \RVV{\citep{bhisitkul2022ubx1325}} investigated how safe and well-tolerated a single IVT of UBX1325 was for people with advanced DME and wet AMD.
The purpose of the \RVV{\citep{huang2022hyperreflective}} review is to assess the significance of HRDs, which can be seen using an OCT, as a discriminating parameter for detecting DME. The study  \RVV{\citep{rubsam2021behavior}} suggests that dexamethasone implants have a better prognosis in individuals with HF than anti-VEGF medications.  Hemolysis has been reported in primates treated with aflibercept via IgG-Fc gamma receptors. Anti-VEGF medications have not been shown to reduce inflammation or retinal cell loss in this illness \RVV{\citep{warner2022design}}. the result showed that the multi-cistronic rAAV2/2 gene therapy can decrease vascular leakage and inflammation via Tie2 receptor activation\\

\noindent \textbf{Summary}\\
\noindent Researchers have examined a large number of circulating, vitreous, and genetic biomarkers to help find diseases and make new treatments. OCT and OCTA imaging methods performed better as compared to fundus biomicroscopy and fundus photography in order to examine the structure and vasculature of the retina. Whereas OCT has been utilized to evaluate a patient's reaction to treatment. It is possible to cure DME; however, the outcomes of such treatments are frequently disappointing. As a result, it is essential to create biomarkers that can assist in the prediction of the treatment response in order to maximize the effectiveness of the treatment for specific patients. The new gene therapies are likely to become an efficient approach for the treatment of DME patients.

\subsection{Age-related Macular Degeneration}
\noindent The aging of the population has been a contributing factor in recent years to the rise in the number of patients diagnosed with ocular disorders. AMD is one of the most prevalent and, if left untreated, can result in complete blindness. The purpose of study \RVV{\citep{heesterbeek2020risk}} is to review previous research on the phenotypic, demographic, environmental, genetic, and molecular risk factors for the development of AMD. The progression of the disease has been measured in different ways. Even though vision loss seems like a good way to measure the progression of AMD in natural history studies or clinical trials, it is often not a good idea to use visual acuity as an endpoint because vision loss can take years to happen. To account for this, most AMD studies have relied on anatomical endpoints to track the disease's development across short periods of time. Geographic atrophy (GA) and choroidal neovascularization (CNV) are the two main anatomical markers used to diagnose late AMD \RVV{\cite{schaal2016anatomic}}. GA, which is also known as dry AMD, is characterized by the loss of photoreceptors, RPE, and choriocapillaris, which results in a gradual loss of vision over the course of time. CFP, FAF, and OCT are the three main imaging modalities utilized for GA detection. On CFP, it can be difficult to spot early signs of GA development and establish the margins of GA in a reliable manner, whereas FAF and OCT imaging are more suited for this purpose and are more likely to produce accurate results. These new vessels expand into the retina, causing subsequent leakage and/or haemorrhage, which can lead to serous RPE detachment, which is accompanied by a rapid loss of vision, and finally causes a scar in the macula that poses a threat to the patient's vision. CFP, OCT, and FA can identify exudative nAMD by fluid leakage and haemorrhaging, however, in some circumstances, a CNV can already be detected before exudation occurs using indocyanine green angiography (ICGA) and OCTA imaging \RVV{\citep{treister2018prevalence}}. This is possible because ICGA and OCTA imaging are more sensitive to CNVs than CFP, OCT, and FAG. Drusen can be a result of aging or an early sign of AMD, depending on their number, size, shape, distribution, and morphology. The goal of the study \RVV{\citep{domalpally2022extramacular}} was to determine the prevalence of drusen outside the macula as well as their contribution to the development of AMD. Drusen size, area, and placement were analysed from the macular grid using 30-degree, wide-angle, colour photos from the third baseline field. Comparisons were made between drusen found outside of the macula and those found inside. It was observed that drusen outside the macula are common in eyes with AMD, and they occur more often as the number of drusen in the macula grows. Extramacular drusen do not provide an additional risk to previously recognized risk factors in the progression of intermediate AMD to late AMD. The study \RVV{\citep{salehi2022retinal}} presented meta-analysis and systematic review, suggested that patients with AMD have significantly reduced values for several OCT measurements, including subfoveal CT, average pRNFL thickness, and average macular GCC thickness, compared to the general population. Quantifying the relative ellipsoid zone reflectivity (rEZR) could be a structural surrogate measure for an early disease development AMD \RVV{\citep{sassmannshausen2022relative}}. Pigmentary abnormalities, the existence of reticular pseudodrusen (RPD), and the volume of the retinal pigment epithelial drusen complex (RPEDC) were examined in relation to the rEZR using linear mixed-effects models. The results of this investigation demonstrated a connection between rEZR and the existence of iAMD high-risk characteristics as well as increasing disease severity. HF seen on OCT scans is associated with ectopic RPE and hence represents a risk factor for the development of advanced AMD \RVV{\citep{cao2021hyperreflective}}. It was observed that HF is not predictor but rather a marker of disease severity. The process of function gain and loss begins with individual RPE cells in the in-layer and extends to all aberrant phenotypes. The presence of evidence for RPE transdifferentiation, which may have been caused by ischemia, lends support to the concept of an epithelial–mesenchymal transition.

\noindent The pathophysiology and etiology of AMD are heavily dependent on inflammation. Humanin G (HNG) is a mitochondrially derived peptide (MDP) that has been shown to be cytoprotective in AMD and to be able to defend against the mitochondrial and cellular stress that is caused by damaged mitochondria in AMD. The purpose of study \RVV{\citep{nashine2022effect}} was to evaluate the hypothesis that the levels of inflammation-related marker proteins are higher in AMD and that treatment with HNG lowers the levels of those proteins. It was observed that HNG functions to decrease inflammatory protein production in stressed or injured cells, which may have a role in the development of AMD. It is important to highlight that HNG does not have any deleterious effects on cells that are healthy and have proper homeostasis.
The study \RVV{\citep{bhandari2022cataract}} was presented to determine if patients who underwent incident cataract surgery were at an increased risk for acquiring late-stage AMD. Late AMD was characterized by the presence of neovascular AMD or geographic atrophy seen on annual stereoscopic fundus scans or as documented by medical records, including intravitreous injections of medication intended to inhibit the effects of vascular endothelial growth factor. It was concluded that participants with up to 10 years of follow-up having cataract surgery did not raise the chance of developing late AMD.  
The objective of the study \RVV{\citep{chua2022association}} was to investigate the correlations between air pollution and self-reported cases of AMD as well as in vivo measurements of retinal layer thicknesses. Greater self-reported AMD was associated with greater exposure to PM\textsubscript{2.5}, while differences in retinal layer thickness were associated with {PM\textsubscript{2.5}}, {PM\textsubscript{2.5}} absorbance, PM\textsubscript{10}, NO\textsubscript{2} and NO\textsubscript{x}. Polypoidal choroidal vasculopathy is common in Asia and is considered to be a form of neovascular AMD. In a similar vein, cardiovascular disease (CVD), which is also a complex condition associated with aging, is a main cause of morbidity and mortality. Previous work \RVV{\citep{ikram2012age} \citep{hu2010neovascular}} has shown that patients with AMD have a higher risk of cardiovascular disease, suggesting a "common soil." Smoking, poor diet, and a lack of physical activity are all risk factors for cardiovascular disease, which also contribute to the development of AMD \RVV{\citep{mauschitz2022age}}.\\
\noindent According to the review of many cohort studies in the general population, high levels of physical activity appear to be protective against the onset of early AMD \RVV{\citep{mauschitz2022physical}}. These findings confirm that physical activity is a modifiable risk factor for AMD and can help guide future efforts to minimize the public health burden of this condition. A number of pharmacologic treatments are available for neovascular AMD; however, there is currently no authorized therapy that appreciably slows the progression of dry AMD. Both dry AMD and neovascular AMD have unmet medical needs related to the development of viable treatment options. In light of these findings, it is clear that innovative methods of drug delivery are required to enhance the pharmacological effect and drug concentration at the target areas. The study \RVV{\citep{jimenez2022novel}} summarised the pathophysiology and the existing therapy options for AMD, concentrating on the developing ocular sustained drug delivery techniques undergoing clinical trials. Although there is currently no cure for AMD, its symptoms can be suppressed. Current treatments for AMD are divided into four categories: device-based, anti-inflammatory drug, anti-vascular endothelial growth factor, and natural product treatment \RVV{\citep{cho2022age}}. All of these treatments come with side effects, but early AMD therapy combined with products has many benefits because it can stop RPE cell apoptosis at safe doses. Death of RPE cells is associated with oxidative stress, inflammation, and carbonyl stress, as well as a lack of essential cell components. Anti-oxidant, anti-inflammatory, and anti-carbonylation properties can be possessed by certain natural products. Candidates for AMD medicines derived from natural products reduce RPE cell death effectively; they have the potential to be utilized as medication for preventing early (dry) AMD. RPE cell transplantation intends to arrest or reverse vision loss by preventing the death of photoreceptor cells. It is regarded as one of the most promising stem cell therapy applications in the field of regenerative medicine. Recent studies have focused on transplanting RPE cells produced from human pluripotent stem cells (hPSC) \RVV{\citep{o2020advancing}}. Early clinical trial data indicate that transplantation of RPE cells produced from hPSCs is safe and can enhance vision in AMD subjects. Unfortunately, the techniques currently employed to generate hPSC-RPE cells for clinical studies are inefficient. Delivering RPE cells on a thin porous membrane for better integration into the retina can be one way to enhance transplantation outcomes. Another way to improve transplantation outcomes is to manipulate the outcome by controlling immune rejection and inflammatory reactions. In article \RVV{\citep{cohn2021subthreshold}}, author summarised the most important findings from pre-clinical studies about how different laser interventions might work to make changes that are good for the RPE, Bruch's membrane, and choriocapillaris. As laser technology has progressed toward short pulse, non-thermal delivery, such as the nanosecond laser, the most important takeaways from clinical trials of laser treatment for AMD have been summarised. Another study \RVV{\citep{csaky2022new}} discussed the different treatment approaches for AMD.\\
\textbf{Summary}\\
\noindent AMD is considered to be the most prevalent and, if left untreated, can lead to total blindness. CFP, FAF, OCT, OCTA  are the imaging modalities utilized for AMD diagnosis and progression tracking. However, OCT has been widely used by ophthalmologists to detect structural changes due to AMD. Advancements in multimodal imaging and functional testing tools, as well as continuous exploration of important disease pathways, have set the stage for future well-conducted randomized trials using nanosecond and other subthreshold short pulse lasers in AMD.
\subsection{Glaucoma}
\noindent The functional and anatomical changes that occur in glaucomatous eyes can be powerfully described using today's technologies for evaluating the disease's activity. However, there is still a need for innovative diagnostic tools that can diagnose glaucoma early and more precisely \RVV{\citep{wu2022measures}}. Glaucoma has been identified by screening tests, and even though therapy was associated with a lower risk of glaucoma development, there is still no evidence that treatment improves visual outcomes and quality of life \RVV{\citep{chou2022screening}}. The study \RVV{\citep{aspberg2021screening}} was conducted to evaluate how population screening affects the rate of blindness caused by OAG. The longest-ever follow-up of an OAG screening project that lasted more than 20 years. According to the findings, the prevalence of cases of bilateral blindness in the population that was tested dropped by 50\%. The study \RVV{\citep{munteanu2022study}} performed an assessment of risk factors and various indicators of symptoms between POAG patients and non-glaucoma patients (NG), as well as between POAG with high intraocular pressure and normal intraocular pressure, in tertiary preventive measures. Only age (F = 2.381, df = 40, p = 0.000) remains statistically significant after controlling for potential confounders such as gender, place of residence, and marital status. The most common forms of pediatric glaucoma and its diagnosis and treatment were reviewed, based on the childhood glaucoma research network (CGRN) \RVV{\citep{karaconji2022approach}}. These include juvenile open-angle glaucoma (JOAG),and primary congenital glaucoma (PCG). In addition to this, other causes of glaucoma linked to, non-acquired ocular anomalies (Peters anomaly, Axenfeld-Rieger anomaly, and aniridia), systemic disease (neurofibromatosis, Sturge-Weber syndrome) were investigated.
Early diagnosis of the structural changes paves the way for earlier therapy and results in slower disease progression. Screening for glaucoma through tonometry has a significant false positive and false negative detection rate. It was observed that screening with an assessment of the optic disc is likely to identify the majority of glaucoma incidences. The study \RVV{\citep{karvonen2020diagnostic}} evaluated the screening capacities of the OCT, laser polarimetry (GDx), and scanning laser ophthalmoscopy (Heidelberg Retinal Tomograph, HRT), and found that all tools were quite similar. Since the accuracy of each of the factors that were evaluated was moderate, screening with these parameters alone does not produce reliable results. The prospective study \RVV{\citep{yu2016risk}} demonstrates that OCT event and trend-based progression analysis programmes compare to linear mixed modeling (without relying on a normative database) and detect progression earlier than SAP. Damage to the RNFL could be detected with OCT prior to the onset of visual field abnormalities on SAP, suggesting that RNFL thickness assessment is a useful screening tool for glaucoma \RVV{\citep{kuang2015estimating}}. Luis et al. \RVV{\citep{vazquez2021recent}} summarized the findings of current studies that concentrate on the relevance of  OCT parameters in the diagnosis and monitoring of glaucoma. It has been shown that the ONH, RNFL, and macular parameters have significant diagnostic ability. According to Wanza et al. \RVV{\citep{mwanza2010reproducibility}}, the maximum allowable difference in RNFL between two visits is 4µm. The thinning that is more than 4µm classified as a statistically significant progressive change from the baseline. The study \RVV{\citep{aksoy2020comparative}} was presented to evaluate the accuracy of SD-OCT segmentation software in differentiating early glaucoma from ocular hypertension and healthy eyes. In addition to this, compassion of macular layer thicknesses between early glaucoma, ocular hypertension, and healthy eyes was performed. It was concluded that analysis of the pRNFL and macular segmentation can work together to provide a more accurate early diagnosis of glaucoma. The efficacy of SD-OCT RNFL thickness measurements in glaucoma diagnoses was evaluated \RVV{\citep{mansoori2011ability}}, and results showed that SD-OCT could be helpful for identifying glaucoma patients in the elderly. The progression of the RNFL loss is more sensitive than the GCIPL loss in patients with early to moderate glaucoma \RVV{\citep{hammel2017comparing}}. However, in a more advanced stage, GCIPL remains above ground, making macular analysis the more promising method for diagnosing progression \RVV{\citep{bowd2017estimating}}. In addition to GCIPL, metrics related to ONH can be used to monitor development in the advanced stages \RVV{\citep{chen2009spectral}}.\\
Despite developments in imaging technology, perimetry still plays a vital role in the diagnosis and management of glaucoma. The review article is to highlight recent developments in perimetry methods and to illustrate improvements in collecting and analyzing data on the visual field \RVV{\citep{prager2021advances}}. The diagnosis and characterization of glaucomatous field damage have been significantly aided by the application of artificial intelligence in research settings. In addition, tablet-based techniques and virtual reality headsets show potential for the screening and remote monitoring of glaucoma patients. Research has shown that the LC plays a crucial role in the pathophysiology of glaucoma development and progression, and is thus considered an anatomic site of glaucomatous optic nerve injury \RVV{\citep{czerpak2021curvature}}. The most significant findings were the decrease in LC thickness, posterior LC displacement, and the presence of localized defects \RVV{\citep{bastelica2022role}}. In vivo, evaluation of LC features in both normal and glaucomatous eyes has been possible with the development of high-resolution OCT devices, most notably enhanced depth imaging OCT (EDI-OCT) and swept-source OCT (SS-OCT). The study \RVV{\citep{kim2022morphologic}} investigated whether the LC curve changes when IOP falls down after eye drops in normal tension glaucoma (NTG) subjects. It was concluded that topical glaucoma treatment resulted in a reduction in IOP from 15.7 2.5 mm Hg at baseline to 11.2 1.7 mm Hg. There are other clinical studies \RVV{\citep{kim2019lamina} \citep{czerpak2022strain} \citep{bastelica2022role} \citep{guan2022structure} \citep{glidai2022microstructural} \citep{mochida2022association} }found in the literature highlighting the significance of LC for glaucoma diagnosis and progression tracking.
Despite the advancement in imaging technology, accurate analysis of the LC is still changeling \RVV{\citep{andrade2022elucidation} \citep{kim2020comparison}}.\\
\noindent Increased IOP and/or glaucomatous optic neuropathy have been linked to a wide range of systemic diseases, including renal disease and hemodialysis, neurologic disorders,  primary familial amyloidosis, endocrine disorders, vascular disease, collagen vascular disease, hematologic disorders, irradiation; systemic viral disease, dermatologic disorders \RVV{\citep{funk2022multiple}}. An evaluation of the systemic illness causing the elevated IOP is necessary. Per et al. presented a study \RVV{\citep{wandell2022systemic}} intended to examine the prevalence of OAG among people in Region Stockholm in relation to other somatic comorbidities. Higher fully adjusted OR (95\% confidence intervals) were found for women and men with, cancer 1.175 (1.120–1.233) and 1.106 (1.048–1.166), hypertension 1.372 (1.306–1.440) and 1.243 (1.179–1.311), diabetes 1.138 (1.074–1.207) and 1.216 (1.148–1.289). It was concluded that glaucoma is more likely to develop in people who have certain somatic disorders, most notably diabetes, hypertension, and cancer. In addition to this, the risk of glaucoma is also higher in neighborhoods with higher socioeconomic status as compared to neighborhoods with lower socioeconomic status. Jun et al. \RVV{\citep{ro2022association}} conducted their research on the risk of OAG in the 12 years that followed the diagnosis of chronic kidney disease (CKD) using a cohort that was representative of the entire country. The results showed that CKD is a major contributor to the development of OAG, and that the risk of OAG increases with the severity of CKD. The purpose of study \RVV{\citep{kolli2022background}} was to determine whether the genetic risk for POAG influences the correlations between cardiopulmonary diseases and glaucoma indicators. The history of common cardiopulmonary conditions and cardiopulmonary measurements were analyzed in the UK Biobank, together with history of glaucoma. The prevalence of diabetes (17.5\% vs 6.5\%), CKD (6.7\% vs 2.0\%) dyslipidaemia (31.2\% vs 18.3\%) were all greater in glaucoma patients than in controls (adjusted p0.0013 for each) within decile 1. Contrast test p-value for difference 0.05 indicates that the extent of the connection between glaucoma and diabetes, CKD, and glycated haemoglobin varies between deciles 1 and 10. The study \RVV{\citep{mauschitz2022retinal}} conducted retinal layers assessment as biomarkers for brain atrophy. They investigated the relationship between segmented retinal layers and various cerebral parameters from  magnetic resonance imaging (MRI). Relationships between retinal measurements and volumetric brain measures, as well as fractional anisotropy (FA) as a marker of microstructural integrity of white matter (WM) were analyzed using multiple linear regression. Inner retinal volumes were correlated with total brain and GM volumes, and even more strongly with WM volumes and FA. It was  that  both the inner and outer retina were linked to hippocampal size, whereas the outer retina was most strongly associated with GM volume.

\noindent Wang et al. \RVV{\citep{wang2022genetic}} reviewed recent advances in the genetics of POAG. The study discussed how recent developments in research methods have led to the discovery of new risk genes, as well as how subsequent biological investigations could be conducted in order to define how the risk that is represented by a genetic sequence variant manifests itself in patients. By analysing transcriptomes from single cells with Smart-Seq2, new genes were found involved in regeneration that substantially increase axon regeneration \RVV{\citep{li2022single}}. Among these, Anxa2 is the most powerful because of the synergistic effect it has with its receptor tPA in Pten-deletion-induced axon rejuvenation. In a clinically relevant model of glaucoma, Anxa2, its downstream effector ILK, and Mpp1 significantly protect RGC somata, axons and preserve vision. Nigus et al. \RVV{\citep{asefa2022bioinformatic}} prioritized the genes that are most likely to be "causal" and to uncover the functional properties and underlying biological pathways of POAG candidate genes. They drew on data from the GERA and UK Biobank cohorts to analyze the genetic risk factors for POAG. Systematic gene-prioritization analyses were performed based on, nearest genes, co-regulation analysis, epigenomic data, transcriptome-wide association studies, and nonsynonymous single-nucleotide polymorphisms. The study found 142 genes that should be prioritized, of which 64 were found to be new for POAG. According to at least four different lines of evidence, the genes that were given the highest priority were BICC1, AFAP1, and ABCA1. Another review study \RVV{\citep{aboobakar2022genetics}} summarized the genetic relationships between various types of glaucoma and the potential roles these genes play in disease pathogenesis. There are other studies \RVV{\citep{hamel2022integrating} \citep{milanowski2022associations} \citep{choquet2022association}\citep{he2022bench}} presented in the literature related to glaucoma risk genes.
\noindent Glaucoma treatment has been challenging, because ocular barriers have inherent mechanics that impede the entry of ophthalmic medicines. Several carriers (inorganic, polymeric, hydrogel, and contact lens-based) with specialized chemical and physical properties have been intensively investigated as potential solutions. The review article \RVV{\citep{patel2022recent}} summarized the latest developments in ocular delivery formulations with a particular emphasis on glaucoma, including the different types of nanocarriers and delivery routes. IOP can be lowered with the use of the new Rho kinase inhibitor netarsudil/latanoprost FDC by enhancing trabecular outflow \RVV{\citep{asrani2020fixed} \citep{brubaker2020one}}. A very promising platform for the treatment of glaucoma and simultaneous protection of the ocular surface would be the combination of hypotensive liposomal formulations with osmoprotective agents \RVV{\citep{gonzalez2022novel}}. Drug delivery systems for the ocular surface, like contact lenses and nanotechnology, are currently under study as potential sustained release(SR) therapeutics. 
There is growing interest in using aqueous gels prepared with hydrophilic polymers (hydrogels) and based on stimuli-responsive polymers for the treatment of numerous ocular disorders \RVV{\citep{akulo2022intravitreal}}. Because of their chemical structure, they are able to incorporate a wide variety of ophthalmic medications, allowing them to achieve their optimal therapeutic doses while also providing more clinically relevant time courses (weeks or months, as opposed to hours and days). This will inevitably result in a reduction in dose frequency, which will improve patient compliance and clinical outcomes. Glaucoma is a chronic disease that may respond well to gel technologies used as drug-delivery methods and as antifibrotic therapy during and after surgery \RVV{\citep{fea2022glaucoma}}. Sakaorat et al. \RVV{\citep{petchyim2022clinical}} investigated bleb-related infections, including their symptoms, causes, treatments, and effects. Pain and redness were the primary symptoms that patients experienced when they had a bleb-related infection. Nearly 25\% of people had experienced some kind of eye injury in the past. Patients who display symptoms and engage in undesired behavior that have the potential to result in bleb infection should be identified, and treatment and education should both be provided to these patients.
The study is the first direct proof that glaucoma can be treated with noninvasive femtosecond laser trabeculotomy (FLT) \RVV{\citep{mikula2022femtosecond}}. The study \RVV{\citep{mikula2021intraocular}} examined the safety and efficacy of FLT in reducing IOP in a perfused anterior segment model. The findings suggested that FLT treatment can result in a considerable reduction in IOP in a perfusion model, which suggests that it could be a viable noninvasive treatment option for POAG. Ex-Press shunt implantation, canaloplasty, and viscocanalostomy, are alternative surgical techniques that show promise in equivalence but need more research to evaluate discrepancy in outcome \RVV{\citep{siesky2022glaucoma}} accurately. In addition to differences in treatment results, social disparities can also be seen to have an effect on clinical care in the form of decreased adherence, choice, and, access. Adherence to glaucoma drugs is a significant issue in the management of glaucoma patients, as up to fifty percent of patients fail to receive the desired treatment advantages. The study \RVV{\citep{zaharia2022adherence}} overarching objective was to draw connections between the various approaches to gauging glaucoma patients' propensity to take their prescribed medications, as well as the interventions designed to improve adherence.\\
\textbf{Summary}\\
\noindent The damage caused by glaucoma is considered to be permanent and cannot be completely remedied. Nevertheless, the advancement of the disease can be decelerated by means of pharmaceutical interventions, laser therapy, or surgical procedures, which may aid in mitigating additional visual impairment. Various modalities have been used to detect structural and functional damage due to glaucoma, which includes fundus, GDx, scanning laser ophthalmoscopy, SAP, OCT, and OCTA. However, OCT is widely used by ophthalmologists for the structural analysis of glaucomtous damage. It has been revealed that the ONH, RNFL, and macular characteristics all have substantial diagnostic ability. LC analysis with OCT has great potential in glaucoma management if some of the current constraints are resolved, particularly those relating to image acquisition.
\section{Technical studies: Review}
\noindent Manual identification of retinal lesions through the scans of fundus/OCT/OCTA is a time-consuming and subjective task, thus resulting in intra-variations. So, automated algorithms have been used in clinical practice to assist the ophthalmologist. Research is ongoing to improve the accuracy and robustness of automated techniques, the following sections discuss the automated methods used for the identification and classification of major ocular diseases. We have divided the literature into three categories based on techniques, such as traditional image processing, ML and DL. 
\subsection{Traditional Image Processing Techniques}
\noindent \textbf{Optic Disc and Cup-based Methods:}
Ophthalmologists widely use fundus images for initial screening of various retinal diseases such as DR, AMD, DME, and glaucoma. The bright circular region in the center of the human retina is characterized as an optic disc (OD) in fundus scan. Locating the OD accurately is a crucial stage in computer-assisted glaucoma and DR diagnosis. OD detection has been a challenge in fundus analysis if there are other bright spots on the retina or if the images were not taken in a very controlled setting. In the paper \RVV{\citep{kose2011statistical}}, simple statistical methods were suggested for finding the OD and macula and determining the diameter of the OD and the distance between it and the macula. The weighted-distance method was used to make the healthy parts of a retinal image bigger. Qureshi et al. \RVV{\citep{qureshi2012combining}} propose an ensemble algorithm that can automatically detect the OD and macula in fundus images. The feature set was based on pyramidal decomposition, edge detection, entropy filter, hough transformation, and uniform sample grid. On the three publicly accessible databases, Diaretdb0, Diaretdb1, and DRIVE, experimental findings and analyses were presented, and the combined algorithm achieved an average Euclidean error of 29.64, 24.26, and 26.80, respectively. The study \RVV{\citep{usman2014robust}} proposed a method based on classic image processing techniques to localize the OD. To eliminate gaps and erroneous regions, a threshold is first applied to the red plane of the fundus picture, and then morphological techniques are performed. The linked component labeling algorithm was employed to assign labels to the objects of the binary image. Adaptive histogram equalization and Laplacian of Gaussian (LoG) kernel were applied to enhance the bright regions within the image. The threshold image is then subjected to morphological opening to eliminate the noisy regions. Quantitative evaluation of the proposed system was performed on publicly available datasets DRIVE, STARE, and DiaretDB and achieved an accuracy of 100\%, 97.50\%, and 95.85\%, respectively. Approximate Nearest Neighbor Field (ANNF) maps are often in computer vision and graphics to solve problems, including noise removal, image completion, and retargeting. Ramakanth et al. \RVV{\citep{ramakanth2014approximate}} extended the application of ANNF maps to include medical image analysis and more particularly, the detection of OD in fundus images. ANNF algorithm feature match was employed to determine the similarity between a reference image of an OD. This gives a list of the patches of image that are closest to the patches in the reference image. For OD detection, a probability map was made from the patche's distribution in the query image. Five publicly accessible databases (DIARETDB0, DIARETDB1, DRIVE, STARE, and MESSIDOR) were used to evaluate the suggested methodology. The study \cite{kao2014automated} proposed a method that employed area which was free from vessels and adaptive Gaussian template for fovea center detection in retinal images. The center of the OD is localized by using the template matching method. Next, the disc–fovea axis was defined by scanning the vessel-free region. Finally, the fovea center was identified using the matching of the fovea template. The centers of the OD and fovea in the various image resolutions were identified using adaptive Gaussian templates. For the DIARETDB0, DIARETDB1, and MESSIDOR databases, the proposed method found the fovea with an accuracy of 93.1\%, 92.1\%, and 97.8\%, respectively. The OD and OC were extracted from fundus images usbased on adaptive thresholding for glaucoma diagnosis \RVV{\citep{issac2015adaptive}}. An automatic  method \RVV{\citep{hu2017optic}}  was proposed that combined color difference information and vessel bends information to determine the OC boundary from fundus images. Xiong et al. \RVV{\citep{xiong2016approach}} proposed OD localization method that can accurately localize the OD even when the retinal image contains pathological abnormalities. Extracted features included were vessel direction, edges, intensity, and luminous region size. The proposed approach achieved an accuracy of 100\% for the DRIVE, 95.8\% for the STARE, 99.2\% for the DIARETDB0, and 97.8\% for the DIARETDB1 database. Wavelet feature extraction combined with optimized genetic feature selection to segment OD for glaucoma diagnosis through fundus images \RVV{\citep{singh2016image}}. Panda et al. \RVV{\citep{panda2017robust}} developed OD localization method incorporating three features; retinal vascular visual cues—global/local vessel symmetry, and vessel component count. The first OD center is determined by utilizing the skeletal image component with the highest concentration of major blood vessels. The proposed technique was effective for ocular diseased with different symptoms, such as bright lesions, hemorrhages, and twisted blood vessels. The study \RVV{\citep{mahmood2022optic}} proposed a technique based on color and blur analysis for accurate detection and localization of the OD. To improve the visibility of the OD, fundus image was transformed into Lab color space. The extended maximum transform and directional blur were used to extract OD candidates accurately. In order to isolate the OD from the rest of the candidates, a radial blur was applied. Zaaboub et al. \RVV{\citep{zaaboub2022optic}} proposed an algorithm for OD segmentation in fundus scans. In the first stage, the OD was located, which was accomplished by performing 1) a preprocessing step, 2) vessel removal, and 3) a geometric analysis that delineate OD position. An OD contour is accurately completed using a candidate. Ten different public databases and one local database were employed to test the algorithm. RimOne and IDRID had an accuracy of 98.06\% and 99.71\%, respectively. \\

\noindent \textbf{Blood Vessels Extraction Methods:}
\noindent Computer-aided pathology systems rely heavily on blood vessel detection in retinal images for early screening and diagnosis of ocular diseases such as retinal detachment, DR, and DME. Numerous studies \RVV{\citep{ravichandran2014fast} \citep{liao2014retinal} \citep{ali2017vessel}  \citep{ali2017vessel} \citep{alhussein2020unsupervised}} were found in the literature that utilized the histograms and enhancement techniques for vessel segmentation. Ravichandran et al. \RVV{\citep{ravichandran2014fast}} developed an enhancement technique that incorporated histogram matching and Gabor filtering. The method first applied a region-based histogram equalizer to the retinal image, then used a 2D Gabor filter to further improve the appearance of the vessels. In the paper \RVV{\citep{liao2014retinal}}, a novel approach was proposed to the enhancement of retinal vessels. Initially, a multi-scale top-hat transformation was used to extract the best high-contrast and low-contrast picture features from an image. The optimal bright image features are then added to the image, and the optimal dim image features are removed, for a preliminary quality enhancement. As shown by the results findings on the DRIVE and STARE databases, the proposed technique efficiently boosted contrast and improved the finer features of the retinal vessels. Ali et al. \RVV{\citep{ali2017vessel}} proposed a method to detect retinal vessels in fundus images by combining automatic thresholding and Gabor Wavelet (GW). The green channel was extracted and then used to generate gabor feature image by utilizing GW. The final vessel output is generated by combining two vessel-enhanced images, each of which has been converted to a binary image by automatic thresholding. The algorithm achieved an accuracy of 94.53\% on the Drive dataset. A method that performed retinal blood vessel analysis using more traditional methods was proposed \RVV{\citep{toptacs2021retinal}}. The model extracted pixels-based features and grouped them into five groups, gradient, morphological, edge detection, statistical, and Hessian matrix. Every pixel is assigned an 18-D feature vector, and it is fed into the neural network. The system accuracy was calculated to be 96.18\% for DRIVE and 94.56\% for STARE. The normalized first and second-order derivatives of a Hessian matrix were used by study \RVV{\citep{yang2014fast}} to segment medical images. The Hessian matrix's eigenvalues stand in for luminance data, while the eigenvector of the smallest eigenvalue reveals the orientation of the lines. A novel Hessian matrix-based vessel enhancement measure was presented in the study \RVV{\citep{jerman2016enhancement}} and addressed the issues with existing Hessian methods. These included insufficient responses to vessels of variable intensities and scales, as well as vessel bifurcation. Using the eigenvalues produced by the Hessian matrix at two different scales, the study \RVV{\citep{alhussein2020unsupervised}} developed an unsupervised segmentation method to extract the thick and thin vessels. CLAHE technique was employed to enhance the contrast of retinal images. For contextual region tuning of CLAHE, a better version of the PSO algorithm was used. A morphological filter and a wiener filter were employed to remove noise. In order to extract thick and thin vessels, the eigenvalues of the Hessian matrix were calculated at two different scales. Global otsu thresholding was performed to intensity-transformed images and enhanced images of thick vessels, while ISODATA local thresholding was applied to enhanced images of thin vessels. Area, eccentricity, and solidity were used as region parameters in a post-processing step. On the publicly available CHASE DB1 and DRIVE datasets, the proposed framework was evaluated, where it showed a sensitivity of 77.76 and 78.51, and an accuracy of 95.05 and 95.59, respectively. The study \RVV{\citep{madathil2022mc}} introduced a Morphological Closing-based Dynamic Mode Decomposition (MC-DMD) method for enhancing the retinal vessels, which is both effective and robust. The proposed algorithm uses the power of mathematical morphology to create the input channel for the DMD system, which separates the retinal images into their vessel and non-vessel features. The proposed method was accessed on three publicly available datasets: DRIVE, STARE, and HRF. 

\noindent The  other techniques for vessel enhancement in retinal images include, visual adaptation model \RVV{\citep{wang2021retinal}}, Bi-orthogonal wavelet transform and bilateral filtering \RVV{\citep{bala2021contrast}}, retinex theory and dark channel prior method \RVV{\citep{zhang2022double}}, luminosity and contrast enhancement \RVV{\citep{kumar2022luminosity}}, morphological operation \RVV{\citep{ashanand2022efficient}}, graph-based method \RVV{\citep{zhao2015correction}}, multiscale fractional anisotropic tensor \RVV{\citep{alhasson20182d}}, and statistical feature-based transformation \RVV{\citep{mahapatraoptimal}}. \\

\noindent \textbf{Retinal Layers and Lesions Extraction Methods:}
\noindent  The automated identification of retinal boundaries is an area of significant research interest due to its ability to offer a reliable, measurable, and unbiased evaluation of retinal lesions. Several automated algorithms for retinal layer segmentation have been suggested in scholarly literature. Those eight inner retinal boundaries were retrieved by Kromer et al. \RVV{\citep{kromer2017approach(99)}}. Before segmenting the layers, median filtering was performed for preprocessing, and curve regularization was then utilized. Duan et al. \RVV{\citep{Duan_2018(100)}} presented a developed model that used groupwise curve alignment to extract the retinal layers in OCT volume. The seven sub-retinal layers were automatically segmented using a high-pass iterative filter. In addition, they introduced a new denoising method tailored specifically for the OCT image \RVV{\citep{Roychowdhury_2013(101)}}. Using gradient information and shortest path search, Yang et al. \RVV{\citep{Yang_2010(102)}} devised a fast and accurate automated segmentation system to extract nine intra-retinal layers. Niu et al. \RVV{\citep{Niu_2014(103)}} developed an algorithmic technique for the automated segmentation of the six retinal layers. This method utilizes correlation smoothness constraint and dual gradient information. The construction of the edge map was followed by the utilization of a convolution operator in order to obtain the gradient map. The removal of outliers was facilitated by the imposition of smoothness constraints on spatial correlation.\\ 
\noindent Several automated segmentation algorithms have been suggested in the literature for extracting retinal layers and subsequently measuring their thickness. Mayer et al. \RVV{\citep{Mayer_2010(109)}} presented an algorithmic approach for the quantification of RNFL in images obtained through SD-OCT. The utilization of gradient and local smoothing techniques was implemented in order to minimize the energy function utilized for the segmentation of retinal layers. The study calculated the average thickness of the RNFL for individuals with normal vision (94.1±11.7μm) and those diagnosed with glaucoma (65.3±15.7μm). 
A study \RVV{\citep{Kafieh_2015(110)}} generated a thickness map of the eleven retinal layers through SD-OCT images of normal individuals without any known ocular abnormalities. Segmentation of retinal layers was based on edge statistics rather than contextual information. The average thickness of the RNFL, GCL-IPL, and GCC in the macula was calculated using a graph-based approach proposed \RVV{\citep{Gao_2014(111)}}. It was estimated that the RNFL thickness of healthy people was 36.5 μm while that of glaucoma patients was 26.7 μm. Three different OCT devices were used to test and compare the Iowa Reference Algorithms from the Iowa Institute for Biomedical Imaging, which are used for automatic intra-retinal layer segmentation and image scaling \RVV{\citep{Terry_2016(112)}}. Twenty-five healthy volunteers were scanned twice for macular volume using a 3D-OCT 1000 (Topcon), Cirrus HD-OCT (Zeiss), and a non-commercial long-wavelength (1040nm) OCT. Using the Iowa Reference Algorithms, the average thickness of 10 intra-retinal layers were calculated for the fovea, inner ring, and outer ring of the ETDRS field of view. The Iowa Reference Algorithms accurately segmented all 10 intra-retinal layers and showed more repeatability than the onboard software. With fixed-AEL scaling, the algorithm gave significantly different thickness values for the three OCT devices ($P<0.05$). 
An automated approach was suggested to segment and estimate the thickness of the ILM, inner outer segment, and RPE layers in an OCT image, and an online platform was made available for this purpose \RVV{\citep{Ometto_2019(113)}}.  Motamedi et al. \RVV{\citep{Motamedi_2019(114)}} aimed to establish normative data for macular RNFL, GCL-IPL, and INL thickness; the obtained measurements for these parameters were 39.53 ± 3.57 µm, 70.81 ± 4.87 µm, and 35.93 ± 2.34 µm, respectively. The present study \RVV{\citep{Kamal_Abdellatif_2019(115)}} aimed to investigate alterations in the thickness of the outer retinal layer in individuals of varying ages, utilizing images obtained through SD-OCT. The subjects included in the study were deemed to be within normal limits.

\noindent Babu et al. \RVV{\citep{Babu_2012(134)}} proposed an algorithm for glaucoma diagnosis with an improved correlation coefficient. In order to measure CDR, the retinal nerve head vitreal boundary (RV) and the choroid nerve head boundary (RC) were separated. The RV and RC choroid boundaries were identified using multilevel thresholding and wavelet transform techniques. The accuracy was 92\%, and the findings came extremely near to matching the values of the gold standard.  Nithya et al. \RVV{\citep{Nithya_2015(135)}} compared segmentation methods for OCT and fundus glaucoma diagnosis. Four normal and eight glaucoma images were included. Fundus and OCT images of the same patient were used to determine CDR. Cup and disc regions in a fundus image were segmented using  Hill climbing, fuzzy c-means clustering, and region growth. RPE and RNFL segmentation was used to determine cup and disc diameter in OCT images. Fundus and OCT CDR results were compared to clinical standards. Fuzzy c-mean clustering had the lowest performance error in experiments. Zhang et al. \RVV{\citep{zhang2015automated}} proposed an automated model to segment and quantify the  CME with macular hole (MH) in 3D OCT scans. The model consisted of three stages, denoising, flattening, and the segmentation of intra-retinal layers. Next, intra-retinal CME was segmented utilizing adaptive boosting and kernel graph cut. Following that, adaptive boosting and kernel graph cut were used for fine segmentation of intra-retinal CME. The model was evaluated on 3D OCT from 18 CME and MH subjects and achieved the accuracy and false positive volume fraction of 84.6\% and 1.7\%, respectively. Sugruk et al. \RVV{\citep{sugmk2014automated}} presented a model for the detection of AMD and DME. The model extracted the RPE layer from the macular OCT scans in order to diagnose AMD. Whereas to diagnose DME, cysts from the macular pathology were extracted. For cases of AMD, they found a success rate of 100\%, while for DME, they found a success rate of 86.6\%. Chiu et al. \RVV{\citep{chiu2015kernel}} proposed a kernel regression based classification model to identify  retinal layer boundaries and fluids within the retina. Then classification estimates were used to refine the extracted retinal boundaries while employing graph theory and dynamic programming framework. The model was evaluated on 110 B-scans from 10 subjects with severe DME pathology and achieved a mean Dice coefficient of 0.78. Wang et al. \RVV{\citep{wang2016machine}} developed a model to identify between AMD, DME, and healthy macula scans. The Correlation-based Feature Subset (CFS) selection algorithm was used to filter the linear configuration pattern (LCP) based OCT images. Overall accuracy for the three classes was 99.3\% for the best model based on the sequential minimum optimization (SMO) approach. Rashno et al. \RVV{\citep{rashno2017fully}} proposed a framework based on their proposed framework is based on neutrosophic transformation and graph-based shortest path search for the extraction of fluid-filled cyst segments from OCT scans. After undergoing a neutrophic transformation, an image was divided into three zones: true, indeterminate, and false. Noise in an image was represented by the indeterminate set, whereas the true set was obtained using their gamma-correction technique. The ILM, RPE, OPL, and ISM layers were extracted using a graph shortest path search. Using ILM and RPE, a target region of interest (ROI) was created, from which fluid-filled regions are automatically extracted using a cluster-based segmentation algorithm. Moreover, they were able to reach a sensitivity of 67.3\% on the Duke Dataset-II, 88.8\% on the Optima dataset, and 76.7\% on their own dataset. Tehmina et al. \RVV{\citep{Khalil2018Access}} developed a technique that segments retinal layers to calculate CDR for glaucoma diagnosis. Delineating ILM and RPE measured cup-diameter-calculation (CDC) and disc-diameter-calculation (DDC), respectively, and employed countor, interpolation, and thickness value estimation techniques.
\subsection{Machine Learning Schemes}
\noindent This section presents the techniques based on classical ML models, such as linear regression, logistic regression, random forest, support vector machine (SVM), and XGBoost, for the identification and segmentation of different significant biomarkers of DR, DME, AMD, and glaucoma.\\

\noindent \textbf{Detection of Retinal Lesions through Fundus Scans:}
\noindent DR is a serious threat to sight and must be diagnosed and treated early to prevent permanent vision loss. MAs are the earliest symptom of DR, and their diagnosis is crucial. The study proposed \RVV{\citep{akram2013identification}} a three-stage methodology to discover MAs by early utilizing filter banks. The technique began by identifying and removing any potential MA candidate regions from the retinal image. The system created a feature vector for each candidate region based on variables such as shape, color, intensity, and statistics to determine the MA region. In order to increase the accuracy of classification, a hybrid classifier incorporated the Gaussian mixture model (GMM), the SVM, and an extension of the multimodel mediod based modeling approach in an ensemble. The model was evaluated on publicly available datasets DIARETDB0 and DIARETDB1. Usman et al. \RVV{\citep{akram2014detection}} introduced a method for identifying and categorizing NPDR lesions. The system that was proposed involved preprocessing, the extraction of candidate lesions, the creation of a feature set, and classification. Candidates for the various NPDR signs (MAs, HMs, and EXs) were extracted. A feature set was created for each lesion based on the characteristics of lesions. The real lesions are found and labeled with the use of a hybrid classifier, which was based on weighted mixture of multivariate m-Mediods and a GMM. Based on the types, number, and locations of lesions, the system categorized retinal images into different stages of NPDR. The proposed model was evaluated on four databsets: DRIVE, STARE, MESSIDOR, and DIARETDB, and achieved accuracy of 95\%, 97.5\%, 98.90, and 95.05\%, respectively. Huda et al. \RVV{\citep{huda2019improved}} proposed a classification model for DR diagnosis based on decision trees, logistic regression, and SVM. Jebaseeli \RVV{\citep{jebaseeli2021prediction}} developed a system that classified DR and analyzed the disease severity with high accuracy. Adaptive Histogram Equalization (AHE) technique was used for image enhancement. After that, Hop Field Neural Network technique simultaneously segmented the boundaries and determined width of thevessels in fundus scan. The model was tested using a local dataset in addition to publicly available datasets (DRIVE, STARE, MESSIDOR, HRF, and DRIONS).
Retinal blood vessel analysis on fundus images can provide a variety of significant biomarkers of retinal disease. DR is one of the ocular diseases that can be detected by analyzing the blood vessels in the retina. Deciphering cardiovascular illness from a retinal fundus image requires a careful examination of the vascular tree. Bifurcations and crosses of blood vessels must be located for an accurate study. Using COSFIRE filters, the study \RVV{\citep{azzopardi2013automatic}} presented a method for detecting automatically vascular bifurcations in segmented fundus images.
The COSFIRE filter's output was determined by taking the geometric mean of the weighted responses of the blurred and shifted Gabor filters that were specifically chosen. The performance of algorithm was done on DRIVE and STARE datasets. Recall of 97.88\% and precision of 96.94\% were achieved on forty fundus scans from the DRIVE data set. Twenty manually segmented images from the STARE dataset had a recall of 97.32\% and a precision of 96.04\%. Manoj et al. \RVV{\citep{manoj2013neural}} proposed a technique that used feature based on orientation gradient vector fields, morphological transformation, and Gabor filter responses to extract the retinal vasculature in order to diagnose retinal disorders. A vector in 9-D feature space describes each pixel in the retinal image, and neural network classifiers are used to categorize those pixels using Feed Forward Backpropagation Neural network, Multi-Layer Perceptron, and Radial Basis Function. The method was evaluated on DRIVE, STARE, and MESSIDOR and achieved an accuracy of 96.23\%, 95.83\%, and 95.41\%, respectively. Strisciuglio et al. \RVV{\citep{strisciuglio2016supervised}} proposed a robust method  based on a set of B-COSFIRE filters selective to segment blood vessels in fundus images. Features were chosen automatically for maximum flexibility, and they can be customized for a variety of other applications. Analyzed and compared the efficacy of several distinct selection approaches based on the principles of machine learning and information theory. \\
Glaucoma affects the RNFL, which results in increased CDR; it is a clinically significant parameter for glaucoma diagnosis and screening. Computational analysis, such as the CDR, cup area, and rim area, is made possible by fundus imaging, greatly MESSIDOR assisting in identifying glaucoma. The fundus image provides the analysis of OC and OD; however, the edges of the OC are not very clear. Because of this, it is very hard to segment the OC accurately, and the performance of OD segmentation also needs to be improved. Usman et al. \RVV{\citep{akram2015glaucoma}} presented a unique feature set-based diagnostic system for automated identification of glaucoma using fundus images. The system was comprised of modules, which are as follows: preprocessing, detection of the region of interest based on autonomously segmented OD, feature extraction, and classification. Robust OD localization method \RVV{\citep{usman2014robust}} was employed. With the use of 2-D MESSIDORGabor wavelet and subsequent thresholding-based vessel segmentation, the vascular pattern is made more visible. Following ROI extraction, many features (CDR, Rim to disc ratio, mean intensity, standard deviation, energy, and gradient) were taken from it to create a detailed representation of the feature space. Local Fisher discriminant analysis (LFDA) was performed to do supervised enhancement of features. The retinal images were classified as normal or glaucoma using the m-Mediods model of normality. The performance of the proposed system was evaluated using publicly available (DRIVE, DiaretDB, Drions, HEI MED, HRF, MESSIDOR) and locally gathered fundus databases. An automated technique \RVV{\citep{mvoulana2019fully}} was proposed for glaucoma diagnosis from fundus scans. First, the method segmented the OD by combining a brightness criterion and a template-matching technique. Next, texture-based and model-based methods were employed to segment the OD and OC accurately. Finally, glaucoma screening is achieved through the calculation of the CDR, which allows for the differentiation between healthy and glaucomatous individuals. A publicly accessible DRISHTI-GS1 dataset was used to evaluate the proposed method and achieved 98\% accuracy. The study \RVV{\citep{mohamed2019automated}} proposed an automatic glaucoma screening model based on superpixel classification. The preprocessing steps were noise removal and illumination correction, then input images were aggregated into superpixels by Simple Linear Iterative Clustering (SLIC). The statistical pixel-level (SPL) technique extracted image attributes from each superpixel based on histogram data and textural information. The extracted features are then fed into SVM to classify each superpixel into OD, OC, blood vessel, and background regions. On RIM-One dataset, the model was tested and achieved an accuracy and sensitivity of 98.6\% and 92.3\%, respectively. Rehman et al. \RVV{\citep{rehman2019multi}} employed region-based statistical and textural features to detect and localize OD in fundus images. Highly discriminative features were selected using the mutual information criterion, and four benchmark classifiers, SVM, RF, AdaBoost, and RusBoost, were compared. The RF classifier showed more competitive results than other classifiers and achieved an accuracy of 99.3\%, 98.8\%, and 99.3\% on the DRIONS, MESSIDOR, and ONHSD datasets, respectively. DME is an ocular condition in which fluid rich in fat drains out of damaged blood vessels and is deposited near the macula, causing blurred central vision. The study \RVV{\citep{akram2014automated}} proposed a novel approach for macula detection while utilizing a rich feature set and a classifier based on the GMM. The method was evaluated on the DRIVE and STARE databases and achieved an accuracy of 100\% and 95.4\%, respectively.
In ophthalmology, a transportable and cost-effective computer-aided diagnosis system can be achieved by the use of the d-Eye lens, which can be attached to a smartphone \RVV{\citep{elloumi2018mobile}\citep{elloumi2021fast}}. Mrad et al. \RVV{\citep{mrad2022fast}} proposed an automated technique for glaucoma screening that is specifically designed for fundus images captured by smartphones was provided (SCFIs). The first challenge was to design an algorithm that achieved a higher level of accuracy even with moderate-quality SCFIs. The second task is to make the detection process computationally cost and recourse effective so that the method can be used on a smartphone. To do so, a central concept was used to infer glaucoma from vessel displacement inside the OD, where the vascular tree may still be well described using SCFIs. So, the vessel tree is broken up into sections and divided into quadrants based on the ISNT. The centroid of the vessel distribution on each quadrant is then determined. After the feature vector was generated, it was fed into a classifier (SVM) to diagnose glaucoma accurately.\\

\noindent \textbf{Detection of Retinal Lesions through OCT Scans:}
\noindent An automatic technique based on graph theory dynamic programming and SVM was presented in the study \RVV{\citep{Srinivasan_2014(105)}}. 
Seven to 10 retinal layers were effectively extracted using the proposed method.
Septiarini et al. \RVV{\citep{septiarini2018automated}}  proposed an RNFL segmentation model by creating a co-occurrence matrix. The model employed 160 and 40 fundus images for  training and testing, respectively, and achieved an accuracy of 94.52\%. The study \RVV{\citep{zang2019automated}} performed the analysis of retinal layers and capillary plexuses from OCT and OCTA scans by segmenting the optic disc and retinal layers. A neural network and graph search technique was combined to segment the OD. The study \RVV{\citep{hassan2016fully}} proposed an automated algorithm for detecting ME while using directional gradients of the candidate OCT scan. Three features from computed gradients were extracted and used to train linear discriminant analysis to classify ME and healthy scans. They tested their proposed system on 30 OCT B-scans and got a sensitivity of 100\% and a specificity of 86.67\%. In the work \RVV{\citep{abhishek2014segmentation}}, an automated segmentation approach was described for detecting intra-retinal layers in OCT images that are significant for edema detection. They found RPE layer and detected the shape of the drusen, and finally, the technique employed a binary classification to distinguish between AMD and DME scans. Results from experiments showed that AMD and DME were classified with an accuracy of 87.5\%. Srinivasan et al. \RVV{\citep{srinivasan2014fully}} developed an algorithm for the detection of AMD and DME based on multiscale histograms of oriented gradient descriptors.  Finally, supervised SVM was used for the classification task. The classifier was successful in identifying all cases of AMD and DME and 86.67\% of normal patients. The quantitative classification of AMD and normal eyes through retinal OCT images was presented \RVV{\citep{farsiu2014quantitative}}. The model semi-automatically segmented the RPE, drusen, and retina. A map of "normal" non-AMD thickness was created by registering and averaging maps of thickness from control participants. Five automated classifiers were generated based on a generalized linear model regression framework. The classifier achieved an area under the curve (AUC) greater than 0.99.
Khalid et al. \RVV{\citep{khalid2017fully}} proposed a model for the diagnosis of retinal epithelial detachment (RE), CSR, and AMD based on multilayered SVM. The model was evaluated on 2819 OCT images (1437 healthy, 640 RE, and 742 CSCR) from 502 patients across two datasets, achieving an accuracy of 99.92\%, a sensitivity of 100\%, and a specificity of 99.86\%. The study \RVV{\citep{hassan2016structure}} presented proposes an automated framework based on SVM to classify ME and CSR from OCT images. A total of 30 labeled images (10 ME, 10 CSR, and 10 healthy) were utilized for training a model while using five features (two derived from cyst fluids inside the retinal layers and three features were extracted from thickness profiles of the sub-retinal layers). A total of 90 TD-OCT images (30 for ME, 30 for CSR, and 30 for healthy) from 73 patients were used to evaluate the algorithm. It correctly identified 88 of 90 cases (97.77\% accuracy, 100\% sensitivity, and 93.33\% specificity). Hassan et al. \RVV{\citep{hassan2016automated}} utilized coherent tensors to develop an automated method to segment and evaluate the subretinal layers in OCT scans. Then, the SVM classifier was employed to make a ME prediction based on the subretinal layers of the candidate images. Seventy-one OCT images were obtained locally from 64 patients, 15 and 49 were ME and healthy subjects, respectively. Overall, the model successfully distinguishes between ME patients and healthy subjects with an accuracy of 97.78\%. The proposed model \RVV{\citep{rathore2021bright}} assists in predicting whether or not DR may be identified based on the number of exudates visible in retinal fundus images. Several procedures have been taken in order to detect exudates, including scaling, removal of blue channels, performing feature extraction with Local Binary patterns (LBP), and classifying the images via SVM. 

\subsection{Deep Learning (DL) Schemes}
\noindent DL computer models have recently made significant advances in various fields, such as computer vision, speech recognition, genomics, drug discovery, and ophthalmology. The DL algorithm can automatically learn complex structures from large data sets without explicit feature extraction. However, in order to achieve generalization  large amount of data is required to train DL model. We have divided the DL literature into different categories, such as the segmentation model, segmentation-based classification model, and  classification model. The segmentation section includes the studies which only extract the different biomarkers from fundus/OCT/OCTA scans for various ocular diseases and the segmentation-based classification model performed classification based on the identified biomarkers. Whereas the classification section includes those studies which performed the classification of scans without any segmentation performed.

\subsubsection{Segmentation}
\paragraph{Retinal Lesions \& Optic Cup/Disc Segmentation from Fundus Scans}
\noindent Sevastopolsky et al. \RVV{\citep{sevastopolsky2017optic}} proposed a universal method for automatically segmenting the OD and OC, which was based on U-Net CNN. CLAHE was used as a preprocessing step to equalize the contrast. The model was evaluated on publicly available databases DRIONS-DB, DRISHTI-GS, and RIM-ONE v.3 and achieved IOU of 0.89, 0.75, and 0.69, respectively. A CNN model was developed and trained to automatically and simultaneously segment the OD, fovea, and blood vessels \RVV{\citep{tan2017segmentation}}. Fundus images were normalized, then three channels were extracted and fed into the model. On Drive dataset, the model correctly classified 92.68\% of the ground truths. The best single-image accuracy was 94.54\%, and the worst was 88.85\%. The study \RVV{\citep{al2018dense}} proposed a DL model for the segmenting OC and OD. The model was based on DenseNet, a fully-convolutional network with a symmetric U-shaped topology that enables pixel-wise classification. The CDR for glaucoma diagnosis is then estimated along two axes using the projected OD and OC boundaries. The model was evaluated on four publicly available datasets, ORIGA, DRIONS-DB, Drishti-GS, ONHSD, and RIM-ONE. The results showed model achieved better segmentation, and it was suggested that the model could be used to detect various other retinal lesions. The majority of current ML segmentation techniques rely on manual segmentation of the disc. The annotation of pixel-level optic disc masks is a time-consuming task that invariably results in inter-subject variance. To address this issue, Xiong et al. \RVV{\citep{xiong2022weak}} proposed an automatic Bayesian U-Net with weak labels and Hough transform-based annotations to segment OD from fundus images. The expectation-maximization approach was alternately applied to estimating the OD mask and updating the weights of the Bayesian U-Net in order to optimize the model. Another study \RVV{\citep{lu2019weakly}} presented a weakly-supervised learning approach based on modified CNN to segment the OD in fundus images. Labels at the image level and bounding box labels were employed to guide segmentation. The enhanced constraint CNN method was combined with the GrabCut method to construct a more refined foreground segmentation map with image-level labels and use them as "GroundTruth" for the subsequent training step. A weak loss function was used to constrain the training network base output size of a modified U-net model. The model was evaluated on RIM-ONE and DRISHTI-GS databases. This DL network contains 22 layers, which had 11 inception modules. Li et al. \RVV{\citep{li2018efficacy}} proposed DL algorithm for recognizing glaucomatous optic neuropathy (GON) from color fundus images. A total of 70000 fundus images were randomly acquired from LabelMe \RVV{\citep{label}}. The results showed model achieved an AUC of 0.986 with a specificity of 92.0\% and a sensitivity of 95.6\%. Sun et al.\RVV{\citep{sun2018optic}} proposed based on deep object detection networks to segment OD from retinal fundus images. To find the OD border by transforming the projected bounding box into a vertical and non-rotated ellipse. The method outperformed state-of-the-art algorithms in OD segmentation on the ORIGA dataset using Faster R-CNN as the object detector.
The study \RVV{\citep{sun2022gnas}} presented a Neural architecture search (NAS) in a two-level nested U-shaped structure. The segmentation model achieved average dice of 92.88\% on the REFUGE dataset. The model was validated on Drishti-GS and GAMMA and obtained a dice of 92.32\% and 92.11\%, respectively. Sharath et al. \RVV{\citep{shankaranarayana2019fully}} proposed a DL framework to estimate monocular retinal depth from a fundus image. To handle the sparsity of labeled data, pretraining the deep network using a pseudo-depth reconstruction technique which was more effective than denoising methods. A fully convolutional guided network that used the depth map and the fundus image to perform OD and OC segmentation. The model was evaluated on three datasets ORIGA, RIMONEr3, and DRISHTI-GS. The study \RVV{\citep{tian2020graph}} used a multi-scale CNN to extract feature maps. For the segmentation task, GCN requires the feature map to be appended to graph nodes. The model was tested on the REFUGE dataset, Dice similarity coefficients (DSC) of the proposed technique for OD and OC were 0.97 and 0.95, respectively. A DL model DDSC-Net (densely connected depthwise separable convolution network) \RVV{\citep{liu2021joint}} was proposed for OD and OC based on multi-category semantic segmentation. To achieve better segmentation results, the model utilized an image pyramid input and a depthwise separable convolutional layer. Model was evaluated on two publicly available datasets, Drishti-GS and REFUGE. The DDSC-Net outperformed GL-Net by 0.70 in disc coefficients on the Drishti-GS dataset and pOSAL by 0.79\% on the REFUGE dataset. Wang et al. \RVV{\citep{wang2019coarse}} developed a coarse-to-fine DL architecture based on a classical CNN, the U-net model, to accurately segment the OD. The network used two distinct sets of inputs during training: color fundus images and their corresponding grayscale vessel density maps. The model fused the data using an overlap technique to locate a local image patch (disc candidate region), which was then used as input into the U-net model for further segmentation. On our dataset of 2978 test images, the model achieved an average of 0.89 for IoU and 0.93 for DSC. Surendiran et al.\RVV{\citep{surendiran2022segmentation}} developed modified recurrent neural networks (mRNN) with fully convolutional network (FCN) for the extraction and segmentation of OD and OC. FCN generated a feature map for the intra- and interslice contexts, whereas RNN paid more attention to the interslice context. A novel method JOINED proposed for multi-task learning for joint OD, OC, and fovea detection \RVV{\citep{he2022joined}}. To make the most of the information provided by the distance from each image pixel to landmarks of interest, a distance prediction branch was built in addition to the segmentation and detection branches. The JOINED pipeline has two stages: the coarse stage and the fine stage. At the coarse stage, a joint segmentation and detection module performed OD/OC coarse segmentation and generated a heatmap, which showed the location of fovea. After that, ROI was cropped for further fine processing, and use the predictions from the coarse stage as extra information to improve performance and speed up convergence. The model was on publicly available GAMMA, PALM, and REFUGE datasets. Although many DL methods have shown promising results in the area of OD and OC segmentation, it remains a difficult problem to segment the OC boundary while also increasing computing efficiency correctly. To address this issue, study \RVV{\citep{wei2022rmsdsc}} proposed a robust Multiscale Feature Extraction with Depthwise Separable Convolution (RMSDSC-Net) that tradeoff between performance and computational cost. The basic building blocks were the Multiscale Input (MSI), Dilated Convolution Block (DCB), Depthwise Separable Convolution Unit (DSCU), and External Residual Connection (ERC). MSI can help to mitigate data loss caused by the network's pooling layers when it comes to having detailed representations of features. The model builds DSCU and DCB modules to improve segmentation performance and computational efficiency by preserving higher-level semantic features while minimizing the loss of spatial information from tiny details in the image. Finally, ERC was set up between the encoding and decoding layers to reduce feature degradation as much as possible. The model achieved Dice Coefficients of (0.978, 0.919) and (0.965, 0.910) for OD and OC segmentation on the DRISHTI-GS and REFUGE databases, respectively. Garifull et al. \RVV{\citep{garifullin2021deep}} developed a model for segmenting DR lesions based on a Bayesian baseline. The model considered the parameters of a CNN as random variables and used stochastic variational dropout approximation to quantify uncertainty. The method achieved an AUC of 0.84 for HMs, 0.641 for EXs, 0.593 for MAs, and 0.484 for microaneurysms on IDRiD dataset.
State-of-the-art models cannot achieve significant segmentation results because of a lack of sufficient pixel-level annotated data during training. To address this shortcoming, Lui et al.\RVV{\citep{liu2019joint}} proposed a semi-supervised conditional GAN-based approach for joint OD and OC segmentation. The model comprised a segmentation net, a generator, and a discriminator, that learned to map between the fundus images and segmentation maps. To further enhance the segmentation performance, both labeled and unlabeled data were used. Extensive trials demonstrated that model attained improved results on ORIGA and REFUGE datasets for segmenting the optic disc and cup. Jiang et al. \RVV{\citep{jiang2019optic}} proposed a multi-label DL model(GL-NET) that combined the GANs. GL-Net had a generator and discriminator. In the generator, skip connections were used to facilitate the fusion of low and high-level feature information, which minimizes the downsampling factor and prevents excessive feature information loss. L1 distance and cross-entropy were used as loss functions to improve segmentation accuracy. The model was verified on DRISHTI-GS1 dataset. Another study \RVV{\citep{son2019towards}}  presented the OD and blood vessel segmentation model based on GAN. Results showed the model achieved better performance in blood vessel segmentation on DRIVE and STARE datasets. However, OD segmentation on DRIONS-DB, RIM-ONE, and Drishti-GS datasets did not provide statistically significant increases in AU-ROC. Table \ref{OD} summarized the various DL model for the segmentation of OD.

\begin{table*}
\centering
\caption{Summarizing the different DL-based studies that segmented the OD. }
\begin{tblr}{
  cell{4}{4} = {},
  cell{4}{6} = {},
  hlines,
  vline{2-6} = {-}{},
}
\textbf{Study}                                                        & \textbf{Techniques}          & \textbf{Dataset}                                     & \textbf{IOU}                                                                                                                                                                                                           & \textbf{Dice}        & \textbf{Accuracy}                                                                                                                                                                                                                                                                                                  \\
{\RVV{\cite{sevastopolsky2017optic}}} & {CLAHE\\U-Net CNN}           & {DRIONS-DB,\\DRISHTI-GS, \\RIM-ONE v.3}              & {0.89, \\0.75, \\0.69}                                                                                                                                                                                                 & -                    & -                                                                                                                                                                                                                                                                                                                  \\
{\RVV{\cite{tan2017segmentation}}}                 & CNN                          & DRIVE                                                & -                                                                                                                                                                                                                      & -                    & 92.68\%                                                                                                                                                                                                                                                                                            \\
{\RVV{\cite{al2018dense}}}                        & DenseNet                     & {ORIGA, \\DRIONS-DB, \\DRISHTI-GS, \\ONHSD\\RIM-ONE} & {\textcolor[rgb]{0.133,0.133,0.133}{}\\\textcolor[rgb]{0.133,0.133,0.133}{0.77,\textcolor[rgb]{0.133,0.133,0.133}{0.94,}}\\0.90,\\\textcolor[rgb]{0.133,0.133,0.133}{0.91,}\\\textcolor[rgb]{0.133,0.133,0.133}{0.90}} & -                    & {\textcolor[rgb]{0.133,0.133,0.133}{}\\\textcolor[rgb]{0.133,0.133,0.133}{93.0\%}\\\textcolor[rgb]{0.133,0.133,0.133}{89.12\%}\\99.69\%\\\textcolor[rgb]{0.133,0.133,0.133}{99.90\%}\\\textcolor[rgb]{0.133,0.133,0.133}{99.69\%}} \\

{\RVV{\cite{sun2022gnas}}}                       & CNN                          & {REFUGE,\\Drishti-GS,\\GAMMA\\~}                     & -                                                                                                                                                                                                                      & {0.92,\\0.92,\\0.92} & -                                                                                                                                                                                                                                                                                                                  \\
{\RVV{\cite{liu2021joint}}}                      & {Depth-wise \\separable CNN} & {Drishti-GS,\\REFUGE}                                & -                                                                                                                                                                                                                      & {0.70,\\0.79}        & -                                                                                                                                                                                                                                                                                                                  \\
{\RVV{\cite{wang2019coarse}}}                  & U-net~                       & Local~                                               & 0.89                                                                                                                                                                                                                   & 0.93                 & -                                                                                                                                                                                                                                                                                                                  \\
{\RVV{\cite{wei2022rmsdsc}}}                     &                              & {DRISHTI-GS,\\REFUGE}                                & {0.97, \\0.91}                                                                                                                                                                                                         & -                    & -                                                                                                                                                                                                                                                                                                                  
\end{tblr}
\label{OD}
\end{table*}

\paragraph{Blood Vessels Segmentation from Fundus Scans}
\noindent The paper \RVV{\citep{wang2015hierarchical}} introduced a supervised approach for retinal blood vessel segmentation by combining CNN and Random Forest. The CNN served as a hierarchical feature extractor that can be trained, while the ensemble Random Forest performed the role of a classifier. The suggested method automatically learned features from the raw images and predicted the patterns using a combination of the benefits of feature learning and classical classifier. Two publicly available databases, DRIVE and STARE, of retinal images were used to evaluate the model and achieved an accuracy of 97.67\% and 98.13\%, respectively. To improve feature recognition in retinal images, a unique method was proposed \RVV{\citep{fang2015retinal}}, which first applied the DL method for vessel segmentation in order to produce the probability map of the image. The multi-scale Hessian response on the retinal image's probability map was then used to detect landmarks. Melinščak et al. \RVV{\citep{melinscak2015retinal}} proposed a CNN-based model for the classification of retinal vasculature. The network has four convolutional and max pooling layers and two fully connected layers. The ReLU activation function was used in convolutional layers, and the softmax activation function was employed in the last fully connected layer. The model achieved an average accuracy of 94.66\% on DRIVE dataset. The study \RVV{\citep{fu2016retinal}} proposed a DL model and utilized the CNNs to generate a vessel probability map. The model was a modification of a holistically nested edge detection (HED) network \RVV{\citep{xie2015holistically}}. Probability maps from output layers were combined to produce a single probability map. Conditional Random Fields (CRF) were utilized for the exact localisation of vascular boundaries. Mean field approximation of CRF distribution yields maximum posterior marginal inference. The method achieved an accuracy of 94.70\% and 95.45\%  on the DRIVE and STARE datasets, respectively. Liskowski et al. \RVV{\citep{liskowski2016segmenting}} proposed a DL model for vessel segmentation. The preprocessing steps were based on global contrast normalization and zero-phase whitening. Data augmentation was performed to increase the number of images. Six distinct CNN models were constructed: PLAIN, GCN, ZCA, AUGMENT, NO-POOL, and BALANCED. The segmentation models were verified on DRIVE, STARE, and CHASE databases. A supervised method \RVV{\citep{jiang2018retinal}} was proposed based on a pre-trained fully CNN through transfer learning. The suggested method reduces the complex challenge of retinal vascular segmentation from full-size picture segmentation to regional vessel element detection. Unsupervised image post-processing techniques were applied to the proposed method to enhance the final result further. Using the DRIVE, STARE, CHASE DB1, and HRF databases, extensive testing had shown an accuracy of 95.93\%, 96.53\%, 95.91\%, and 96.62, respectively. Hu et al. \RVV{\citep{hu2018retinal}} proposed a model based on CNN and fully connected CRFs. There are essentially two phases to the segmentation procedure. First, a multiscale CNN architecture with an enhanced cross-entropy loss function was presented to generate the inter-image probability map. To acquire more specific knowledge of the retinal arteries, a multiscale network was built by merging the feature map of each intermediate layer. The proposed cross-entropy loss function concentrates on learning the difficult cases and pays less attention to losing small amounts of data on the easier samples. Second, the final binary segmentation result was achieved by applying CRFs, which used spatial context information by considering the interactions among all of the pixels in the fundus images. The DRIVE and STARE public datasets were used to test the efficacy of the proposed method and achieved an accuracy of 95.33\% and 96.32\%, respectively.
The study \RVV{\citep{wang2019dense}} presented a model based on Dense U-net and the patch-based learning strategy for retinal vessel segmentation. Training patches were obtained using a random extraction strategy, the Dense U-net was utilized as the training network, and a random transformation technique was employed to augment the training data. The segmented image can be recovered using an overlapping-patches sequential reconstruction technique. The DRIVE and STARE public datasets were used to test the effectiveness of the model, which achieved an accuracy of 95.11\% and 95.38\%, respectively.
By balancing losses with stacked deep FCN, Park et al. \RVV{\citep{park2020m}} proposed a novel conditional generative adversarial network termed M-GAN for performing retinal vascular segmentation. For enhanced segmentation, a M-generator with deep residual blocks was included, while an M-discriminator with a greater in-depth network facilitates more rapid adversity-based model training. In particular, to facilitate scale-invariance of vessel segmentations of varying sizes, a multi-kernel pooling block was included between the stacked layers. The M-generator utilizes down-sampling layers to collect relevant data for feature extraction and up-sampling layers to create segmented retinal blood vessel pictures from the collected data. The pre-processing step utilized automated color equalization (ACE) to improve the visibility of the retinal vessels, and post-processing with a Lanczos resampling approach to smooth the vessel branching that reduced false negatives. To verify the proposed method, DRIVE, STARE, HRF, and CHASE-DB1 datasets were used and achieved an accuracy of 97.06\%, 98.76\%, 97.61\%, and 97.36\%, respectively.
The study \RVV{\citep{boudegga2021fast}} proposed U-shaped DL design with lightweight convolution blocks used to maintain good performance while decreasing computational complexity. In order to improve the quality of the retinal image and the information gleaned from the blood vessels, a series of preprocessing and data augmentation techniques was proposed. The proposed method was evaluated on the DRIVE and STARE databases. It was shown to produce a better compromise between the retinal blood vessel identification rate and the detection time, with an average accuracy of 97.80\% and 98\% in 0.59 s and 0.48 s per fundus image, respectively.
Wang et al. \RVV{\citep{wang2021fine}} proposed a model for fine retinal vascular segmentation by integrating Nest U-net and patch-learning. Training samples that included fine retinal vessels were generated efficiently with the help of a custom extraction approach, giving them a significant advantage in the segmentation of these specialized structures. The model sent high-resolution feature maps directly from the encoder to the decoder network. The model was learned with k-fold cross-validation, predictions were made using testing samples, and the final retinal vasculature was reconstructed using a sequential approach. The proposed model was evaluated using the DRIVE and STARE datasets and achieved an accuracy of 95.12\% and 96.41\%, respectively.

\noindent According to the properties of the retinal vessels in fundus scans, a residual CNN-based retinal vessel segmentation model was presented \RVV{\citep{xu2021retinal}}. The encoder-decoder network structure was built by joining the low-level and high-level feature graphs, and dilated convolution was integrated into the pyramid pooling. In addition to this, an improved residual attention module and deep supervision module were also used. The results from data sets DRIVE and STARE demonstrate that algorithm can segment the entire retinal vessel, along with related vessel stems and terminals. The accuracy achieved on DRIVE and STARE datasets was 95.90\% and 96.88\%, and specificity was 98.85\% and 97.85\%, respectively.
Deng et al. \RVV{\citep{deng2022retinal}} proposed a model for vessel segmentation based on multi-scale attention with a residual mechanism D-Mnet (Deformable convolutional M-shaped Network) and an improved PCNN (Pulse-Coupled Neural Network) model. The model was based on the encoder-decoder network structure. In order to boost the efficiency of retinal blood vessel segmentation, the network integrates an enhanced PCNN model, bringing together the benefits of supervised and unsupervised learning. Publicly available databases, DRIVE, STARE, CHASE-DB1, and HRF were used to conduct comparative verification, the model achieved an accuracy of  96.83\%, 97.32\%, 97.14\%, and 96.68\%, respectively. Kar et al. \RVV{\citep{kar2022retinal}} proposed a segmentation model based on a generative adversarial network (GAN) and several loss functions to detect retinal vessels accurately. CLAHE method was used in the preprocessing stage to improve the contrast of blood vessels. The GAN architecture combined a segmentation network (the generator) and a classification network (the discriminator). The inception module detects fine vessel segments by extracting multi-scale properties of vessel segments at various scales. The discriminator is made up of two layers of stacked self-attention networks and a layer of position-wise fully linked feed-forward networks that infer a binary output. The transformer's attention mechanism can effectively discriminate and store global and local information. The DRIVE, STARE, CHASE DB1, HRF, ARIA, IOSTAR, and RC-SLO databases were used to test the robustness and effectiveness of the proposed method. 
Chen et al. \RVV{\citep{chen2022pcat}} proposed Patches Convolution Attention-based Transformer UNet (PCAT-UNet), which was a U-shaped network with a convolution branch based on transformer. Skip connections were employed to fuse the deep and shallow features. The model captures the global dependency connection and the features of the underlying feature space, overcoming the difficulties of insufficient retinal microvessel feature extraction and low sensitivity caused by easily predicting pixels as background. PCAT-UNet was evaluated using three publicly accessible retinal vasculature datasets DRIVE, STARE, and CHASE DB1. Results for accuracy were 96.22\%, 97.96\%, and 98.12\%, and results for sensitivity were 85.76\%, 87.03\%, and 84.93\%, respectively. The study presented \RVV{\citep{zhang2022edge}} framework that provided new edge-aware flows into U-Net encoder-decoder architecture to steer retinal vascular segmentation, which makes the segmentation more sensitive to the capillaries' fine edges. Using characteristics taken from the encoder path, edge-gated flow with gated convolution learns to highlight the vessel edges and then exports the resulting edge prediction. To further improve the segmentation results, the edge-downsampling flow extracted the edge features from the edge prediction output and re-feeds them into the decoder path. On the publicly available DRIVE, STARE, and CHASEDB1 datasets, the proposed technique outperforms the state-of-the-art U-Net baseline by 0.0056, 0.0026, and 0.0047, respectively. Most segmentation approaches still have certain shortcomings in effective fine vessel detection, however, mainly as a result of information loss issues produced by many pooling operations and insufficient process issues of local context features by skip connections. In order to solve this problem, a novel retinal vascular segmentation network named ResDO-UNet \RVV{\citep{liu2023resdo}} was proposed based on the encoder-decoder architecture to offer an automatic and end-to-end detection strategy using fundus photographs. Together with a depth-wise over-parameterized convolutional layer (DO-conv), a residual DO-conv (ResDO-conv) network was proposed to serve as the network's backbone in order to obtain robust context features, which would improve feature extraction. Furthermore, a pooling fusion block (PFB) was developed to implement nonlinear fusion pooling, which used the benefits of max pooling and average pooling layers to mitigate the information loss that results from performing numerous pooling operations. Meanwhile, an attention fusion block (AFB) was employed as a solution to the problem of insufficient processing of local context information by skip connections. The model was evaluated on DRIVE, STARE, and CHASE\_DB1 datasets.
In the study \RVV{\citep{qu2023tp}}, a vessel segmentation model (TP-Net) based on fundus images was proposed, which consisted of three modules, i.e., main-path, sub-path, and multi-scale feature aggregation module (MFAM). The main path's responsibility is to identify the retinal vascular trunk, while the branching path is to track vessel information accurately. Retinal vascular segmentation was improved by combining the predictions from the two pathways using MFAM. In the main-path, a three-layer lightweight backbone network was designed based on retinal vessel features. Then a global feature selection mechanism (GFSM) was developed for the automated selection of features that are significant for the segmentation task. An edge feature extraction approach and an edge loss function were proposed in sub-path, which improved the network's ability to capture edge information. Finally, MFAM combined the prediction by main and sub-paths, which can eliminate disturbances from the background while still keeping edge features, leading to improved vascular segmentation. The proposed TP-Net was tested using the DRIVE, STARE, and CHASE-DB1 datasets. A new lightweight segmentation model Wave-Net \RVV{\citep{liu2023wave}}, was proposed for accurate vascular segmentation in the fundus images. The skip connections of the original U-Net were replaced with a detail enhancement and denoising block (DED) to improve the precision segmentation. DED reduced the impact of the semantic information loss problem in thin vessels and learned more about micro structures and features. In addition, it helped to reduce the impact of the semantic gap issue. Additionally, a multi-scale feature fusion block (MFF) was created for multi-scale vessel identification to fuse cross-scale contexts. The model was evaluated on DRIVE, CHASEDB1, and STARE datasets and achieved the F1 score of 0.8254, 0.8349, and 0.8140, respectively. 

\noindent Table \ref{bloodvessel} summarises the studies that performed segmentation of blood vessels in fundus images. There are various other DL models, such as Bridge-Net \RVV{\citep{zhang2022bridge}}, CSAUNet \RVV{\citep{huang2022csaunet}}, DilUnet \RVV{\citep{huang2022csaunet}}, Staircase-Net \RVV{\citep{sethuraman2022staircase}}, DCCMED-Net \RVV{\citep{budak2020dccmed}}, CcNet \RVV{\citep{feng2020ccnet}}, MD-Net \RVV{\citep{shi2021md}}, NFN+ \RVV{\citep{wu2020nfn+}}, DF-Net \RVV{\citep{yin2022df}} found in the literature for segmentation of retinal vessels in fundus images.

\begin{table*}
\centering
\caption{Summarizing the DL models for the segmentation of blood vessels in fundus images }
\arrayrulecolor{black}
\begin{tabular}{l!{\color{black}\vrule}l!{\color{black}\vrule}l!{\color{black}\vrule}l!{\color{black}\vrule}l!{\color{black}\vrule}l} 
\hline
\textbf{Study}                                                                                                                            & \textbf{Techniques}                                                          & \textbf{Dataset}                                                               & \textbf{Accuracy}                                                                                                                                & \textbf{Specificity }                                                                                                                                                                                                                                                                                                                                                                     & \textbf{Sensitivity}                                                                                                                                                                                                                                                                                                                                                                                                             \\ 
\cline{1-5}\arrayrulecolor{black}\cline{6-6}
\begin{tabular}[c]{@{}l@{}}\textbf{}\\\textbf{\RVV{\cite{wang2015hierarchical}} } \\\end{tabular}            & CNN Random
  Forest                                                          & \begin{tabular}[c]{@{}l@{}}DRIVE, \\STARE\end{tabular}                         & \begin{tabular}[c]{@{}l@{}}97.67\% \\98.13\%\end{tabular}                                                        & \begin{tabular}[c]{@{}l@{}}97.33\% \\97.91\%\end{tabular}                                                                                                                                                                                                                                                                                                 & \begin{tabular}[c]{@{}l@{}}81.73\% \\81.04\%\end{tabular}                                                                                                                                                                                                                                                                                                                                        \\ 
\arrayrulecolor{black}\cline{1-5}\arrayrulecolor{black}\cline{6-6}
\begin{tabular}[c]{@{}l@{}}\textbf{}\\\textbf{\RVV{\cite{melinscak2015retinal}}} \\\end{tabular} & CNN                                                                          & DRIVE                                                                          & 94.66\%                                                                                                                          & -                                                                                                                                                                                                                                                                                                                                                                                         & -                                                                                                                                                                                                                                                                                                                                                                                                                                \\ 
\arrayrulecolor{black}\cline{1-5}\arrayrulecolor{black}\cline{6-6}
\begin{tabular}[c]{@{}l@{}}\textbf{}\\\textbf{\RVV{\cite{fu2016retinal}} } \\\end{tabular}              & \begin{tabular}[c]{@{}l@{}}CNN \\CRF\end{tabular}                            & \begin{tabular}[c]{@{}l@{}}DRIVE,\\STARE\end{tabular}                          & \begin{tabular}[c]{@{}l@{}}94.70\% \\95.45\%\end{tabular}                                                        & -                                                                                                                                                                                                                                                                                                                                                                                         & \begin{tabular}[c]{@{}l@{}}72.94\% \\71.40\%\end{tabular}                                                                                                                                                                                                                                                                                                                                        \\ 
\arrayrulecolor{black}\cline{1-5}\arrayrulecolor{black}\cline{6-6}
\begin{tabular}[c]{@{}l@{}} \\\textbf{\RVV{\cite{jiang2018retinal}}} \\\end{tabular}        & AlexNet                                                                      & \begin{tabular}[c]{@{}l@{}}DRIVE, \\STARE,\\CHASE\end{tabular}                 & \begin{tabular}[c]{@{}l@{}}\\95.93\%, \\96.53\%, \\95.91\%, \\96.62\%\end{tabular} & \begin{tabular}[c]{@{}l@{}}\\98.32\% \\97.98\% \\97.70\% \\98.26\%\end{tabular}                                                                                                                                                                                                                                             & \begin{tabular}[c]{@{}l@{}}71.21\% \\78.20\% \\72.17\% \\76.86\%\end{tabular}                                                                                                                                                                                                                                                                                    \\ 
\arrayrulecolor{black}\cline{1-5}\arrayrulecolor{black}\cline{6-6}
\begin{tabular}[c]{@{}l@{}}\textbf{\RVV{\cite{hu2018retinal}}} \\\end{tabular}                          & \begin{tabular}[c]{@{}l@{}}\\CNN \\CRFs \\~\end{tabular}                       & \begin{tabular}[c]{@{}l@{}}DRIVE, \\STARE\end{tabular}                         & \begin{tabular}[c]{@{}l@{}}95.33\% \\96.32\%\end{tabular}                                                        & \begin{tabular}[c]{@{}l@{}}77.72\%  \\75.43\%\end{tabular}                                                                                                                                                                                                                                                                                                & \begin{tabular}[c]{@{}l@{}}97.93\% \\98.14\%\end{tabular}                                                                                                                                                                                                                                                                                                                                        \\ 
\arrayrulecolor{black}\cline{1-5}\arrayrulecolor{black}\cline{6-6}
\begin{tabular}[c]{@{}l@{}} \\\textbf{\RVV{\cite{wang2019dense}}} \\\end{tabular}             & \begin{tabular}[c]{@{}l@{}}Dense U-net, ~\\Patch-based learning\end{tabular} & \begin{tabular}[c]{@{}l@{}}DRIVE, \\STARE\end{tabular}                         & \begin{tabular}[c]{@{}l@{}}95.11\% \\95.38\%\end{tabular}                                                        & \begin{tabular}[c]{@{}l@{}}\textcolor[rgb]{0.133,0.133,0.133}{97}\textcolor[rgb]{0.133,0.133,0.133}{.}\textcolor[rgb]{0.133,0.133,0.133}{36}\textcolor[rgb]{0.133,0.133,0.133}{\%} \\\textcolor[rgb]{0.133,0.133,0.133}{97}\textcolor[rgb]{0.133,0.133,0.133}{.}\textcolor[rgb]{0.133,0.133,0.133}{22}\textcolor[rgb]{0.133,0.133,0.133}{\%}\end{tabular} & \begin{tabular}[c]{@{}l@{}}~\textcolor[rgb]{0.133,0.133,0.133}{79}\textcolor[rgb]{0.133,0.133,0.133}{.}\textcolor[rgb]{0.133,0.133,0.133}{86}\textcolor[rgb]{0.133,0.133,0.133}{\%} \\\textcolor[rgb]{0.133,0.133,0.133}{~}\textcolor[rgb]{0.133,0.133,0.133}{79}\textcolor[rgb]{0.133,0.133,0.133}{.}\textcolor[rgb]{0.133,0.133,0.133}{14}\textcolor[rgb]{0.133,0.133,0.133}{\%}\end{tabular}  \\ 
\arrayrulecolor{black}\cline{1-5}\arrayrulecolor{black}\cline{6-6}
\begin{tabular}[c]{@{}l@{}}\textbf{\RVV{\cite{park2020m}}} \\\end{tabular}                            & \begin{tabular}[c]{@{}l@{}}GAN \\ACE \\Lanczos resampling\end{tabular}       & \begin{tabular}[c]{@{}l@{}}\\DRIVE, \\STARE,  \\HRF, ~\\CHASE-DB1\end{tabular} & \begin{tabular}[c]{@{}l@{}}97.06\% \\98.76\% \\97.61\%, \\97.36\%\end{tabular}   & \begin{tabular}[c]{@{}l@{}}98.36\% \\99.38\% \\- \\-\end{tabular}                                                                                                                                                                                                                                                                                         & \begin{tabular}[c]{@{}l@{}}83.46\% \\83.24\% \\- \\-\end{tabular}                                                                                                                                                                                                                                                                                                                                \\ 
\arrayrulecolor{black}\cline{1-5}\arrayrulecolor{black}\cline{6-6}
\begin{tabular}[c]{@{}l@{}}\textbf{\RVV{\cite{boudegga2021fast}}} \\\end{tabular}                 & \begin{tabular}[c]{@{}l@{}}U-net~\\Patch-learning\end{tabular}               & \begin{tabular}[c]{@{}l@{}}DRIVE \\STARE\end{tabular}                          & \begin{tabular}[c]{@{}l@{}}95.12\% \\96.41\%\end{tabular}                                                        & \begin{tabular}[c]{@{}l@{}}98.69\% \\99.45\%\end{tabular}                                                                                                                                                                                                                                                                                                 & \begin{tabular}[c]{@{}l@{}}\\80.60\% \\82.30\% \\~\end{tabular}                                                                                                                                                                                                                                                                                                                                  \\ 
\arrayrulecolor{black}\cline{1-5}\arrayrulecolor{black}\cline{6-6}
\begin{tabular}[c]{@{}l@{}}\textbf{\RVV{\cite{deng2022retinal}}} \\\end{tabular}                      & \begin{tabular}[c]{@{}l@{}}D-Mnet \\PCNN\end{tabular}                        & \begin{tabular}[c]{@{}l@{}}\\DRIVE, \\STARE,\\CHASE\_DB1 ,\\HRF\end{tabular}   & \begin{tabular}[c]{@{}l@{}}96.83\%, \\97.32\%, \\97.14\% \\96.68\%\end{tabular}  & -                                                                                                                                                                                                                                                                                                                                                                                         & -                                                                                                                                                                                                                                                                                                                                                                                                                                \\ 
\arrayrulecolor{black}\cline{1-5}\arrayrulecolor{black}\cline{6-6}
\begin{tabular}[c]{@{}l@{}}\textbf{\RVV{\cite{kar2022retinal}}} \\\end{tabular}                        & \begin{tabular}[c]{@{}l@{}}GAN \\CLHAE\end{tabular}                          & \begin{tabular}[c]{@{}l@{}}\\DRIVE,\\CHASE\_DB1, \\HRF,  \\ARIA\end{tabular}   & \begin{tabular}[c]{@{}l@{}}97.42\% \\98.73\% \\97.73\% \\96.28\%\end{tabular}    & -                                                                                                                                                                                                                                                                                                                                                                                         & -                                                                                                                                                                                                                                                                                                                                                                                                                                \\ 
\arrayrulecolor{black}\cline{1-5}\arrayrulecolor{black}\cline{6-6}
\begin{tabular}[c]{@{}l@{}}\textbf{\RVV{\cite{chen2022pcat}}} \\\end{tabular}                         & \textcolor[rgb]{0.125,0.125,0.125}{PCAT-UNet}                                & \begin{tabular}[c]{@{}l@{}}\\DRIVE, \\STARE,\\CHASE\_DB1\end{tabular}          & \begin{tabular}[c]{@{}l@{}}96.22\% \\97.96\% \\98.12\%\end{tabular}                              & -                                                                                                                                                                                                                                                                                                                                                                                         & \begin{tabular}[c]{@{}l@{}}85.76\% \\87.03\%\\84.93\%\end{tabular}                                                                                                                                                                                                                                                                                                               \\ 
\arrayrulecolor{black}\cline{1-5}\arrayrulecolor{black}\cline{6-6}
\begin{tabular}[c]{@{}l@{}}\textbf{\RVV{\cite{zhang2022edge}}} \\\end{tabular}                       & \textcolor[rgb]{0.125,0.125,0.125}{U-Net}                                    & \begin{tabular}[c]{@{}l@{}}\\DRIVE, \\STARE,\\CHASEDB1\end{tabular}            & \begin{tabular}[c]{@{}l@{}}97.01\% \\96.91\% \\98.11\%\end{tabular}                              & -                                                                                                                                                                                                                                                                                                                                                                                         & -                                                                                                                                                                                                                                                                                                                                                                                                                                \\
\arrayrulecolor{black}\cline{1-5}\arrayrulecolor{black}\cline{6-6}
\end{tabular}
\arrayrulecolor{black}
\label{bloodvessel}
\end{table*}

\paragraph{Retinal Lesions Segmentation from OCT Scans}

\noindent OCT has been utilized extensively for scanning the retina to detect various ocular diseases. Significant indicators for a wide variety of ocular disorders can be found in the retinal layers. Multi-scale, end-to-end CNN architecture was proposed to delineate choroidal borders \RVV{\citep{Sui_2017(107)}}. The method was successful when applied to data on various global and local scales. Pixel data was used to update the appropriate graph-edge weight immediately.  Results from testing the system on 912 OCT images showed that it performed best when using learned graph-edge weights. Gopinath et al.\RVV{\citep{gopinath2017deep}} proposed a model that combined CNN with a Long Short Term Memory (LSTM) to extract the retinal layers through an OCT image. The model's pixel-wise mean absolute error was 1.30 ± 0.48. Fang et al. \RVV{\citep{Fang_2017(116)}} presented a framework (CNN-GS) that integrated the CNN and graph search techniques for automated layer-by-layer retinal delineation. In order to learn how to classify retinal images, CNN was used to extract features from a specific layer of the retina. In addition, the probability maps produced by CNN were employed using a graph search approach to identify the boundaries of the retina.The supervised model \RVV{\citep{xiang2018automatic}} was developed for segmenting layers and neovascularization. For the neural network classifier, spatial features (3), gray-level features (7), and layered-like features (14) were extracted. Multiple-scale bright and dark layer detection filters were employed to enhance retinal layer pathologies. To refine the retinal layers, graph search algorithm was utilized, and the weights of nodes were computed based on extracted layers. To validate the model, 42 SD-OCT images of AMD patients were used. Hu et al. \RVV{\citep{hu2019automatic}} proposed a model that combined multiscale CNN (MCNN) and graph search to extract the retinal layer in OCT images. Initially, multiscale features of the retinal layer were extracted in order to generate probability maps. To lessen the likelihood of the network incorrectly identifying the background as a target, the model uses location information to differentiate between foreground and background pixels. Finally, an enhanced graph search technique was used to delineate the retinal layers using probability maps. Mariottoni et al. \RVV{\citep{Mariottoni_2020(117)}} introduced an algorithm that was capable of determining the thickness of the RNFL without the need for segmentation. The algorithm was trained using conventional RNFL thickness values obtained from SD-OCT images. In the study \RVV{\citep{wei2020segmentation}}, the priority of the mutex relationship among retinal layers was considered, and introduced new loss function as mutex dice loss (MDL). In addition to this, a novel FCN-based model was proposed that utilized the depth max pooling (DMP) to segment fluids and retinal layers in SD-OCT images. The Shortest Path (DL-SP) algorithm \RVV{\citep{mishra2020automated}} was proposed for automatically identifying the retinal layers responsible for drusen and reticular pseudodrusen (RPD) in OCT images. The U-net model was used to generate probability maps and then combined with pixel-to-pixel edge weights, which were measured using the gradient in the z-direction. The model was evaluated on 1000 images and achieved absolute mean differences for RPD and dursen of 0.75±1.99 pixels (2.92±7.74 μm) and 1.53±1.47 pixels (5.97±5.74 μm), respectively. 
A segmentation model DeepRetina \RVV{\citep{li2020deepretina}}, was proposed to segment the retinal layers. Xception65 extracted feature maps and then fed them into atrous spatial pyramid pooling module to get multiscale information. The method was validated using 280 OCT volumes (40 B-scans per volume) and achieved IOU and sensitivity of 0.90 and 92.15\%, respectively. The paper \RVV{\citep{li2021multi}} presented a novel two-stage approach that uses a graph convolutional network (GCN) to identify all nine retinal layers and the OD in OCT images. Multi-scale global reasoning module integrated into the U-shaped neural network between the encoder and the decoder to use the network's prior knowledge of anatomy. The method was validated on Duke SD-OCT dataset, dice score, and pixel accuracy of 0.820 ± 0.001 and 0.830 ± 0.002, respectively. The work \RVV{\citep{sousa2021automatic}} presented a method for the segmentation of the ILM, RPE, and BMO in OCT images of healthy and Intermediate AMD subjects. U-Net and DexiNed, two DL networks were used, and the results showed an average absolute error of 0.49, 0.57, and 0.66 for ILM, RPE, and BM, respectively. In the paper, \RVV{\citep{he2021structured}}, a unified DL framework was introduced that directly performed modeling of the distribution of the surface positions. A single feed-forward operation generated surfaces that are topologically accurate, continuous, and smooth. An embedded residual recurrent network (ERR-Net) \RVV{\citep{hu2021embedded}} was developed based on a graph search for coarse-to-fine retinal layer delineation. In addition to resolving the gradient issue introduced by depth, the ERR-Net also encapsulates the image's global spatial structure. Graph search was used to refine the retinal boundaries. The model was evaluated on Duke, Open University of Miami , and AREDS2 datasest. Parra et al. \RVV{\citep{parra2022loctseg}} introduced a novel FCN architecture, dubbed LOCTSeg, to segment different diagnostic markers in OCT images. LOCTSeg was a lightweight model designed to balance performance and efficiency. Two publicly available benchmarking datasets were used to assess the performance of the model, AROI (1136 images) and HCMS (1715 images). The evaluation showed that the model achieved increased Dice score by 3\% on the AROI dataset and by 1\% on the HCMS dataset. The study \RVV{\citep{viedma2022oct}} proposed Mask R-CNN for segmenting retinal layers from OCT images. A CNN model was used for feature extraction from images and generated feature maps which were fed into the region proposal network (RPN). After that, bounding boxes (called anchors) were generated that were distributed over each feature map. The RPN then separates these anchors into two groups: foreground class (positive anchors), which are located in areas that reflect features relating to the objects, and background class (negative anchors), which are located outside of these objects. The study \RVV{\citep{man2023multi}} investigated different U-net models that combined with VGG and ResNet to segment the retinal layers, and compared their accuracy. Results showed that VGG16 and U-net (VGG16-Unet) performed better than the U-net and U-net++ model.

\noindent The study \RVV{\citep{wilkins2012automated}} presented a CNN model based on U-net autoencoder architecture to detect intraretinal fluid (IRF) in OCT image. The OCT images were collected from 2006 to 2016 at the Ophthalmology Department, University of Washington, 934 B-scans were used for training and 355 B-scans were used for the validation purposes. Karri et al. \RVV{\citep{karri2017transfer}} proposed an algorithm for identified retinal pathologies in OCT images. A pre-trained CNN, GoogLeNet, was fine-tuned to increase its prediction capabilities, and salient responses were identified during prediction to comprehend the properties of the learned filters. Subjects with dry AMD, DME, and normal were considered during the study.
Roy et al. \RVV{\citep{roy2017relaynet}} proposed a DL model (ReLayNet) to segment the retinal layers and fluid masses through OCT scans. The proposed model was evaluated on the Duke publicly available dataset. ReLayNet was able to provide more accurate estimates of layer thickness than graph-based comparing techniques. RelayNet extracted the ILM, NFL-IPL, INL, OPL, ONL-ISM, ISE, OS-RPE, and cumulative retinal fluids with dice coefficients of 0.99, 0.90, 0.94, 0.87, 0.84, 0.93, 0.92, 0.90, 0.99, and 0.77 respectively. Schlegl et al. \RVV{\citep{schlegl2018fully}} developed a semantic segmentation model to extract IRF and sub-retinal fluid (SRF) from 1200 OCT volumes acquired from Zeiss Cirrus and Heidelberg Spectralis OCT machines. The model extracted IRF and SRF with a mean accuracy of 94.0\% and 92.0\%, respectively. Agari et al. \RVV{\citep{asgari2019multiclass}} developed a encoder-decoder model for solving the multitask problem of drusen segmentation. To segment RPE and BM, instead of training a multiclass model, a single decoder was employed for each target class. To further enhance the regularization, links between each class-specific branch and the decoder were developed. To enhance OCT image segmentation performance, Y-Net \RVV{\citep{farshad2022net}} was proposed, an architecture that fuses frequency domain characteristics with the image domain. Y-net performed better than U-net, and obtained an increase in fluid segmentation dice score by 13\% and our overall dice score by 1.9\%.
Hsu et al. \RVV{\citep{hsu2022automatic}} proposed a DL model that segmented the IRF, SRF, and ellipsoid zone (EZ) in OCT images, in addition to this, also correlated the extracted features with visual acuity. The modified U-net model was trained on manually annotated 127 scans from 50 patients and validated on 38 scans from 16 patients. For IRF and SRF, the model obtained values of 0.80 and 0.89 for Srensen-Dice coefficients, respectively. The study \RVV{\citep{philippi2023vision}} employed a transformer-based technique to detect and isolate retinal lesions in SD-OCT scans automatically. The approach combined the data-efficient training of CNNs with the efficient long-range feature extraction and aggregation capabilities of Vision Transformers. Swin UNEt TRansformers (Swin-UNETR) \RVV{\citep{hatamizadeh2022swin}} was used, which was a segmentation network tailored to the unique challenges of medical image analysis. A private dataset consisted of 3842 SD-OCT images was used to evaluate the method. Specialists at the Franziskus Eye Center in Muenster manually classified the images. The Unet3+ achieved highest mean dice score of 0.508, whereas Swin-UNETR-24 obtained the second-best score of 0.457. Wang et al. \RVV{\citep{wang2022automated}} proposed a novel technique for segmenting 10 retinal layers in OCT images, including intraretinal fluid. A fan filter was utilized to minimize the impact of vessel shadows and fluid regions in an OCT image, hence improving the linear information pertaining to retinal borders. Random forest classifier was used to predicate the retinal boundaries. By combining the unique techniques of boundary redirection (SR) and similarity correction (SC), the model was able to perform boundary tracking and identify the retinal layers. On average, the proposed method utilized OCT images from 415 healthy subjects and 482 DME patients.\\

\begin{table*}
\centering
\caption{Summarizing the Segmentation Models based on DL Techniques for Retinal Lesions from OCT Scans}
\begin{tblr}{
  width = \linewidth,
  colspec = {Q[177]Q[146]Q[138]Q[137]Q[156]Q[183]},
  cell{2}{3} = {},
  cell{2}{4} = {},
  cell{5}{4} = {},
  cell{6}{4} = {},
  cell{7}{3} = {},
  cell{7}{4} = {},
  cell{9}{2} = {},
  cell{9}{4} = {},
  cell{9}{5} = {},
  hlines,
  vline{2-6} = {-}{},
}
\textbf{Study}                                              & \textbf{\textbf{Techniques}}                                             & \textbf{\textbf{Images}}                              & \textbf{\textbf{Machine}}                      & \textbf{Retinal lesion}                              & \textbf{Evaluation}                                                         \\
{\RVV{\cite{Sui_2017(107)}}}          & Mutliscale CNN                                                           & {912 OCT B-scans\\Normal subjects:42\\ME patients:31} & Heidelberg Spectralis~                         & Choroidal boundaries                                 & -                                                                           \\
{\RVV{\cite{gopinath2017deep}}}   & CNN LSTM                                                                 & -                                                     & -                                              & Retinal layers                                       & Mean absolute error: 1.30 ± 0.48.                                           \\
{\RVV{\cite{xiang2018automatic}}}    & Graph searchMultiscale CNN                                               & ~ 42~AMD scans~                                       & Cirrus HD-OCT 4000                             & {Retinal layers and \\neovascularization}            & \textcolor[rgb]{0.133,0.133,0.133}{}                                        \\
{\RVV{\cite{mishra2020automated}}}  & U-netProbability maps                                                    & {1000 images\\Normal: 20\\AMD: 25}                    & Heidelberg Spectralis                          & {Drusen and \\Reticular pseudodrusen (RPD)}          & {0.75±1.99\\5.97±5.74~\\~}                                                  \\
{\RVV{\cite{li2020deepretina}}}          & {Xception65,\\Atrous spatial \\pyramid pooling}                          & 280 OCT volumes                                       & Cirrus HD-OCT                                  & Retinal layers                                       & {IOU: 0.90\\Sensitivity: 0.92}                           \\
{\RVV{\cite{li2021multi}}}              & Graph CNN                                                                & 1.12 radial OCT B-scans2. Duke SD-OCT dataset         & DRI OCT-1 Atlantis                             & Retinal layers                                       & {Dice score:~0.820 ± 0.001\\Pixel accuracy: 0.830 ± 0.002,}                 \\
{\RVV{\cite{schlegl2018fully}}}    & CNN                                                                      & 1200 OCT volumes                                      & {Zeiss Cirrus and \\Heidelberg Spectralis OCT} & IRF and SRF                                          & {Mean accuracy~\\IRF:~94\% \\SRF: 92\%} \\
{\RVV{\cite{philippi2023vision}}} & {\textcolor[rgb]{0.133,0.133,0.133}{CLAHE}\\Unet3+\\Vision Transformers} & 3842 SD-OCT                                           & {Spectralis SD-OCT,\\Heidelberg Engineering.}  & {IRF, SRF, and~Pigment epithelium\\detachment (PED)} & Mean dice: 0.50                                                             \\
{\RVV{\cite{hsu2022automatic}}}        & U-net model                                                              & 127 scans                                             & -                                              & {IRF, SRF, and \\Ellipsoid zone (EZ)}                & {IRF:~0.80~\\SRF: 0.89}                                                     
\end{tblr}
\end{table*}

\vspace{-0.3cm}
\subsubsection{Segmentation driven classification methods}

\paragraph{Techniques Based on Fundus Scans}
\noindent The study \RVV{\citep{lim2015integrated}} proposed a model CNN-FE to extract feature-enhanced inputs that highlight disc pallor without a degree of vessel kinking and blood vessel obstruction in fundus image. Pixel-level probability maps constructed by CNN went through a process of robust refinement, which takes into account information already known about the retinal morphology. In addition, confidence was estimated on the validity of the segmentation by analyzing the probability maps. Finally, the extracted cup and disc border were utilized to estimate CDR. MESSIDOR and SEED-DB  datasets were used to evaluate the model. The overall screening performance of CNN-FE was higher (AUC = 0.847) than that of the reconstruction-based method (AUC = 0.838). A DL model for directly screening for glaucoma using fundus images based on image-relevant information \RVV{\citep{fu2018disc}}. Global image stream, segmentation-guided network, local disc region, and disc polar transformation streams were defined as four deep streams on different levels. Finally, the probabilities of each stream's output were combined to produce a final output. The model was evaluated on two glaucoma datasets (SCES and SINDI), and the results showed better performance than other state-of-the-art methods.
An improved U-net CNN model was proposed to segment the OD and OC from the fundus image \RVV{\citep{joshua2019segmentation}}. The DRISHTI-GS and RIM-ONE v.3 datasets were used to evaluate the model. Another DL model using Gradient-weighted Class Activation Mapping (Grad-CAM) \RVV{\citep{kim2019medinoid}} was proposed for glaucoma diagnosis based on OD localization. The model was tested on fundus images from Samsung Medical Center (SMC) and achieved accuracy, sensitivity, and specificity of 96\%, 96\%, and 100\%, respectively. Sreng et al. \RVV{\citep{sreng2020deep}} proposed a model for glaucoma diagnosis through fundus images. DeepLabv3+ architecture and encoder module with multiple CNN were employed for segmentation. For glaucoma classification, the model was tested on RIM-ONE, ORIGA, DRISHTI-GS1 and ACRIMA datasets with an accuracy of 97.37\%, 90.00\%, 86.84\%, and 99.53\%, respectively. Tulsani et al. \RVV{\citep{tulsani2021automated}} presented a novel method for detecting glaucoma by employing segmentation of the OD and OC in fundus scan. For the segmentation task, a custom UNET++ model was developed by tuning the hyperparameters and a custom loss function. The employed loss function was useful for addressing the class imbalance that arises due to the small size of the ONH. Based on the identification of clinical features, the proposed method was 96\% accurate at classifying images as either glaucomatous or healthy. Training times were decreased than state-of-the-art models and achieved  Intersection over Union (IOU) scores (0.9477 for OD and 0.9321 for OC) using the improved model. The model was evaluated on RIM-ONE, DRIONS-DB, and ORIGA, and it was able to achieve an accuracy of 91\%, 92\%, and 90\%, respectively. Abdel et al. \RVV{\citep{abdel2022tweec}} proposed a deep convolutional (TWEEC) network that extracted anatomical information of OD and blood vessels. The spatial retinal images and wavelet subbands were fed into the model as inputs. TWEEC model achieved accuracy for the spatial and wavelet inputs of 98.78\% and 96.34\%, respectively. A novel multi-task strategy \RVV{\citep{hervella2022end}} for identifying glaucoma while segmenting the optic disc and cup was proposed. The model achieved an increase in performance by utilizing pixel-level and image-level labels during training. Biomarkers like the CDR was extracted from the segmentation maps that were already predicted with the diagnosis. The model designed concurrent segmentation and classification that maximizes the use of shared parameters. In order to minimize the need for loss weighting hyperparameters, a multi-adaptive optimization technique was employed during training. CDR-based classification achieved an area under the curve of 94.18\% on the REFUGE dataset. The study \RVV{\citep{nawaz2022efficient}} developed a DL model for glaucoma diagnosis, the EfficientNet-B0 feature extractor was used to compute the deep features from the suspect samples. The features computed by EfficientNet-B0 are then fed into the EfficientDet-Bi-directional D0's Feature Pyramid Network (BiFPN) module, where they fused many times using a top-down and bottom-up approach. Finally, the anticipated class of glaucoma lesions inside that confined region was predicted. To demonstrate model generalizability, cross-dataset validation was performed on the High-Resolution Fundus (HRF) and RIMONE datasets. An OD localization and Glaucoma Diagnosis Network (ODGNet) was presented by study \RVV{\citep{latif2022odgnet}}. Initially, a visual saliency map combined with shallow CNN localized OD from images. In the second step, pre-trained transfer learning models (AlexNet, ResNet, and VGGNet) diagnosed glaucoma. The model was evaluated on ORIGA, HRF, DRIONS-DB, DR-HAGIS, and RIM-ONE publicly available datasets. The results showed that ODGNet tested on ORIGA for glaucoma diagnosis achieved accuracy, specificity, sensitivity, and AUC of 95.75\%, 94.90\%, 94.75\%, and 97.85\%, respectively. Touahri et al. \RVV{\citep{touahri2022improved}} proposed a glaucoma diagnosis model through fundus images that first segment the OD and OC and then classify them into normal or glaucomatous scans. To begin, the OD region was segmented in the fundus images to create a ROI. To get the fine-grained segmentation, a U-Net model was developed. The model was validated using the publicly available REFUGE dataset. Roshini et al. \RVV{\citep{roshini2022automatic}} proposed MultiResUNet architecture for glaucoma diagnosis based on CDR estimation. The results showed that MultiResUNet achieved a mean accuracy of 97.2\%.
Context encoding network (CE-Net) \RVV{\citep{wang2022optic}} architecture was developed for segmentation of the OD in diabetic retinal images. The model consisted of three module, 1) an encoder for features extraction, 2) a context extractor, and 3) a decoder. The context extractor module consisted of a residual multi-kernel pooling (RMP) and improved dense atrous convolutional block. The model was validated on Indian Diabetic Retinopathy Image Dataset (IDRID). A model \RVV{\citep{zhang2022automated}} proposed for DR diagnosis based on lesion detection through fundus images. Inception V3 model was adopted for classification, whereas the grading of DR was performed by identification of different lesions. The Kaggle DR dataset was used for the training and testing of model. Another DL model \RVV{\citep{li2018automated}} was proposed for the detection of DR (PDR, DME). A total of 106,244 nonstereoscopic retinal images were used to test the model. For external validation, 35,201 images of 14,520 eyes from population-based cohorts of Malays, Caucasian Australians, and Indigenous Australians were used. When tested on the independent, multiethnic data set, the AUC, sensitivity, and specificity were all found to be 0.95, 92.5\%, and 98.5\%, respectively. A fully patch-based CNN model \RVV{\citep{zago2020diabetic}} was developed for DR diagnosis by performing the retinal lesion localization. The use of strides enhances lesion localization by a factor of 25. Only 28 fundus images (from DIARETDB1) annotated at the pixel level were utilized in the training process for the model and tested on Messidor dataset. The system achieved sensitivity and an AOUC of 94.0\% and 0.912, respectively. The study \RVV{\citep{qomariah2021segmentation}} presented a unique DLnetwork (MResUNet) that adapts UNet by replacing its identity mapping residual units with modified residual units to segment microaneurysm for DR diagnosis. During training, the mean weighted loss function was employed to handle imbalanced pixels of background and microaneurysms. Based on experimental results, the model outperformed than autoencoder, FCN16, FCN8, and UNet in terms of sensitivity on the IDRID and DiaretDB1 datasets. The paper \RVV{\citep{kumari2022deep}} presented image-processing techniques to extract the four key features microaneurysms, blood vessels, hemorrhages, and exudates from raw fundus images, and then employed a CNN for automatic identification of DR. When compared to other models, DenseNet-16 provides the best accuracy, when tested on DRIVE database. Jiwane et al. \RVV{\citep{jiwane2022detecting}} developed a DL model (ResNet50) for the detection of DR based on soft exudate and hard exudate along with OD. The paper \RVV{\citep{murugan2022micronet}} presented CNN model to train MA and non-MA patches, and a majority voting method was employed to identify MA patches. Retinopathy Online Challenge (ROC) data was used to assess the effectiveness of the provided technique. A three-class semantic segmentation model \RVV{\citep{selccuk2022automatic}} was proposed to extract the exudates and hemorrhage. Also, a color space transformation was done, and the classic U-Net algorithm was employed so that high performance was achieved in images with low contrast. The results showed that the Dice and Jaccard similarity indices for the segmentation performance were calculated to be close to 0.95. In order to detect and grade DR through fundus images, the study \RVV{\citep{parthiban2022efficientnet}} introduced a Wavelet Neural Network (EN-CSOWNN) model trained with EfficientNet and Chicken Swarm Optimization. In order to identify diseased areas in an image, a customized U-Net-based segmentation model was employed. Additionally, feature vectors were derived using the EfficientNet model, and class labels were assigned using the wavelet neural network model. Ultimately, the CSO approach was used to optimize the  model's classification performance by adjusting the model's initial parameters. The model was validated on MESSIDOR dataset an accuracy of 98.60\%. Tomaz et al. \RVV{\citep{bisneto2020generative}} developed a DL model based on GAN to segment OD for glaucoma diagnosis through fundus images. ROIs segmented by the GAN were characterized using taxonomic indices. These indices were based on the diversity of species and the frequency of individuals, or the range of pixel values and the number of pixels with a given value, respectively. The model was evaluated on RIM-ONE and Drishti-GS public databases and achieved 77.9\% accuracy. With modifications and changes, the model obtained 100\% accuracy and a ROC curve of 1.
\paragraph{Techniques Based on OCT Scans}
A DL technique was presented to segment retina surfaces in OCT volume and diagnose AMD \RVV{\citep{Shah_2017(139)}}. The training data was used to cultivate a set of features and a transformation. Normal and diseased image surfaces were learned using the same CNN. A total of 40 OCT volumes were used to validate the model, 20 volumes from each group. The suggested method outperformed graph-based optimum surface segmentation using convex priors (G-OSC). Shah et al. \RVV{\citep{Shah_2018(141)}} developed a CNN model to detect intermediate AMD using segmented multiple retinal surfaces through OCT scans. In order to classify B-scan images into "healthy" and "intermediate AMD," a single CNN was trained to segment all three retinal layers in a single pass. The model was validated on 3000 B-scans acquired from 50 OCT volumes. Saha et al. \RVV{\citep{saha2019automated}} developed a DL method for the automated detection OCT biomarker and classification of early AMD. The model automatically detected and classified HFs, hyporeflective foci inside the drusen, and subretinal drusenoid deposits from OCT B-scans. A total of 19584 OCT B-scans with at least one eye diagnosed with early or intermediate AMD  were included in the dataset, images were acquired from the Doheny Eye Centers. The model detected the subretinal drusenoid deposit with an accuracy of 86\%. The accuracy of detecting HFs and hyporeflective foci was 89\% and 88\%, respectively. Fauw et al. \RVV{\citep{de2018clinically}} developed a DL model that segments retinal layers from 3D OCT scans and then employs the extracted information for the diagnosis of retinal diseases. The segmentation network was trained on 877 images, and the classification network was trained on 14,884  maps and achieved an accuracy of 96.4\% on the validation set. Chen et al. \RVV{\citep{Chen_2019(145)}} developed a DL method for screening early glaucoma that utilized features from fundus and EDI-OCT images. Both textual and structural elements from each modality were integrated. Once the OC was segmented from the fundus image using brightness compensation, CDR and textural features were obtained. Each pixel in an OCT image was labeled as being on the anterior LC surface or the background using a region-aware method and a residual U-Net architecture. After extracting features of LC deformation using an improved templated local binary pattern, the LC depth and width of the BMO's were calculated.
A CNN model \RVV{\citep{raja2020extraction}} was proposed based on CDR estimation for glaucoma diagnosis. To estimate the CDR, first ILM and RPE layers were extracted using CNN, and the graph searched was to refine the layers. Afterward, missing areas were filled by linear interpolation. Finally, cup and disc borders were determined in order to calculate the CDR. The model used the Armed Forces Institute of Ophthalmology (AFIO) dataset and achieved average  specificity, sensitivity, and accuracy of 94.07\%,  94.6\%, and 94.68\%, respectively. Hassan et al. \RVV{\citep{hassan2020rag}} proposed a hybrid convolutional framework (RAG-FW) that extracted multiple retinal lesions (such as IRF, SRF, HE, drusen, and CA) from OCT scans and utilized them for grading of retinopathy. RAG-FW was tested on 43,613 multi-vendor OCT scans and performed better than state-of-the-art solutions by getting 14.15\% better at extracting retinal fluids from Duke-II, 2.02\% better at classifying retinopathy from Zhang, and 1.24\% better from BIOMISA datasets. Hina et al. \RVV{\citep{raja2020clinically}} proposed  a hybrid convolutional network (RAG-NETv2) for glaucoma diagnosis and grading by utilizing extracted RNFL, GC-IPL regions. For segmentation, encoder-decoder architecture was employed, atrous convolution, skip connection, and pyramid pooling techniques allowed to retain fine details of retinal layers. Afterward, the thickness profiles of extracted regions were computed and fed as a feature vector to the SVM for grading of OCT scan. The model was trained and tested on the publicly available AFIO dataset, and achieved a mean DC score of 0.8697 for extracting the regions, the F1 score, and accuracy of 0.9577  and 91.17\% for glaucoma diagnosing and grading, respectively. In the study \RVV{\citep{smitha2022detection}}, a GAN-based model was proposed for the automated segmentation and classification of OCT-B images for the purpose of diagnosing AMD and DME. The handcrafted Gabor features were integrated into the method in order to improve retina layer segmentation, and non-local denoising was utilized in order to get rid of speckle noise. The model showed better results for OCT image segmentation and classification, with an F1-score of 0.79 and an accuracy of up to 92.42\%. Deep ensemble learning model \RVV{\citep{moradi2023deep}} was proposed for early AMD diagnoses based on retinal layer segmentation in OCT images. In order to automatically annotate 11 retinal borders, the model combined a graph-cut method with a cubic spline. After the images had been refined, they were fed into a deep ensemble model that used a Bagged Tree with deep learning classifiers. Our boundary refinement-based segmentation model has a much lower overall error rate than OCT Explorer segmentation (1.7\% vs 7.8\%, p-value = 0.03). 

\begin{table*}
\centering
\caption{Summarizing the Segmentation Driven Classification Model based on DL Techniques}
\begin{tblr}{
  width = \linewidth,
  colspec = {Q[256]Q[169]Q[77]Q[133]Q[148]Q[150]},
  cell{2}{3} = {},
  cell{3}{2} = {},
  cell{3}{3} = {},
  cell{4}{3} = {},
  cell{5}{3} = {},
  cell{6}{3} = {},
  cell{7}{3} = {},
  cell{8}{3} = {},
  hlines,
  vline{2-6} = {-}{},
}
\textbf{Study}                                                & \textbf{\textbf{Techniques}}                    & \textbf{Modality} & \textbf{Dataset}                           & \textbf{Retinal lesion}                              & \textbf{Evaluation}                                                                                                                          \\
{\RVV{\cite{lim2015integrated}}}         & {Pixel-level probability \\maps\\CNN}           & Fundus            & {MESSIDOR,\\SEED-DB}                       & {OD, OC\\CDR}                                        & AUC: 0.847                                                                                                                                   \\
{\RVV{\cite{joshua2019segmentation}}} & U-net                                           & Fundus            & {DRISHTI-GS,\\RIM-ONE v.3}                 & {OD, OC,\\CDR}                                       & .-                                                                                                                                           \\
{\RVV{\cite{kim2019medinoid}}}           & {Gradient-weighted\\Class Activation \\Mapping} & Fundus            & {Samsung Medical \\Center}                 & OD                                                   & {Accuracy:~96\%\\Sensitivity:~96\%\\Specificity: 100\%,\textcolor[rgb]{0.133,0.133,0.133}{}} \\
{\RVV{\cite{sreng2020deep}}}           & DeepLabv3+                                      & Fundus            & {RIM-ONE, \\ORIGA, \\DRISHTI-GS1,\\ACRIMA} & -                                                    & {Accuracy:~\\97.37\%, \\90.00\%, \\86.84\%,~\\99.53\%\\~}                    \\
{\RVV{\cite{tulsani2021automated}}} & UNET++ model                                    & Fundus            & {RIM-ONE,\\DRIONS-DB, \\ORIGA}             & OD, OC                                               & {Accuracy:\\91\%, \\92\%,\\90\%.}                                                            \\
{\RVV{\cite{abdel2022tweec}}}          & {Wavelet Transformation\\CNN\\~}                & Fundus            & RIM-ONE v.2                                & {OD,\\Blood vessels}                                 & Accuracy: 98.78\%                                                                                                            \\
{\RVV{\cite{hervella2022end}}}      & {Segmentation maps\\CNN}                        & Fundus            & REFUGE                                     & OD, OC, CDR                                          & AUC: 0.94                                                                                                                \\
{\RVV{\cite{saha2019automated}}}       & CNN                                             & OCT               & Local dataset                              & {HFs\\Hyporeflective foci,\\Subretinal drusenoid\\~} & {Accuracy:\\89\%,\\88\%,\\86\%\\~}                                                           \\
{\RVV{\cite{hassan2020rag}}}         & {Hybrid convolutional\\framework (RAG-FW)}      & OCT               & {Duke-II, Zhang,~\\BIOMISA}                & {IRF, SRF, HE, \\drusen, and CA}                     & -                                                                                                                                            \\
{\RVV{\cite{raja2020extraction}}}       & RAG-NET                                         & OCT               & BIOMISA                                    & ILM and RPE                                          & {Accuracy:~ 94.68\%\\Specificity:~94.07\%\\Sensitivity:~94.6\%}                              
\end{tblr}
\end{table*}

\subsubsection{Classification}
\paragraph{Fundus Scans}
\noindent The study \RVV{\citep{ahn2018deep}} suggested that using DL algorithms in conjunction with fundus photography can be an effective method for differentiating between normal and glaucoma subjects, even in the early stages of the disease. From Kim's Eye Hospital, fundus images (1,542 images) were acquired from both healthy and glaucomatous eyes; out of a total 754 were used for training, 324 for validating, and 464 for testing. A logistic regression and CNN was developed; in addition to this GoogleNet Inception v3 model was also fine-tuned using the same datasets. The fundus image is a 3D array (240x240x3), but for the purpose of performing logistic regression, the images were flattened into a one-dimensional array. CNN model was consisted of two convolutional layers with 2020 and 4040 patch sizes, 1 stride, and 16 and 32 depths were utilized. Patch size 22 and stride 2 were used for max pooling. Fully connected layers have 32 and 64 hidden units. Convolutional and fully linked layers employed 0.5 dropout rate to avoid overfitting. The training accuracy of the logistic model was 82.9\%, the validation accuracy was 79.5\%, and the test accuracy was 77.2\%. Transfer-learned On training data, the GoogleNet Inception v3 model obtained an accuracy and AUROC of 99.7\% and 0.99, while on validation data, it reached 87.7\% and 0.95, and on test data, it reached 84.5\% and 0.93. The AUROC and accuracy of the CNN were 0.98 and 92.2\% on the training data, 0.95 and 88\% data, and on the validation data, 0.94 and 87\% on the test data, respectively. 
The paper \RVV{\citep{zhao2019direct}} presented a semi-supervised model for glaucoma detection based on CDR estimation without segmentation of the OD and OC. The method directly regresses the CDR value based on the feature of the OHD using MFPPNet through fundus image. The proposed technique was tested on Direct-CSU and public ORIGA glaucoma datasets and improved average accuracy of 0.063\% and the correlation of about 0.726 with measurements taken before human specialists manually segmented the optic disc/cup. On a dataset of 421 fundus images, estimated CDR values were tested for glaucoma screening and achieved an AUC of 0.905. ImageNet-trained models (VGG16, VGG19, InceptionV3, ResNet50, and Xception) were trained for automatic glaucoma assessment using fundus images \RVV{\citep{diaz2019cnns}}. The Xception model achieved an average AUC of 0.9605 with a 95\% confidence range. Gheisari et al. \RVV{\citep{gheisari2021combined}} proposed a CNN and RNN that extracted the spatial features in a fundus image and temporal features from a fundus video. Combined CNN and RNN were used to train with 1810 images and 295 videos. The average F-measure for CNN basic and combined model was 79.2\% and 96.2\%, respectively. Other studies \RVV{\citep{raghavendra2018deep} \citep{gomez2019automatic}} reported the DL model for glaucoma diagnosis without any explicit segmentation of biomarkers. 
\noindent The study \RVV{\citep{gulshan2016development}} used DL to develop an algorithm for automated detection of DR and DME in retinal fundus images. Between May and December 2015, a panel of 54 US-based licensed ophthalmologists and ophthalmology senior residents graded a total of 128 and 175 retinal images for DR, DME, respectively. For EyePACS-1, the model had an AUC of 0.991 (95\% CI, 0.988-0.993). Another automated DL model \RVV{\citep{gargeya2017automated}} was developed for the detection of DR. Seventy-five thousand and one hundred thirty-three publicly available fundus images from diabetic patients were used to train and test the model respectively. Whereas validation was performed on MESSIDOR 2 and E-Ophtha databases. Tests performed on the MESSIDOR 2 and E-Ophtha databases yielded an AUC of 0.94 and 0.95, respectively. Wang et al. \RVV{\citep{wan2018deep}} proposed a DR classification model, with transfer learning of AlexNet, VggNet, GoogleNet, and ResNet. The model was tested on a publicly available Kaggle dataset and achieved a classification accuracy of 95.68\%. Qummar et al. \RVV{\citep{qummar2019deep}} proposed DL ensemble approach for DR detection and employed five CNN models, Resnet50, Inceptionv3, Xception, Dense121, and Dense169. The results showed that the model detected all the stages of DR when tested on Kaggle dataset. A DL method was proposed for feature extraction from fundus images and SVM-based classification of DR \RVV{\citep{qomariah2019classification}}. The paper employed a CNN method for DR classification of fundus images. Pre-trained CNN models, AlexNet, VGG-16, and SqueezeNet) resulted in a 93.46\%, 91.82\%, and 94.49\% accuracy in classification, respectively. The study \RVV{\citep{doshi2020diabetic}} proposed and investigated the usage of multiple down-scaling techniques prior to submitting image data to a DL network for classification. Multi-Channel Inception V3 architecture with a unique self-crafted preprocessing phase was employed. Jordi et al. \RVV{\citep{de2020deep}} presented a DL interpretable classifier for DR detection through fundus scans, it also performed grading. The classifier was able to provide an explanation for the classification findings by giving a score to each point in both the hidden space and the input space. These scores, generated by pixel-wise score propagation model, represented how significantly each pixel aided in the overall classification. 
Four different transfer learning algorithms (VGG16, ResNet50, InceptionV3, and DenseNet121) were used to detect DR from fundus images \RVV{\citep{sheikh2020smartphone}}.
DenseNet121 model yielded the best results for making predictions. Karki et al. \RVV{\citep{karki2021diabetic}} developed a DL model based on EfficientNet for the classification of the DR. In addition to this, model performed grading of images, into mild, moderate, severe, or PDR. A quadratic kappa score of 0.924377 was attained by the best model on the APTOS test dataset after the models were trained using various datasets. The work \RVV{\citep{deepa2022ensemble}} presented a CNN ensemble (MPDCNN) model for accurate fundus image-based DR identification and grading. In the first phase, each input image was split into four patches and fed into one of two pre-trained CNN models (InceptionV3 and Xception). Prior knowledge is derived from the pertinent characteristics that are located in the shallow-dense layers of CNN models. The model was taught the crucial details from DR images by combining features from shallow and dense layers. In the second step, combined probability vectors from four patches were utilized to train the network classifier. DR classification accuracy was enhanced by using the ensemble method with multi-stage DL model and achieved an accuracy of 96.2\% with fivefold cross-validation. A hybrid method \RVV{\citep{butt2022diabetic}} was proposed for finding and classifying DRbased on fundusimages. Model employed transfer learning, based on GoogleNet and ResNet-18 architectures, to find features that can be put together to make a hybrid feature vector. A number of classifiers were used to classify fundus images into binary and multiclasses based on the extracted feature vector. The model was trained and tested on APTOS. Kaggle, and The Aravind Eye Hospital in India datasets. For binary classification, the proposed a maximum accuracy of 97.8\%, and for multiclass classification, it had an accuracy of 89.29\%. The paper \RVV{\citep{muthukannan2022optimized}} developed a DL model (CNN-MDD) to detect early-stage AMD. Maximum entropy transformation was applied, and then images were fed into CNN which optimized using a flower pollination optimization algorithm (FPOA) for feature extraction. A Multiclass SVM classifier was employed to identify the disease from the CNN's output. The model was tested on Ocular Disease Intelligent Recognition (ODIR) dataset and had specificity, precision, accuracy, and recall of 95.21\%, 98.30\%, 95.27\%, and 93.3\%, respectively. The study \RVV{\citep{bhimavarapu2023deep}} presented the DL with improved activation function for DR diagnosis from fundus images that reduced loss and processing time. Models was trained and evaluated using the DIARETDB0, DRIVE, CHASE, and Kaggle datasets. On the Kaggle dataset, the ResNet-152 model achieved the highest accuracy of 99.41\%.  A lightweight CNN \RVV{\citep{lu2023automatic}} used transfer learning to classify the DR and DME simultaneously. The model's average accuracy, precision, recall, and specificity after five rounds of cross-validation were 96.66\%, 96.85\%, 99.32\%, and 96.63\%.  In the article \RVV{\citep{adak2023detecting}}, significant parameters of fundus images were captured using transformer-based learning models for a more nuanced understanding of DR severity. To determine the severity of DR from fundus photographs, transformers were employed and used four models: the Vision Transformer (ViT), Data-Efficient Image Transformers (DeiT), Bidirectional Encoder representation for image Transformer (BEiT), and Class-Attention in Image Transformers (CaiT). The model was tested on used the publicly available APTOS-2019 dataset. The work presented a CNN ensemble model for accurate fundus image-based DR identification and grading. The other studies \RVV{\citep{rakhlin2018diabetic} \citep{fellah2023diabetic} \citep{moin2023diabetic} \citep{swarnalatha2023detection} \citep{elmoufidi2023diabetic}} proposed DL model for DR classification through fundus images.\\
\vspace{-0.3cm}
\paragraph{OCT Scans}
\noindent A CNN model \RVV{\citep{An_2019(142)}} was developed for glaucoma diagnosis through fundus and OCT images. The parameters OD, RNFL deviation map, macular GCC thickness map, and RNFL thickness map were all calculated by commercial software and were used to train the model. Another study feature-based and feature agnostic techniques \RVV{\citep{Maetschke2019PLOSONE}} for glaucoma diagnosis. A feature method based on 22 parameters (calculated by the OCT machine) and a traditional machine learning classifier was used. An unsegmented OCT volume was classified as normal or glaucomatous using a feature-agnostic framework built on 3D CNN for glaucoma diagnosis. To acquire depth information, they used 3D convolution, which allowed them to locate the significant area for diagnosis. The features were derived from raw data by using Class Activation Maps. The method failed to detect glaucoma in OCT images from patients over the age of 65 or those with advanced disease. In order to digitally stain the neuronal and connective tissues of ONH, a DL framework called DRUNET \RVV{\citep{Sripad2018BOE(144)}} was presented, which was a U-Net-derived fully convolutional neural network and takes advantage of skip connections. The peripapillary sclera and LC were both successfully separated by the algorithm. The model could extract both global (spatial) and local (texture) characteristics of ONH tissues. A shortcoming of the proposed architecture is that it was only trained on 100 OCT images obtained from healthy and glaucoma participants.

\noindent Muhammad et al. \RVV{\citep{Muhammad_2017(140)}} proposed DL model based on AlexNet and utilized transfer learning. The OCT software's measurements of features like RNFL and GCIPL thickness were fed into the CNN. CNN's feature extraction was utilized to train a random forest classifier. Depending on the parameters used, the model's accuracy was anywhere from 63.7\% to 93.1\%. Kermany et al. \RVV{\citep{kermany2018identifying}} proposed a DL model to diagnose DME, CNV, DRUSEN and normative retinal OCT. They used 108,312 OCT images from 4,686 patients  for training and 1,000 scans from 633 subjects for testing and achieved accuracy, sensitivity, and specificity ratings of 96.6\%, 97.8\%, and 97.4\%, respectively. Feng et al. \RVV{\citep{li2019deep}} proposed a deep classification model for choroidal neovascularization (CNV), DME, and DRUSEN through OCT images. To identify retinal OCT images, the proposed approach used an ensemble of four classification model instances, all of which were based on an enhanced ResNet50. The dataset consisted of 21,357 retinal OCT scans that were gathered from 2,796 adult patients at Shanghai Zhongshan Hospital and Shanghai First People's Hospital between 2014 and 2019. The model classification accuracy for the B-scan was 0.973 (95\% 0.971–0.975), the CI, 0.971-0.975), sensitivity was 0.963 (95\% and the 95\% CI, 0.983–0.987). Butola et al. \RVV{\citep{butola2020deep}} proposed a CNN model (LightOCT) to classify OCT images into classes normal, AMD, and DME. LightOCT had a two-convolutional-layer and a fully-connected-layer, and achieved accuracy  greater than 96\%. A DL model \RVV{\citep{jin2022multimodal}} was proposed based on feature-level fusion (FLF) method that combined the OCT and OCTA images for the assessment of CNV in AMD. The model was tested on two external datasets and achieved an accuracy of 95.5\% and an AUC of 0.9796 on multimodal data.

\noindent A novel uncertainty guided semi-supervised model \RVV{\citep{sedai2019uncertainty}} was proposed based on student-teacher methodology. Limited labeled samples and a large number of unlabeled images were used for training. First, using Bayesian deep learning, a teacher segmentation model was trained using the labeled data. In order to generate soft segmentation labels and an uncertainty map for the unlabeled collection, the trained model was employed. After the data were softly segmented, the uncertainty of the teacher model was assessed, and the pixel-wise confidence of the segmentation quality was used to update the student model. Normal OCT scans were reconstructed using an adversarial network that was trained with little supervision \RVV{\citep{wang2021weakly}}. The network then reconstructed the abnormal (disease) images at the inference stage, using the difference between the input and reconstructed scans to identify lesion pathologies. Das et al. \RVV{\citep{das2020unsupervised}} proposed an unsupervised framework using the GAN to perform fast and reliable super-resolution without the requirement of aligned low and high-resolution pairs. Adversarial learning identified mapping priors to obtain the spatial, color, and texture information in the high-resolution scans. Automated AMD diagnosis using the generated images yields an improved classification accuracy of 96.54\%.  Das et al. \RVV{\citep{das2020data}} proposed a semisupervised GAN-based classifier for automated diagnosis using limited labeled data. The two main components of the framework were the generator and the discriminator. The adversarial learning between them aids in the development of a generalizable classifier for the prediction of degenerative retinal illnesses like AMD and DME.


\noindent Research trends shifted towards the analysis of OCT volume for detecting various retinal diseases; the study \RVV{\citep{rasti2018automatic}} proposed a model for the classification of abnormal macula through 3D-OCT. The technique evaluated intraretinal layers and lesions without the use of denoising, segmentation, or retinal alignment operations. A two-stage plan was used to separate abnormal cases from the control group based on adaptive feature learning and diagnostic scoring. Initially, the cumulative characteristics of 3-D volumes were extracted using a wavelet-based CNN model for generating B-scan CNN codes in the spatial-frequency domain. The second step involved using the derived features to score the existence of anomalies in the 3D OCT. The technique was tested on two independent retinal SD-OCT datasets using the five-fold cross-validation (CV) method. The first group is composed of 30 normal participants and 30 patients with DME 3-D OCT scans acquired with a Topcon instrument. The second set of data was from the Heidelberg device and included 45 subjects, each class (AMD, DME, and normal) contained 15 subjects. The results showed that in the two-class classification problem (dataset1), the suggested method achieved an average precision of 99.33\%. When used for the three-class classification problem (dataset2), the model achieved an average precision of 98.67\%. The study \RVV{\citep{hassan2018multilayered}} introduced a multilayered CNN structure tensor Delaunay triangulation that extracted nine retinal and choroidal layers and the macular fluids. The retrieved retinal information was used for the automated diagnosis of maculopathy and the reliable reconstruction of the 3D macula of the retina. The model was validated on 41,921 OCT images collected from different vendors and achieved mean accuracy of 95.27\% for extracting retinal layers. Whereas, for extracting fluid, the reported mean dice coefficient was 0.90,  and the overall accuracy  for maculopathy diagnosis was 96.07\%.  Mantel et al. \RVV{\citep{mantel2021automated}}  proposed a DL model to identify and localize AMD biomekers such as IRF, SRF, and pigment epithelium detachment (PED). Cubic volumes of SD-OCT were collected from 117 AMD eyes, then manual annotation of the retinal lesions was performed. A 3D-FCN \RVV{\citep{li2019segmentation}} based on U-Net was proposed to segment the retinal fluid OCT images. The model was evaluated on the local dataset (75 volumes), achieved Kappa coefficient of 98.47\%, accuracy rate of 99.56\%, and F1 score of retinal fluid was 95.50\%. In the paper \RVV{\citep{mukherjee2022retinal}}, a 3D deep neural network  was proposed that segmented the retinal layers ensuring accuracy and smoothness. The model was made up of two separate but complementary networks: (1) 3D UNet that performed multi-class voxel labeling of retinal layer surfaces, and (2) 3D convolutional-autoencoder, which limits the 3D UNet's output and compels it to estimate a smooth contour. 

\vspace{-0.3cm}
\section{Advanced Deep Learning Schemes}
This section presents a brief introduction to advance DL techniques and state-of-the-art methods for the identification and classification of retinal lesions.
\vspace{-0.3cm}
\subsection{Meta-Learning \& Multi-task Learning}
\noindent Meta-learning is a subfield of DL that focuses on the problem of learning how to learn or learning from previous learning experiences. The goal of meta-learning is to enable a model to quickly adapt to new tasks using only a small amount of data by leveraging the knowledge gained from previously seen tasks. There are several different approaches to meta-learning, each with its own set of advantages and disadvantages.
\begin{itemize}
\item One common approach is to use a neural network as the model and train it on a variety of tasks in a way that the parameters of the network are updated so that they can be used to adapt to new tasks quickly. This is done by defining a loss function for the meta-learning process that tries to minimize the difference between the parameters of the model after adapting to a new task and the parameters of the model after adapting to similar tasks in the past.
\item Metric-based meta-learning, which learns a distance metric in the space of task representations, such that new tasks can be quickly adapted to by finding the most similar tasks to the new task in the learned metric space.
\item Model-based meta-learning methods, which learn a model of the task-generating process and can use this model to adapt to new tasks quickly.
\item Optimization-based meta-learning methods, which learn an optimization algorithm that can quickly adapt to new tasks by using the gradients of the loss function with respect to the model parameters.
\end{itemize}

All of these methods have been successfully applied to a variety of different problems, such as few-shot and one-shot learning, where a model must quickly adapt to new tasks with limited data, reinforcement learning, and other areas. Overall, the main idea behind meta-learning is to train a model on a variety of tasks such that it can quickly adapt to new tasks using the knowledge gained from the previous tasks. This allows the model to improve its learning efficiency and generalization performance.
The meta-learning strategy involves training the model's parameters explicitly so that good generalization performance can be achieved on a new task using a short number of gradient steps and a small amount of training data \RVV{\citep{finn2017model}}. For automated diagnosis of DR in fundus images, an anomaly characterization algorithm \RVV{\citep{matta2023meta}} was developed. A few-shot learning solution in which CNN trained for common conditions was combined with an unsupervised probabilistic model for detecting rare conditions. CNNs often assume that images with the same anomalies were similar, even though they were trained to look for differences. The algorithm achieved an average AUC of 0.938. 

\subsection{Few-shot learning}
Few-shot learning (FSL) model is able to learn and recognize new classes with only a small number of examples. The goal is to train models that can generalize well to new classes, even when only a small number of examples are available for these classes. There are two main approaches for few-shot learning:
\begin{itemize}
    \item Meta-learning: This approach involves training a model on a large number of similar tasks so that it can learn to adapt quickly to new tasks. The model learns a general way of learning rather than memorizing the specific training examples.
    \item Transfer learning: This approach involves using a pre-trained model on a large dataset and fine-tuning it on the few-shot task. The idea is that the model has already learned useful features from the large dataset that can be useful for the few-shot task.
\end{itemize}
FSL is often used in tasks such as image classification, where there is a small number of examples per class. Some other examples of real-world applications that use few-shot learning are medical imaging, rare species identification, and speech recognition. Recently, some new techniques have been proposed for few-shot learning, such as few-shot learning with attention mechanisms, memory-augmented neural networks, and metric-based learning. These methods have shown promising results in various few-shot learning benchmarks. Yoo et al. \RVV{\citep{yoo2021feasibility}} proposed a model based on FSL using GAN  to diagnose rare ocular diseases in OCT images. Before training the classifier, GAN model was built to turn normal OCT images into pathological OCT images for disease. Inception-v3 was trained using a training dataset, and then the final model was tested on a separate test dataset. The study \RVV{\citep{mai2021few}} modeled the problems caused by a lack of labeled data as a Student-Teacher learning with a knowledge distillation (KD). Kim et al. \RVV{\citep{kim2017few}} proposed a model for glaucoma diagnosis in fundus images using FSL. The study \RVV{\citep{murugappan2022novel}} proposed model DRNet, based on FSL and attention, that performed grading and detection of DR. To build the attention mechanism to preserve visual representations, the network makes use of aggregated transformations and class gradient activations. The model was evaluated on APTOS2019 dataset and achieved 99.73 \% accuracy, 99.82\% sensitivity for DR detection, 98.18\% accuracy, 97.41\% sensitivity for DR grading. Gulati et al. \RVV{\citep{gulati2022detection}} used FSL on the  iris dataset to find out hemorrhages or Microaneurysms disease in images. 

\subsection{Incremental learning}
\noindent Incremental learning model continually updates its knowledge from newly acquired data, without being retrained from scratch. This allows the model to continuously learn and improve its performance over time, making it suitable for scenarios where the data is constantly changing. The model parameters update incrementally instead of retraining the entire model on all the data again. This helps to reduce computational resources required to train the model, as well as allow the model to learn from new data continuously. The approach typically involves dividing the incoming data into mini-batches and using these mini-batches to update the model parameters using gradient descent or a similar optimization technique. It can be used in various tasks such as classification, regression, and clustering. It is especially useful in scenarios where the data is too large to fit into memory or where the data is streaming and needs to be processed in real-time. To improve classification accuracy, the study \RVV{\citep{meng2020adinet}} presented model "Attribute Driven Incremental Network" (ADINet) that combined class label prediction with attribute prediction inside an incremental learning framework. Knowledge distillation (KD) was used for image classification to preserve the information of base classes. For improved accuracy in attribute prediction, weights were assigned to each image attribute based on their relative importance. They came up with the concept of attribute distillation (AD) loss to preserve the data of base class attributes despite the advent of new classes. There is only a small performance hit when repeating this incremental learning process numerous times. Hassan et al. \RVV{\citep{hassan2021incremental}} introduced a novel incremental cross-domain adaption technique that can be used by any deep classification model to gradually learn pathological abnormalities in OCT and fundus imaging with only a small number of training examples. By using a Bayesian multiobjective function, the proposed technique not only ensures that the candidate classification network retains its prior learned knowledge during incremental training but also that it understands the relationships between previously learned pathologies and recently introduced disease categories so the model can effectively recognize them during the inference stage. The model achieved an overall accuracy and F1 score of 98.26\% and 0.98, respectively, when tested on six public datasets. In the study \RVV{\citep{he2021incremental}}, an incremental learning-based model was proposed for the DR lesion segmentation that distills the knowledge of the previous model in order to enhance the current model. A probability-map alignment scheme was proposed to combine the previous map and the current map. The scheme dealt with the special class background in the context of segmentation. Using the scheme, it was easy to calculate the optimized value for the model-based weight. The idea of "knowledge distillation" was used to move the information from the probability map to the current model. 

\subsection{Contrastive learning}
\noindent Contrastive learning (CL) is a self-supervised learning method that aims to learn a feature representation of data that separates positive pairs from negative pairs \RVV{\citep{tian2019contrastive}}. The idea is to maximize the similarity between positive pairs while minimizing the similarity between negative pairs. This is typically done by designing a contrastive loss function and optimizing it using gradient descent \RVV{\citep{tan2022retinal}}. The learned representation can then be used for downstream tasks, such as classification or clustering, without the need for labeled data. It has been applied to various domains, including computer vision and natural language processing, and has shown promising results. Numerous DL-based methods have been proposed, and they perform better than human analysis at diagnosing retinal disorders. Cross-entropy is widely employed as a loss function in conventional DL model training. But it has recently been found that this loss function has certain drawbacks, such as a poor margin that might cause erroneous findings, sensitivity to noisy data, and hyperparameter variability. To fix these problems, contrastive learning has been gaining popularity. Islam et al. \RVV{\citep{islam2022applying}} proposed a supervised CL model for detecting DR from fundus images. For image enhancement, CLAHE was used, and Xception model was employed as the encoder for representation learning. The SCL of the model was interpreted by projecting a 128-D embedding space into a 2-D plane using the t-SNE method. Two publicly available datasets, APTOS and Messidor-2 were used for training and testing of the model. For DR (Binary classification), the model achieved an AUC of 98.50\% and an accuracy of 98.36\% and on the APTOS 2019 dataset. Whereas for five-stage grading, it gained AUC of 93.81\% and an accuracy of 84.36\%. Tian et al.\RVV{\citep{tian2019contrastive}} developed CL-driven methodologies to constrain the model's knowledge to learn the difference between new anchor examples and previously acquired positive and negative examples. With the goal of identifying retinal biomarkers in OCT images, a novel contrastive uncertainty network (CUNet) \RVV{\citep{liu2022contrastive}} was developed. To improve the network's capacity for distinguishing between distinct classes of retinal biomarkers, CUNet employed a proposed CL strategy to strengthen the feature representation of biomarkers. To further enhance the network's sensitivity to the fuzzy boundaries of retinal biomarkers, bounding box uncertainty was proposed and integrated with the conventional bounding box regression. In the study \RVV{\citep{kaplan2022contrastive}}, a variational autoencoder (VAE)-based technique was developed for the generation of OCT images of the retina using CL. In the second step, disease-specific OCT images were generated by applying VAEs to the learned embeddings. It was found that the diseases were effectively partitioned in the embedding space, and the suggested method successfully produced high-quality images with high-detail spatial resolution. Alam et al. \RVV{\citep{alam2022contrastive}} proposed a CL-based framework with neural style transfer (NST) augmentation to generate models with improved representations for detecting DR in fundus scans. The EyePACS dataset was used to train and evaluate the model, and clinical data from the University of Illinois, Chicago (UIC) was used for testing. To improve a U-Net embedding capacity to segment retinal vessels in fundus image, a model \RVV{\citep{xu2022local}} was presented that uses a local-region and cross-dataset CL strategy without introducing complex network structures. The main goal was to distinguish the characteristics of pixels that are easily confused with their neighbors within the same local region. The model took full advantage of the global contextual information of the entire dataset that improves the features by employing a memory bank method. The model was evaluated on DRIVE and CHASE-DB1 datasets. In order to provide lesion-aware scanner-independent screening and grading of retinopathy, the study \RVV{\citep{hassan2023angular}} introduced a novel self-supervised segmentation-driven classification pipeline that used a proposed angular contrastive distillation approach to extract retinal lesions. To further improve the proposed framework's diagnostic capabilities, a novel co-attention mechanism was incorporated. The mechanism allowed the underlying network to concentrate on retinal abnormalities and effectively grade retinal diseases without requiring ground truth labels. The model was tested on seven publicly available datasets acquired using four different scanners, where it achieved a 9.22\% improvement in mean IOU for extracting retinal lesions and a 10.71\% improvement in F1 score as compared to state-of-the-art solutions for grading retinopathy. In order to identify and segregate the biomarkers in OCT scans using only image-level annotations, a weakly supervised network called TSSK-NET \RVV{\citep{liu2023tssk}} was proposed. The method is a Teacher-Student network with Self-supervised CL and Knowledge distillation-based anomaly localization. Initially, a unique pre-training technique based on supervised CL was proposed to teach the model morphology of normal OCT images. Second, a module for fine-tuning was built, and a novel hybrid network was proposed. The model employed supervised CL for learning features and cross-entropy loss for learning classes. To further enhance performance, it was proposed to combine these two losses in order to preserve the various morphologies and improve the encoding representation of features. Finally, a knowledge distillation-based anomaly segmentation method was utilized that was effectively integrated with the prior model to relieve the difficulty of insufficient supervision. In the study \RVV{\citep{holland2023clustering}}, the AMD progression through OCT images was analyzed in a self-supervised feature space. CL was used to pretrain an encoder, which then projects images from longitudinal time series to positions in feature space. This enables the construction of disease trajectories that were subsequently denoised, and divided into clusters. These clusters were associated with OCT biomarkers and were discovered in two datasets encompassing time series of 7,912 patients scanned over a period of eight years.

\subsection{Domain Adaptation}
\noindent Domain adaptation is an advanced DL technique that enables models trained on one domain (source domain) to generalize well to another domain (target domain) with different distributions \RVV{\citep{liu2019cfea}}. This is especially useful in scenarios where annotated data is scarce in the target domain. The main idea behind domain adaptation is to align the feature representation of the source and target domains so that the model trained on the source domain can effectively generalize to the target domain. This can be achieved by various methods, including fine-tuning the model using a small amount of labeled target data, adversarial training, and feature reconstruction. The goal is to reduce the domain shift, or the difference between the source and target domains so that the model can perform well on both domains. In order to facilitate unsupervised domain adaptation to segment retinal vessels in fundus images, Zhuang et al. \RVV{\citep{zhuang2019domain}} derived the asymmetrical maximum classifier discrepancy (AMCD) strategy from maximum classifier discrepancy. The model was trained using labeled data and then tested with unlabeled data from the target domain. The three classifiers were trained in an asymmetrical fashion, with one main classifier using only the source examples and the other two assist classifiers being utilized to maximize the discrepancy on target samples. The model was validated on DRIVE, STARE, CHASE-DB1, and IOSTAR eye vessel segmentation datasets. Lui et al. \RVV{\citep{liu2019cfea}} proposed an unsupervised domain adaptation model Collaborative Feature Ensembling Adaptation (CFEA) that collaborated adaptation through both adversarial learning and ensembling weights. By ensembling weights during training, the model not only achieved domain-invariance but also maintained an exponential moving average of the previous predictions, leading to improved prediction for the unlabeled data. Multiple adversarial losses enable the extraction of domain-invariant features to confound the domain classifier and simultaneously benefit the ensembling of smoothing weights, all without performing annotation of any sample from the target domain. 
Song et al. \RVV{\citep{song2020domain}} proposed a domain-adaptive multi-instance learning with attention technique for DR grading. Labeled examples are generated through cross-domain to eliminate irrelevant instances. A multi-instance learning with attention technique was created to collect the spatial information of highly suspicious lesions and performed DR grading. The model achieved an average accuracy of 76.40\% and an AUC value of 0.749 when tested on the Messidor dataset. Wang et al. \RVV{\citep{wang2021unsupervised}} developed an unsupervised domain adaption model based on faster-RCNNs for lesion detection in multi-device retinal OCT images. Both the shift at the image level and the shift at the instance level were minimized together to reduce the domain shift. In order to synchronize the changes across all of the levels, the model used a combination of a domain classifier and a Wasserstein distance critic. Using OCT image data from two separate devices, the model achieved an average accuracy improvement of more than 8\% compared to the technique without domain adaptation and surpassed the performance of other comparable domain adaptation methods. Another study \RVV{\citep{yang2020unsupervised}} proposed an unsupervised domain adaptation framework for lesion detection in OCT images. To compel the network to learn device-independent features, the model developed global and local adversarial discriminators. A non-parameter adaptive feature norm was then presented for the global adversarial discriminator in order to stabilize classification in the target domain. The study proposed a Collaborative Adversarial Domain Adaptation (CADA) model \RVV{\citep{liu2022cada}}  based on domain adaptation with multi-scale inputs and multiple domain adaptors employed in feature and output space. The information loss caused by the network's pooling layers used for feature extraction can be mitigated with multi-scale inputs. CADA is an interactive paradigm that enables collaborative adaptation through adversarial learning and weight ensembling. Model accomplished domain invariance and model generalizability by using adversarial learning at multi-scale outputs from distinct network layers and retaining an exponential moving average (EMA) of the historical weights during training. Multiple adversarial losses in the encoder and decoder layers direct the extraction of domain-invariant features without annotating a single sample from the target domain. The model outperformed on REFUGE, Drishti-GS, and Rim-One-r3 datasets. An innovative approach \RVV{\citep{madadi2022domain}} for glaucoma diagnosis based on learned representations that included both domain-invariant and domain-specific in order to extract generic and domain-specific information. Low-rank coding was employed for aligning source and target distributions, as well as progressive weighting was used for the correct transfer of source domain information and mitigation of negative knowledge transfer to the target domain. The model was evaluated on OHTS, ACRIMA and RIM-ONE. Zhang et al. \RVV{\citep{zhang2022convolutional}} proposed an unsupervised domain-adaptive segmentation (CAE-BMAL) model to extract the OD and OC. Initially, a convolutional autoencoder was used to boost the source domain, allowing the model to generalize better. Then, to mitigate the effects of the complex environment on segmentation, a boundary discrimination branch based on adversarial learning was be introduced. The model was validated on three datasets, Drishti-GS, RIM-ONE-r3, and REFUGE. The study \RVV{\citep{cao2022collaborative}} proposed a unified weakly-supervised domain adaptation framework for the DR diagnosis. The model comprised three parts: domain adaptation, progressive discriminator for individual instances, and multi-instance learning with attention. The method utilized multi-instance learning and an attention mechanism to model the connection between patches and images in the target domain. Additionally, it used a combined learning approach that takes into account data from both the source and the target domains. Results on the Messidor dataset showed that the model had an average accuracy and AUC of 94.90\% and 0.76 for binary-class and  95.80\%,0.74 for multi-class classification, respectively. The model achieved 88.70\% accuracy on Eyepacs dataset. Chen et al. \RVV{\citep{chen2022segmentation}} proposed a segmentation-guided domain-adaptation model for adapting images from various OCT machines into a single image domain. It eliminates the time-consuming processes of manually labeling new datasets and retraining the existing network. The study \RVV{\citep{hou2023self}} deals with image quality enhancement of fundus images in a completely unsupervised manner, neither paired nor high-quality photos were used.
\subsection{Attention-based Models}
\noindent The study \RVV{\citep{fang2019attention}} proposed a novel lesion-aware DL model (LACNN) for retinal OCT image classification. At first, the OCT image was used to create a soft attention map using the lesion detection network. A classification network is then used to assign relative importance to the various convolutional layers based on the attention map. An improved U-Net model \RVV{\citep{liu2021automatic}} based on an attention mechanism was developed to identify the fluid region. By bringing together high-level and low-level information, skip connections improved the accuracy of the segmentation outcomes. The loss function is a combination of the weighted binary cross-entropy loss, the dice loss, and the regression loss (to prevent the problem of converging fluid areas). Another study \RVV{\citep{liu2021one}} presented a one-stage attention-based method for retinal OCT image segmentation and classification. Mishra et al. \RVV{\citep{mishra2021perturbed}} proposed a model for the classification of macular OCT images using Multilevel perturbed spatial attention to extract context-aware diagnostic features. The model classified AMD, DME, and CNV. Preprocessing techniques like region of interest extraction, denoising, and retinal flattening were not required with the proposed end-to-end trainable architecture. Liu et al. \RVV{\citep{liu2020mdan}} proposed an enhanced nested U-Net architecture (MDAN-UNet) for end-to-end segmentation of OCT images. The model was evaluated on two publicly available benchmark datasets, Duke DME and the RETOUCH datasets. The study \RVV{\citep{sun2020automatic}} presented a model to classify OCT volume for diagnosing macular diseases. The model consisted of three modules:  feature extractor from B-scan, 2D map generation, and volume-level classifier. The model was trained and used to construct a 2D map for OCT volume, whereas volume-level classifiers (SVM) classified 2D feature maps. The results of a five-fold cross-validation of the model showed an average of 98.17\% accuracy, 99.26\% sensitivity, and 95.65\% specificity.  The model was trained and tested on a publicly available dataset \RVV{\citep{kermany2018large}}. In terms of accuracy, the model achieved 97.79\% on the training data and 95.6\% on the testing data. Rendering en face OCT of arbitrary retinal layers was made possible by real-time segmentation in combination with high-speed OCT volume acquisition, which can be used to improve the success rate of high-quality scans and give surgeons immediate feedback during image-guided procedures. In the study \RVV{\citep{borkovkina2020real}}, researchers used three tiers of optimization to successfully segment the eight retinal layers in real time using OCT. First, a simplified neural network architecture; second, a novel neural network compression approach using TensorRT; and third, dedicated GPU hardware to speed up calculations. The U-NetRT compressed network offered 21 times faster inference than regular U-Net inference with no loss of accuracy. Kumar et al. \RVV{\citep{kumar2022optical}} developed a model that incorporated Attention and Transfer Learning  to classify the  CNV, DME, drusen from the OCT images. An innovative end-to-end multiscale attention-gated network (MAGNet) was proposed \RVV{\citep{cazanas2022multiscale}} for detecting and segmenting retinal layers and macular cystoid edema in OCT images. In order to deal with class imbalance, the MAGNet utilizes a weighting loss methodology and a FCN model that uses attention gates at different scales to perform segmentation. All B-scans were center-cropped along their longest axis to reduce their dimensions and create 496-by-496-pixel squares as part of the preprocessing phase. The model used  Duke and HCMS datasets for training and testing and achieved the mean dice score of 0.92 ± 0.03. Li et al. \RVV{\citep{li2023magf}} a multiscale attention-guided fusion network (MAGF-Net) was proposed for vessel segmentation in fundus images. A multiscale attention (MSA) block was proposed for the construction of the backbone network in order to capture multiscale contextual variables. To obtain global multiscale contextual information, a feature enhancement (FE) block was designed and integrated into the bottleneck layer. Attention-guided fusion (AGF) blocks were created to combine characteristics from various network levels in order to maximize the usefulness of both channel information from deep layers and spatial information from shallow layers. For further data retention throughout the downsampling process, a hybrid feature pooling (HFP) block was utilized. The model was validated on three public datasets: the CHASE-DB1, the DRIVE, and the STARE. The model achieved F1 and accuracy of 0.8329 and 96.77\% on DRIVE, 0.8307 and 95.78\% on STARE, and 0.8364 and 96.\%49 on CHASE-DB1, respectively. A multi-scale residual attention network (MRANet) \RVV{\citep{yi2023segmentation}} based on U-Net was developed to segment retinal vessels in fundus scans. Initially, a multi-level feature fusion (MLF block) block was introduced to collect blood vessel information more effectively. Then, variable weights of each fused feature were learned through the use of attention blocks, which can preserve more useful feature information while lowering interference from redundant features. Next, a multi-scale residual connection block (MSR block) was created to extract features more effectively. Finally, network overfitting was mitigated by incorporating a DropBlock layer within the network. The model achieved an accuracy rate of 96.98\% and an AUC performance value of 0.98 on the DRIVE dataset, and 97.55\% and 0.98 on the CHASE DB1 dataset, respectively.
\vspace{-0.12cm}
\section{Datasets}
\noindent This section provides information about the datasets of fundus and OCT images. There are various datasets, private and publicly accessible, are present in the literature for both modalities. Here we are only describing the publicly available datasets. The publicly available dataset was originally created to serve as a testing dataset against which various detection algorithms could be evaluated.
\subsection{Fundus Datasets}
\noindent There are various fundus-based datasets that are publicly available, which are used to detect various retinal abnormalities. Table \ref{tabfundus} summarizes the features of publicly available fundus datasets. 
\noindent \textbf{DRIVE} \RVV{\citep{DRIVE}} dataset consists of 40 fundus images and segmented blood vessels.\\

\noindent \textbf{DIARETDB0} dataset \RVV{\citep{DiaRetDb0}} contains 30 CFPs, 20 of which are considered normal and 110 of which exhibit DR (hard and soft EXs, MAs, HMs, and neovascularization). There are a total of 89 CFPs in the DIARETDB1 \cite{DiaRetDb1} database; 84 show at least mild NPDR signs of the DR, while five are considered normal and show no signs of the DR.\\ 

\noindent \textbf{STARE} (Structured Analysis of the Retina) \RVV{\citep{STARE}} dataset conists of ~400 raw fundus images with labels of 13 categories and also segmentation annotation of blood vessels, arteries, and optic nerve for 40, 10, and 80 images, respectively. \\

\noindent \textbf{DRIONS-DB} \RVV{\citep{DRIONS-DB}} consisting of 110 unlabeled images with two different segmentation of the OD for each image. The Messidor project \cite{Messidor} objective was to conduct a comparative evaluation of various segmentation algorithms developed for the purpose of detecting lesions in fundus images. More specifically, there are 1200 fundus images in this dataset, each of which has been labeled with a medical diagnosis. The medical experts have presented two diagnoses for each image, retinopathy grade and risk of DME. It has been determined to classify DR into six levels of severity, four levels of exudation, and three levels of hemorrhage and to specify the number of microaneurysms for each image. \\

\noindent \textbf{RIGA} \RVV{\citep{RIGA}} is a dataset including 750 retinal fundus images for glaucoma analysis. The dataset includes the OC and OD ground truth for each image; however, the diagnosis of glaucoma is not provided. ORIGA30 \RVV{\citep{zhang2010origa}} consists of 482 and 168 fundus images of healthy and glaucoma, respectively, as well as the segmentation of the disc and cup. This dataset was accessible to the public and downloadable in 2010, but it does not appear to be accessible to the public now.\\ 

\noindent \textbf{RIMONE} \RVV{\citep{RIMONE}} first made available to the public in 2011. To test the CDR, 159 stereo fundus images were given in 2015, with two ground truth segmentations of OD and OC. These photos represented healthy individuals and glaucoma patients. The dataset was recently revised and improved for a deep learning context in 2020. The new data set includes 313 and 172 images from healthy and glaucoma patients, respectively. \\

\noindent \textbf{Drishti-GS} \RVV{\citep{Drishti-GS}} is a dataset including OD and OC segmentation for glaucoma assessment. It is comprised of 101 monocular fundus images (31 normal,70 glaucoma images), divided into training and test sets, with four segmentation of the OD and OC for the training set.\\ 

\noindent \textbf{ACRIMA} \RVV{\citep{diaz2019cnns}} has 705 fundus photos with labels (309 normal and 396 glaucomatous images). Two glaucoma specialists were involved for the annotations, and no other clinical evidence was considered while providing labels for the images.\\ 

\noindent \textbf{G1020} \RVV{\citep{G1020}} is a huge dataset of retinal fundus images for glaucoma diagnosis, containing 1020 images (724 healthy and 296 glaucoma). There was segmentation of the OD and OC as well as labeling of the images.
 \\

\noindent \textbf{REFUGE} \RVV{\citep{REFUGE}} dataset was released, which includes 1200 fundus images annotated with clinical glaucoma diagnoses and ground truth segmentation of the OC and OD. \\

\noindent \textbf{PAPILA} \RVV{\citep{kovalyk2022papila}} dataset includes data from 244 patients (333 health, 155 glaucoma images). Each file contains organized data on a single patient's clinical history, as well as segmentation of the OD and OC in both eyes. \\

\noindent \textbf{Kaggle Diabetic Retinopathy (Kaggle-DR)} \RVV{\citep{Kaggle-DR}} a total of 88,702 CFPs are available in the   dataset, split between 35,126 training samples and 53,576 test samples. EyePACS contributed the images, which were taken using a wide range of devices in a variety of settings at numerous primary care clinics in California and abroad. Images of both the left and right eyes were taken at the same resolution for each individual. Clinicians used the Early Treatment DR Study (ETDRS) scale to assess the severity of DR in each image. \\

\noindent \textbf{GAMMA} \RVV{\citep{GAMMA}} dataset includes 3D OCT and 2D fundus images from 300 patients. Each image in the dataset was annotated with glaucoma grade, macular fovea coordinates, and an optic disc/cup segmentation mask from the fundus image.

\begin{table*}

\caption{Summarizing the feature of publicly available fundus dataset. The abbreviations are Class Label: Classification Label, N: healthy images, GLA: glaucoma images, MiDR: Mild DR, MDR: Moderate, PDR: Proliferative DR, SER: Severe DR, HEXs: hard exudates, and SOEXs: soft exudates }
\label{tabfundus} 
\centering
\resizebox{\textwidth}{!}{
\begin{tabular}{lllllll}
\hline\noalign{\smallskip}
Dataset & Images& Categories &Diseases&  Class Label & Segmentation label\\
\noalign{\smallskip}\hline\noalign{\smallskip}
DRIVE \cite{DRIVE} & 40 & - & DR & - & Blood vessels\\
DIARETDB0 \cite{DiaRetDb0}& 130 & N:20, DR:110 & DR& Yes& MAs, HMs, HEXs, SOEXs    \\
DIARETDB1 \cite{DiaRetDb1}& 89  & N:5, PDR:84 & DR& Yes &MAs, HMs, HEXs, SOEXs\\

Drions-DB \cite{DRIONS-DB}& 110 &- & - & No & ONH\\
Messidor\cite{Messidor}& 1200 &- & DR, DME &Yes & -\\
RIGA\cite{RIGA}& 750 &- & Glaucoma & Yes & OC, OD\\
RIMONE \cite{RIMONE}& 485 &N:313, GLA: 172 & Glaucoma & Yes & OC, OD\\
STARE\cite{STARE}& 400 & -& AMD, PDR,CNV &Yes & Blood vessels, artery, optic nerve\\
Drishti-GS \cite{Drishti-GS}& 101 &N:31, GLA: 70 & Glaucoma & Yes & OC, OD\\
(Kaggle-DR) \cite{Kaggle-DR}& 35126 &N: 25810, MiDR: 2443, MDR: 5292, PDR: 708, SER: 873  & DR & Yes & -\\
ACRIMA \cite{diaz2019cnns}& 705 &N:309, GLA: 396 & Glaucoma & Yes & -\\
G1020 \cite{G1020}& 1020 &N:724, GLA: 296 & Glaucoma & Yes & OC, OD\\
REFUGE \cite{REFUGE}& 1200 &N:1080, GLA: 120 & Glaucoma & Yes & OC, OD\\
PAPILA \cite{kovalyk2022papila} & 488 &N: 333, GLA: 155 & Glaucoma & Yes & OC, OD\\
GAMMA \cite{GAMMA}& 300 & - & Glaucoma & Yes & OC, OD, fovea\\
\noalign{\smallskip}\hline
\end{tabular}
}
\end{table*}
 
\subsection{OCT Datasets}
\noindent The publicly available OCT datasets are the Zhang \cite{kermany2018identifying}, Duke-1 \cite{farsiu2014quantitative}, Duke-2 \cite{chiu2015kernel}, Duke-3 \cite{srinivasan2014fully}, Rabbani \cite{rasti2017macular}, BIOMISA \RVV{\citep{hassan2018biomisa}}, and AFIO \RVV{\citep{raja2020data}}. The table \ref{tab:datasets} provides an overview of the OCT datasets that are publicly available.

\noindent \textbf{Zhang dataset} \RVV{\citep{kermany2018identifying}} dataset is one of the most extensive OCT datasets that are freely accessible to the public. The images were acquired from Spectralis OCT, Heidelberg Engineering, Germany at Beijing Tongren Eye Center, the Shanghai First People's Hospital, the California Retinal Research Foundation, Medical Center Ophthalmology Associates, and the Shiley Eye Institute of the University of California San Diego between 1 July 2013 and 1 March 2017. It has a total of 108,309 OCT images and was primarily developed with the intention of screening for DME, DRUSEN, CNV, and other normal images. Among the total of 109,309 scans analyzed, 51,140 of them exhibit a normal retina, while 11,348 scans indicate the presence of DME symptoms, 8,616 shows DRUSEN, and 37,205 depicts CNV symptoms.   \\

\noindent \textbf{Duke-1 \RVV{\citep{farsiu2014quantitative}}} is consisted of 38400 BScans from 269 AMD patients and 115 normal subjects. Images were collected from the Bioptigen system. Total retina (TR, between the ILM and the inner aspect of Bruch's membrane) mean, and standard deviation maps for individual subjects are also provided in the dataset. In addition to this, the dataset contains the subject-specific mean and standard deviation thickness maps of the RPE and drusen complex.\\

\noindent \textbf{Duke-2 dataset} \RVV{\citep{chiu2015kernel}} dataset was created by the Vision and Image Processing (VIP) group at Duke University and is currently accessible to the public. The Duke-II was designed to address DME pathologies of varying severity levels, ranging from mild to moderate to severe. The dataset comprised 610 OCT scans obtained from a cohort of ten subjects diagnosed with DME.\\

\noindent \textbf{Duke-3 dataset \RVV{\citep{srinivasan2014fully}}} was also created by researchers in Duke University's VIP lab. Fifteen controls, fifteen participants with dry AMD, and fifteen with DME all underwent volumetric Spectralis SD-OCT scans. \\

\noindent \textbf{Rabbani dataset \RVV{\citep{rasti2017macular}}} consisted of 4,141 OCT images (including 50 normal OCTs, 48 OCTs of dry AMD, and 50 OCTs of DME) that were acquired at Noor Eye Hospital in Tehran. In this data set, the lateral and azimuthal resolutions were not uniform across patients, although the axial resolution is 3.5 m with a scan dimension of 8.9 7.4 mm2. Therefore, the number of A-scans varies between 512 and 768 scans, and the number of B-scans per volume varies between 19, 25, 31, and 61.\\

\noindent \textbf{BIOMISA dataset \RVV{\citep{hassan2018biomisa}}} was created by the lab at the National University of Science and Technology, Pakistan, for the purpose of investigating retinal layers, lesions, and identifying normal and abnormal retinal conditions like DME, CSR, AMD, and Glaucoma. There are a total of 5,324 scans from 99 people in the dataset, broken down as follows: 657 scans of dry AMD, 2,195 scans of DME, 407 scans of wet AMD, 1,161 scans of CSR, and 904 scans of normal eyes.\\

\noindent \textbf{AFIO \RVV{\citep{raja2020data}}} was created by the lab at the National University of Science and Technology, Pakistan; it contains OCT and fundus images. The images were captured using the TOPCON 3D OCT-1000 camera attached to an OCT machine. There are a total of 50 images in the dataset, including both normal and glaucomatous conditions. Each OCT image has an accompanying annotated fundus image. Glaucoma experts provided labels for the CDR measured from fundus images. The optic nerve head (ONH) is the focal point of OCT scans. An ophthalmologist manually annotated the ILM and the RPE.

\begin{table*}[t]
    \centering
    \small
    \caption{A detailed summary of the publicly available OCT datasets, Zhang \RVV{\citep{kermany2018identifying}}, Duke-1 \RVV{\citep{farsiu2014quantitative}}, Duke-2 \RVV{\citep{chiu2015kernel}}, Duke-2 \RVV{\citep{chiu2015kernel}}, Duke-3 \RVV{\citep{srinivasan2014fully}}, Rabbani \RVV{\citep{rasti2017macular}}, BIOMISA \RVV{\citep{hassan2018biomisa}}, and AFIO \RVV{\citep{raja2020data}} }
    \begin{tabular}{ccccccccccc}
        \hline
        Dataset &Scanner & Label& Total Scans& Subjects &  Drusen& DME & Normal& CNV& CSR &Glaucoma\\\hline
        Zhang  & Spectralis& Yes & 109,309 & -& 8,866 & 11,598 & 51,390 &37,455&- &-  \\

        Duke-1 & Bioptigen& No  & 38,400 & 384 & 26,900 & -&11,500&-&-&- \\

        Duke-2 & Spectralis & No & 610 & 10 & - & 610 & -  & -  & -&-  \\ 
        
        Duke-3  & Spectralis & No & 3,231&  45 & 723& 1,101 &1,407&-&-&- 

    \\

        Rabbani  & Spectralis & No & 4,142 &148 & 969 &862 & 2,311&-&-&-  \\
                
        BIOMISA  & Topcon & No  &  5,324 & 99 & 657 &2,195& 904 &407&1,161&- 

 \\

        AFIO  & Topcon & Yes & 50 & 26 & -& - & 18&-& -& 32   \\

        \hline
    \end{tabular}
    \label{tab:datasets}
\end{table*}

\section{Discussion and Future directions}
\noindent Researchers are increasingly investigating the genetic and genomic factors that contribute to the development and progression of eye diseases. Advances in genomics and personalized medicine may lead to new diagnostic and therapeutic approaches that can target specific genetic mutations or risk factors. Nanotechnology-based formulations are also gaining popularity. The usage of aqueous gels made from hydrogels is also getting attention for the treatment of a wide range of ocular disorders. The next generation of healthcare professionals needs to be trained to address SDOH in clinical care to improve clinical outcomes. \\
\noindent We have reviewed the most commonly used imaging technologies, which include CFP, FA, FAF, OCT, and OCTA, for detecting different ocular disorders. FA (Fluorescein Angiography) is an invasive technique that employs an intravenous dye injection to capture dynamic images of retinal circulation, blood flow, and leakage. FAF (Fundus Autofluorescence) detects natural autofluorescence emitted by molecules in the retinal pigment epithelium (RPE) to assess metabolic health and RPE integrity. These imaging modalities are valuable for diagnosing and monitoring a range of retinal diseases, including vascular disorders and RPE-related conditions. CFP is a non-invasive that takes pictures of the inside of the eye, including the retina, optic disc, and blood vessels. Fundus imaging provides a wide-angle view of the retina, making it useful for general screening and documenting retinal diseases. Fundus scans are very good at detecting significant changes in the retina, but they may not be able to see small changes. However, OCT is a non-invasive imaging technique that provides 3D structural analysis of the retina. OCT scans have higher resolution than fundus photographs, which means that we can analyze more minor changes in the retina. OCT's higher spatial resolution enables earlier detection of eye diseases when treatment is more likely to be efficacious. OCT  technique can be used to diagnose and monitor a variety of eye diseases, such as DR, AMD, DME, and glaucoma. OCTA is also gaining popularity, it provides visualization of retinal and choroidal blood flow and microvascular networks by combining OCT with motion contrast technique.  It allows for depth-resolved imaging, enabling the assessment of both superficial and deep retinal vasculature, perfusion patterns, identification of vascular abnormalities, and evaluation of capillary networks without the need for invasive dye injection or contrast agents. OCTA does not require the use of dye injection, making it more comfortable for patients and reducing the potential risks associated with invasive procedures. OCTA imaging can be repeated over time, allowing for a longitudinal analysis of changes in retinal blood flow. It also offers the potential for quantitative analysis, enabling the measurement of parameters such as vessel density and blood flow velocity. OCTA is often used to diagnose and monitor eye diseases such as diabetic retinopathy and macular edema. It's important to note that while OCTA has these advantages, it does not replace the structural information provided by OCT. Both imaging modalities have their unique strengths and applications, and they are often used together to provide a comprehensive assessment of retinal health. The choice between OCT and OCTA depends on the specific clinical question and the information needed for diagnosis and management. Fundus photography is currently widely employed by an ophthalmologist. Still, it is less commonly used than OCT and OCTA because it produces 2D scans that don't provide as much detailed information about the retina. FA is also still widely used, but it is less commonly used than OCT and OCTA because it is more invasive. FAF is not as widely used as the other techniques because it does not provide as much information about the retina and blood vessels. The choice of imaging modality depends on the specific clinical question, the suspected condition, and the expertise of the healthcare professional.
In many cases, multiple imaging modalities may be used in combination to provide a more comprehensive assessment of the retina and its associated structures. However, the limited availability of ophthalmologists and constrained accessibility to retinal image-capturing systems in developing nations hinder the timely identification of glaucoma. Current trends indicate a shift towards mobile-assisted screening systems for the detection of ocular diseases, as they offer greater cost-effectiveness and durability.

\begin{figure}[t]
	\centering
	\includegraphics[width=8cm,height=10cm,keepaspectratio]{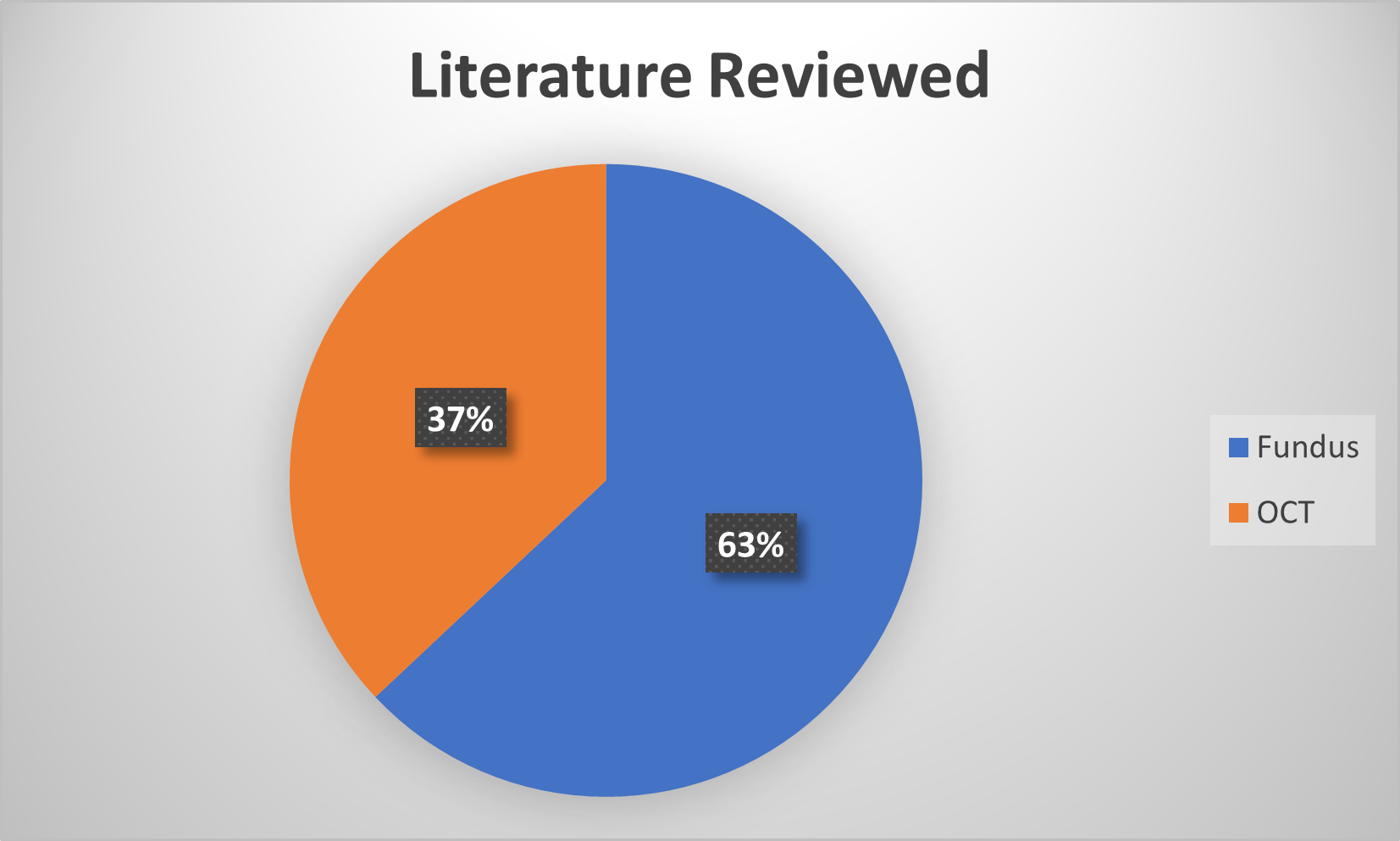}
	\caption{\small Distribution of fundus and OCT studies reviewed in this survey. Fundus-related studies have a higher publication frequency compared to studies based on OCT analysis.}
	\label{chart1}
\end{figure}

\noindent The identification and quantification of various biomarkers are required for the diagnosis and progression tracking of ocular diseases. Manual segmentation can offer high accuracy when performed by experienced specialists. They can precisely outline the boundaries of retinal lesions and differentiate them from normal tissues. However, manual segmentation is subjective and can vary among different observers. Automated detection algorithms, on the other hand, can provide consistent and reproducible results, reducing inter-observer variability. Manual segmentation can be time-consuming and labor-intensive, especially for large datasets or complex cases. Automated detection algorithms can process images rapidly, providing efficient and time-saving solutions for screening or large-scale analysis. Automated methods also allow for the potential integration of telemedicine and remote screening programs, improving access to care. The integration of manual segmentation into routine clinical practice can be challenging due to the time and expertise required. Automated detection algorithms can be integrated into existing imaging systems, facilitating their implementation in clinical workflows and making them more accessible for wider usage. However, automated methods may have some false positives or false negatives, their performance has been improving with advancements in AI. Numerous automated studies have been published exploring the utility of fundus photography and OCT in diagnosing and monitoring various ocular conditions. It was witnessed that a substantial number of published studies focusing on fundus studies compared to OCT studies. Figure \ref{chart1} presents the number of studies found in the literature concerning both OCT and fundus imaging for detecting ocular diseases, highlighting the comprehensive research conducted in these complementary imaging modalities. There is a large number of studies presented in the literature that identified the significant biomarkers and performed diagnosis of ocular disorders with high accuracy. The main areas of concern in fundus images analysis are inconsistent image quality, blurry or otherwise unclear backgrounds, and pixels that seem like blood vessels. Fundus image analysis is expanding into new research areas, including multimodal imaging fusion, combining genetic and clinical data, longitudinal studies, and incorporating other medical imaging modalities. These areas provide opportunities for further advancements and exploration. While research advancements are promising, the translation of fundus image analysis algorithms into clinical practice and widespread adoption may still be limited. Integration with existing healthcare systems and addressing regulatory and ethical considerations are crucial for practical implementation. OCT analysis is complex, as the images can be affected by various artifacts, such as motion artifacts caused by patient movement during the scan, shadowing artifacts due to tissue irregularities or structures in the path of the OCT beam, speckle noise, and axial or lateral resolution limitations. Correcting or mitigating these artifacts requires advanced algorithms and techniques.

\begin{figure}[h]
	\centering
	\includegraphics[width=8cm,height=10cm,keepaspectratio]{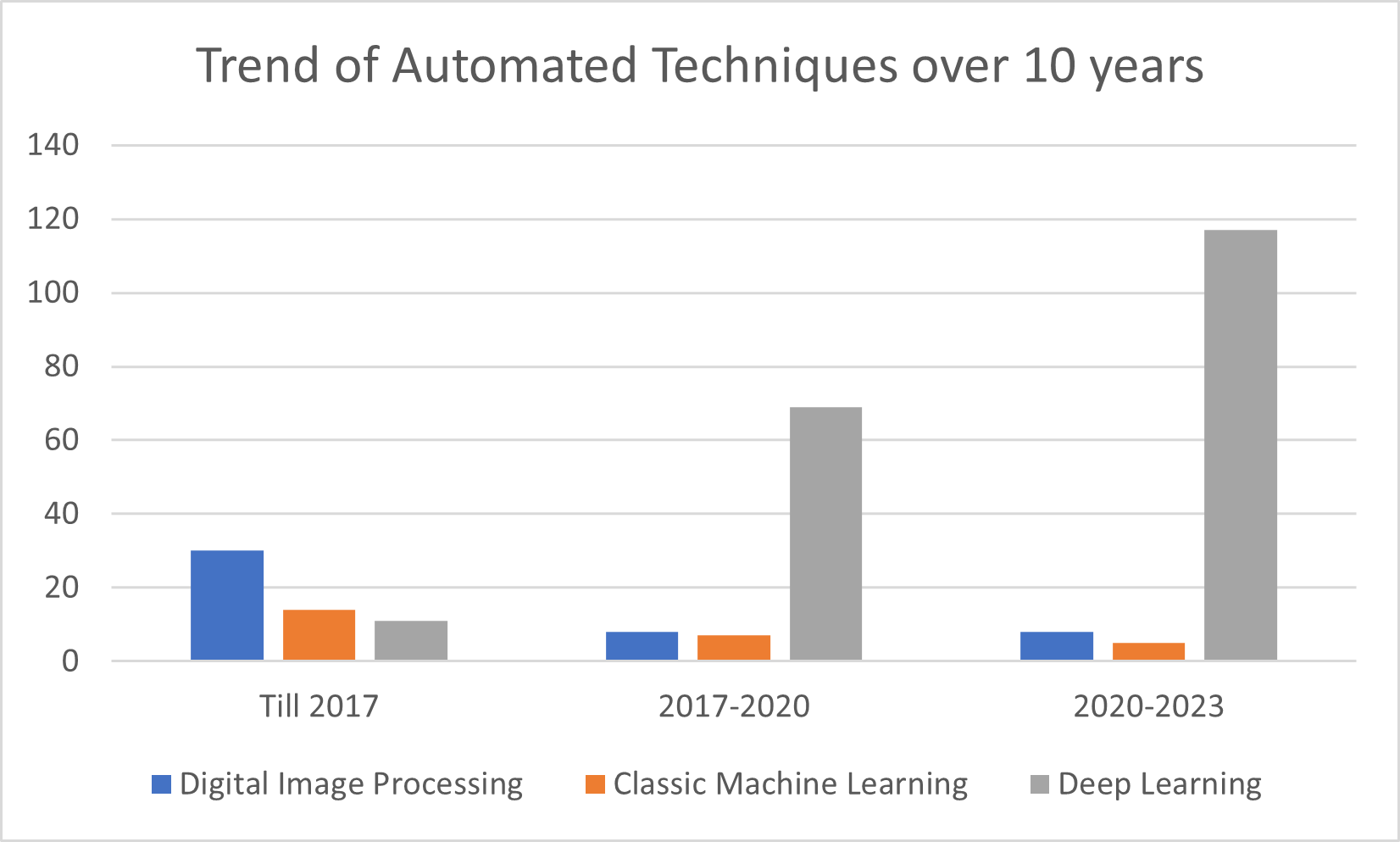}
	\caption{The chart illustrates the transformative trend in automated techniques over the last decade, showcasing the shift from traditional image processing methods to the adoption and prominence of deep learning approaches. It highlights the increasing reliance on deep learning algorithms for tasks such as the identification, quantification, and classification of biomarkers for various retinal diseases.}
	\label{chart2}
\end{figure}

\noindent In this comprehensive review, we thoroughly examined the utilization of digital image processing, classic machine learning, and deep learning techniques in the segmentation of retinal lesions and the classification of ocular diseases. Figure \ref{chart2} illustrates the distribution of different techniques employed in retinal analysis, specifically digital image processing (DIP), classic machine learning (CML), and DL methods. It visually represents the shift in the prominence of these techniques over time, with DL techniques gaining increasing popularity in recent years. We explored the advancements in each approach, highlighting their strengths and limitations.  One of the limitations of image processing techniques is that they require manual parameter tuning, which can be time-consuming and subjective. The selection of appropriate parameters may vary across datasets and researchers, leading to potential inconsistencies in the analysis. Image processing methods may not adapt well to variations in image quality, such as variations in contrast, illumination, or noise characteristics. These techniques often rely on predefined rules or assumptions that may not be universally applicable. Image processing techniques may struggle to handle complex cases, such as overlapping or intertwined structures, subtle abnormalities, or cases with severe ocular pathologies. So, ML-based techniques have gained popularity, but the have limitations in terms of manual feature engineering, which can be time-consuming and may not capture all relevant information in complex domains like retinal analysis. Furthermore, CML models may struggle with generalization to unseen or diverse data, leading to decreased performance in real-world scenarios. Additionally, CML algorithms are sensitive to the choice of hyperparameters and can be prone to overfitting if the model complexity is not properly controlled. In recent years, deep learning (DL) has emerged as a dominant force in the field of retinal analysis, surpassing other machine learning (ML) techniques in popularity and performance. One major advantage is DL's ability to automatically learn hierarchical representations from raw data, eliminating the need for manual feature engineering. Additionally, DL models can be trained end-to-end, allowing them to learn the entire task directly from input to output, without relying on intermediate steps or handcrafted rules. This streamlines the training and inference processes, making DL models more efficient. Furthermore, DL models have demonstrated superior scalability and performance for the segmentation and classification of retinal lesions. With ongoing advancements in the field of retinal analysis, DL is poised to revolutionize the future of diagnosis, treatment, and monitoring of retinal diseases, offering transformative potential for improved patient care and outcomes. Areas of focus include multi-modality fusion for a comprehensive understanding of retinal pathologies, real-time applications for timely diagnosis, transfer learning and domain adaptation for improved generalization, integration with electronic health records for personalized medicine, uncertainty estimation for risk assessment, and collaborative/federated learning for robust and secure analysis. These advancements have the potential to revolutionize retinal analysis, leading to better diagnosis, personalized treatment, and improved patient outcomes. 

\noindent Early detection and appropriate treatment are crucial for preserving vision and halting the progression of diseases. This review highlights the current trends in both clinical and technical domains regarding the detection of significant biomarkers for retinal diseases such as diabetic retinopathy (DR), diabetic macular edema (DME), age-related macular degeneration (AMD), and glaucoma. Imaging techniques utilized for the identification, diagnosis, and management of these diseases were also reviewed. Fundus imaging, optical coherence tomography (OCT), OCT angiography (OCTA), fundus autofluorescence (FAF), and fluorescein angiography (FA) are the most widely employed imaging techniques. Fundus imaging is commonly used for screening, while OCT modalities are preferred for diagnosis and monitoring disease progression. Angiography is gaining popularity in clinical practice for generating 3D angiograms to visualize retinal and choroidal vasculature pathologies. Each technique has its advantages and disadvantages, and the choice depends on the specific eye disease and the healthcare professional's judgment. Portable devices are also gaining attention due to the limited ophthalmologist-to-patient ratio, particularly in rural and developing areas. Research is now focused on enhancing the performance and robustness of these portable devices. Exciting directions in the field include adaptive optics, genetics and genomics, and telemedicine. Deep learning (DL) has emerged as a powerful approach, surpassing traditional methods by automatically learning complex features from data, resulting in improved accuracy and robust analysis of retinal images. However, it is important to note that digital image processing techniques are still relevant and can complement advanced analysis approaches like machine learning (ML) or DL by serving as a preprocessing step or providing feature extraction. Combining different techniques can leverage their strengths and mitigate limitations, leading to a more accurate and robust fundus and OCT image analysis.

\section*{Acknowledgements}
This work has been funded by Khalifa University, \RVV{ Ref: CIRA-2021-052, and ASPIRE, Ref: AARE-2020}

\section*{Statements and Declarations}
All the authors declare that there are no competing interests related to this article.

\end{document}